\documentclass[aps, prd, amsmath, floats, floatfix, onecolumn, nofootinbib]{revtex4-2}
\usepackage{graphicx}
\usepackage{latexsym}
\usepackage{color}
\usepackage{amsfonts,amsbsy}
\usepackage[utf8]{inputenc}
\usepackage[mathscr]{euscript}
\usepackage{ amssymb }
\usepackage{amsthm}
\usepackage{amsmath}
\usepackage[english]{babel}
\usepackage[T1]{fontenc}
\usepackage[utf8]{inputenc}
\usepackage{mathtools}
\usepackage{physics}
\usepackage{hyperref}

\newcommand{\be}{\begin{eqnarray}}
\newcommand{\ee}{\end{eqnarray}}
\graphicspath{ Fig 2\ }

\theoremstyle{definition}

\def\beq{\begin{eqnarray}}
\def\eeq{\end{eqnarray}}

\def\det{\,\mbox{det}\,}

\def\be{\beta}

\def\B1{ Bianchi-I }

\begin{document}

\title{Generalization of conformal Hamada operators}

\author{Les\l{}aw Rachwa\l{}}
\email{grzerach@gmail.com}

 \author{P\'{u}blio Rwany B. R. do Vale}
 \email{publiovale@gmail.com}

\affiliation{Departamento de F\'{i}sica - Instituto de Ci\^{e}ncias Exatas, Universidade Federal de Juiz de Fora, 33036-900, Juiz de Fora, MG, Brazil}

\begin{abstract}
The six-derivative conformal scalar operator was originally found by Hamada in its critical dimension of spacetime, $d=6$. We generalize this construction to arbitrary dimensions $d$ by adding new terms cubic in gravitational curvatures and by changing its coefficients of expansion in various curvature terms. The consequences of global scale-invariance and of infinitesimal local conformal transformations are derived for the form of this generalized operator. The system of linear equations for coefficients is solved giving explicitly the conformal Hamada operator in any $d$. Some singularities in construction for dimensions $d=2$ and $d=4$ are noticed. We also prove a general theorem that a scalar conformal operator with $n$ derivatives in $d=n-2$ dimensions is impossible to construct. Finally, we compare our explicit construction with the one that uses conformal covariant derivatives and conformal curvature tensors. We present new results for operators built with different orders of conformal covariant derivatives.

\end{abstract}

\maketitle

\section{Introduction}
\label{s1}

The theory of General Relativity (GR) is a successful relativistic
theory of gravitation that has given us a new understanding of space
and time, since then renamed as spacetime, and gravity in this setup
can be interpreted as the curvature of the latter. This general Einstein's
theory showed us that the spacetime itself may be viewed as having
many various unusual characteristics and also possessing specific
sets of symmetries, different from the symmetries present on the flat
space (Euclidean space or Minkowskian spacetime), which was earlier
usually considered as the background arena for all non-gravitational
physics. We are talking here about a collection of aspects and properties
of spacetimes which show that they are more flexible and dynamical
than previously thought of. Therefore, we can say in that theoretical
ensemble that ``gravity'' is a byproduct of the presence of such
a tangible and elastic spacetime, which by itself is a prominent object
in General Relativity.

A fundamental feature of this new understanding of space and time
is known as diffeomorphism invariance, which can be interpreted simply
as ``an invariance of the laws of physics between
distinct coordinate systems'' , which is also known
as covariance of these physical laws with respect to general coordinate
transformations. When the latter ones are assumed to be differentiable
mappings between different coordinatizations of the same manifold, then we speak about diffeomorphisms. The
symmetry with respect to arbitrary diffeomorphisms (arbitrary changes
of coordinate systems and reference frames) is typically taken as
the only symmetry of Einsteinian gravitational theory. However, this
symmetry is not the maximal one of a possible consistent relativistic
gravitational theory. The diffeomorphism symmetry of gravity can be
extended in many ways. One of the directions of such extensions is to
invoke another very important symmetry in GR, that is the conformal
symmetry. This is a symmetry and invariance (of a possible gravitational
theory and coupled to it matter theory) with respect to so called
conformal (or Weyl) transformations on the metric structure on the
differential manifold (and corresponding rescalings of matter fields
present additionally there). We will emphasize about such
symmetry below.

The conformal symmetry has a significant role in physics mainly due
to its applications that cover several areas, such as condensed matter,
statistical mechanics, high energy physics, cosmology and quantum
field theory. This theoretical
tool was first introduced and studied by Hermann Weyl \cite{Weyl:1917gp,Weyl:1919fi}. The conformal
(Weyl) transformation of the metric tensor
\begin{equation}
g_{\alpha\beta}\rightarrow e^{2\Omega\left(x\right)}g_{\alpha\beta}\label{eq: rescalings}
\end{equation}
produces another metric with respect to diffeomorphisms but the metric
being within the same conformal class (equivalence class with respect
to conformal rescalings). Above, the conformal parameter is represented
by $\Omega\left(x\right)$. The conformal rescaling of the metric as in (\ref{eq: rescalings}) in this form is valid for any spacetime of any dimension $d$. The transformations in (\ref{eq: rescalings})
scale the non-conformally invariant infinitesimal line element from GR according
to
\begin{equation}
ds^{2}\to e^{2\Omega(x)}ds^{2},
\end{equation}
while the coordinates $x^\mu$ and their differentials $dx^\mu$ remain untouched under such transformations. Among such changes in physical scales to measure length distances and time intervals,
there are not only global transformations, where the conformal parameter
is constant, i.e., $\Omega\left(x\right)\rightarrow\Omega$ (so called scale transformations), but
also local Weyl transformations\footnote{In this case the parameter $\Omega(x)$ may depend on both the location
in space and in time.}. A theory invariant with respect to this transformation is known
as conformally invariant (Weyl symmetric) or a theory with Weyl
invariance, or simply with Weyl symmetry. A special class
of theories, which preserve conformal symmetry on the full quantum
level is given by a set of conformal field theories (CFT's in short) \cite{DiFrancesco:1997nk}.
Such CFT's arose originally in some lower dimensional field theory
models of statistical mechanics \cite{Belavin:1984vu,Polchinski:1987dy}. Being the subset of general quantum
field theories (QFT) they come however with their own specific set
of methods and rules owing to the presence of fully realized conformal
symmetry, which strongly constrains the resulting quantum dynamics.
Additionally the formalism of general and abstract CFT's makes heavy
use of mathematical physics in the form of conformal algebras and
conformal operators (both primary, secondary and also of higher orders)
and also of conformal correlation functions. The construction of primary
operators of the CFT algebras as conformally covariant operators is
the central issue of any CFT, being conceptually quite different from
the main problem of a general QFT. The former is the problem that some
aspects of it for higher derivative scalar operators coupled to non-trivial
background geometry we want to discuss in this article.

It is understood in spacetime physics that upon the transformations
as proposed by Weyl in (\ref{eq: rescalings}), only angles remain
invariant, while sizes, magnitudes and scalar products between vectors
could change. Hence, due to these properties and their consistency
with relativity, the invariance with respect to conformal transformations
could be an additional symmetry fully consistent with the theory of relativistic
gravitation \cite{Rachwal:2018gwu}. In original Einstein's theory, the dynamics of the gravitational field does not enjoy conformal symmetry. Here, we would like to consider a possibility of adding the requirements of local conformal symmetry of the gravitational action in a slightly different theory. Such a symmetry is fully realized and embodied in the
gravitational theory known as Weyl gravity (conformal gravity) \cite{Weyl:1917gp,Weyl:1919fi,Weyl:1923yve,Weyl:1929fm,Bach:1922zz,Bach:1922z}, which in $d=4$ spacetime dimensions
is based on the square of the Weyl tensor in the gravitational action,
in opposition to the mere presence of the Ricci scalar in the action
of Einstein's gravity. This is why conformal gravity can be seen as
the first example of a different (or modified) classical gravitation,
but still being in agreement with main concepts and ideas of General
Relativity. Simply, in Weyl gravity the group of local gravitational gauge symmetries
is bigger than just of diffeomorphisms, because it is extended by
local conformal group factor.

However, Weyl's gravitational theory was forgotten for a while due
to the great success of Einstein's theory of gravitation, which only
includes symmetries in relation to the diffeomorphism transformations.
Along Einstein's theory's apogee, Weyl's theory was left out, for
being considered as a non-physical theory. Fortunately, as the time
passed by, Weyl's theory returned to the theoretical physics mainstream \cite{Deser:1970hs,Englert:1976ep,Kaku:1977pa,Fradkin:1981iu,Fradkin:1985am,Pawlowski:1994kq,Mannheim:1988dj,Barrow:1988xh,deBerredo-Peixoto:2003jda,Mannheim:2011ds,Jizba:2014taa,Jizba:2019oaf,Ghilencea:2018dqd}.
The return of conformal symmetry and conformal methods \cite{Fulton:1962bu,Penrose:1964ge,Hawking:1976fe,penrosebook}
occurred precisely in the general theory of relativity, later also
enriched in quantum field theory, where it became so much appreciated \cite{Deser:1970hs,Bekenstein:1974sf,Deser:1983rq}.
This curious fact happened mainly because of the applications of conformal
methods (but not so much of Weyl's theory) in gravity, high energy
physics and field theory; some of these applications exerted crucial
roles in the development of renormalization methods and its importance
to the renormalization group (RG) flows \cite{Jack:1990eb,Cappelli:1990yc} both in quantum field theories and also in models of statistical physics. Other recent developments in conformal gravity were done in \cite{Maldacena:2011mk,tHooft:2014swy,tHooft:2016uxd}, for black hole physics in \cite{Bambi:2016yne,Bambi:2017ott}, and also generally for the use of conformal methods to resolve spacetime singularities in \cite{Modesto:2016max,Bambi:2016wdn}.

Related to this usage of methods of conformal symmetry there were
also significant studies on conformal anomalies, which analysis of related quantum effects
can be found in \cite{Deser:1976yx,Deser:1993yx,Erdmenger:1997wy,Erdmenger:1996yc,Shapiro:1994st,deBarros:1997he,Asorey:2003uf}, and other applications
that follow on them are discussed in \cite{Carneiro:2004rt,Shapiro:1996rk,Rachwal:2018gwu}.
In particular, the conformal anomaly-generated gravitational effective
action in curved spacetime \cite{Shapiro:2001rh,Balbinot:1999vg}, is of considerable interest due to numerous
physical applications in black hole physics, cosmology, supersymmetry,
string theory and statistical mechanics (see \cite{Duff:1993wm} 
for the review) and also because of the hope to use such an action
as an insight for the full theory of quantum gravity \cite{Odintsov:1990rs,Antoniadis:1991fa,Deser:1993yx,Shapiro:1994st,Deser:1996na}.
The effective action for the gravitationally coupled conformal scalar
field \cite{Vilkovisky:1992pb,Barvinsky:1994hw,Barvinsky:1994cg,Barvinsky:1995it} is a nonlocal functional but all the
nonlocalities are related with the Green functions of a local conformal
operator acting on scalar fields. Previously, the cases of the second
order and the fourth order conformal operators were investigated for
this purpose in the literature \cite{Riegert:1984kt,Fradkin:1983tg,Buchbinder:1985ym}. The first time
two-derivative scalar operator with conformal coupling to background
geometry was considered in the works of Penrose in $d=4$ spacetime
dimensions \cite{Penrose:1965am,Penrose:1960eq} and later in a case of generalized
dimensions by \O{}rsted \cite{oersted}. In turn, the four-derivative
scalar conformal operator was constructed by Fradkin and Tseytlin
in their search for supergravitational theories in $d=4$ \cite{Fradkin:1982xc} and then
immediately generalized by Paneitz in 1982 \cite{paneitz}, and also independently
by Eastwood and Singer \cite{Eastwood:1985eh}. As remarked later by Branson \cite{paneitz} this, what is
now called Paneitz operator, played a significant role in various
applications in differential geometry, not only conformal differential
geometry \cite{branson0}, in $d=4$ and also in higher dimensions \cite{chang} and both in Euclidean  \cite{kengo}
and in Minkowskian \cite{Park:2010ksh} signature of the metric. This significance of conformal scalar operators in differential geometry and theoretical physics is visible at the present moment too.

A great example about the applications of conformal transformations
on operators with higher derivatives can be seen in \cite{Shapiro:1994ww},
which shows us briefly that such operators are useful to investigate
quantum gravitational effects. There are also other applications as
well: with an ability to provide consistent renormalization of quantum
divergences \cite{Meissner:2007xv}, an opportunity to use group renormalization methods to
study asymptotic behavior of couplings on curved backgrounds \cite{Asorey:2005hk}, and finally with a possibility
of building candidate Grand Unified Theories (GUT) models of particle
physics with higher derivative operators \cite{Pawlowski:1994kq}. 

Conformal symmetry and conformal methods are also ubiquitously  found
in condensed matter physics \cite{Ginsparg:1988ui,Ludwig:1992tm}, where, for example, conformal field theories
are used to solve interaction problems in many-body physics \cite{DiFrancesco:1997nk}. The formalism
of CFT's offers a possibility to describe a large set of quantum field
theories reaching a non-Gaussian fixed point (FP) of renormalization group
flows that can be moreover solved exactly around such FP's. However,
in condensed matter not all of the tools of general CFT's are often exploited since
the general and abstract methods of CFT were developed in a slightly
different context (of purely mathematical models of 2-dimensional
field theories, cf. \cite{Riva:2005gd}). In \cite{Ludwig:1992tm}, for example, a well known Tomonaga
theory in the Kac-Moody CFT language is used to show how the properties
of Fermi surfaces can be understood on the purely algebraic CFT base.

In general, CFT's are used to describe the situation of RG flows in generic QFT's with running coupling parameters near a fixed point
of renormalization group flow. This opens up for vast applications of the CFT's and conformal operators in theoretical high energy physics. There, at the fixed point (FP), where the RG flow stops one
surely meets scale invariance which can be easily promoted to full
conformal invariance on the quantum interacting level \cite{Iorio:1996ad}.  In such circumstances,
the reason for quantum RG running of couplings is absent, namely there
are also no perturbative UV divergences and the theory is completely
UV-finite. In general, it is quite difficult to find models of such
finite interacting theories when the gauge or gravitational interactions
are included consistently on the quantum level, but one can consult
the examples found in \cite{Modesto:2014lga,Modesto:2015lna,Modesto:2015foa,Modesto:2016max,Koshelev:2017ebj,Rachwal:2018gwu}. As one can see there the
absence of UV divergences is closely related to the quantum conformality
and the theory which sits at the FP of RG can be cast in the form
of a CFT (possibly on a curved gravitational background like studied in this paper). When the FP
is non-trivial and the theory around it is still with non-vanishing
interactions, so this is a non-Gaussian FP, simply due to the mere fact
of its existence, the theory is said to be asymptotically safe since
it does not have anymore problems with UV divergences \cite{weinberg}. The values
of couplings are non-zero at such FP, but still the theory is finite and without problems in the UV regime, that is it can be defined at arbitrarily high energy scales. In other words, conformality is reached asymptotically in the UV.

One can also see that the correlation functions of quantum fields
in the vicinity of FP show the features of complete scale invariance.
A convenient description of these theoretical phenomena is also in terms of anomalous dimensions of relevant
operators defined near the FP. They physically provide the tools for the analysis of
critical exponents in the broad area of physics of critical phenomena
usually studied in the IR regimes \cite{Weinberg:1976xy}, while in QFT models of fundamental interactions these are usually anomalous dimensions of operators defined in the vicinity of UV FP's of RG flows. Moreover, it is known in statistical physics that the particular microscopic
(UV) details of the models are irrelevant for the situation near IR
FP's, which are used to describe critical phenomena as studied in condensed
matter since some different physical models might belong to the same universality
classes each one characterized by one and the same IR FP, where all the correlation
lengths diverge. Similarly, in the QFT domain of applicability different
QFT models may be in the same UV universality class from the point of view of the UV FP
of quantum RG flows \cite{Rummukainen:1998as,Kallosh:2013hoa}, regardless of their distinguishing infrared details (like mass deformation, etc.). There in QFT, the fact that correlation lengths diverge
is a consequence of the presence of conformal symmetry in the UV regime.

When the CFT
description of the theory near UV FP is obtained (with central charges of CFT algebras and with the set of
primary conformal operators of these algebras, their commutation relations, their spectra,
and also their anomalous scaling dimensions or conformal weights, so in total with all CFT data),
then one can construct a new perturbation calculus around the UV FP, so called conformal
perturbation theory \cite{henkel}. In it one deforms the conformal symmetry by adding new deformation operators or by giving vacuum expectation values to some non-conformally invariant quantum fields, and then one also
allows for perturbation of the conformal structure and an RG running
of new couplings in front of these new deformation operators. These operators have to be relevant deformations from the point of view of the UV FP, must be built out of conformal primary operators of the CFT, and moreover are required to be such that the RG flow is back generated
when one goes down with energies away from the UV FP. However, to achieve this formidable
task, first one needs to have a full description of the situation
at the UV FP, including also the description of all primary conformal
operators. This is why the study of higher derivative scalar conformal
operators\footnote{By coupling matter fields to dynamical quantum background geometry
one sees that higher derivative gravitational terms are naturally
generated \cite{utiyama} in spacetime dimensions $d>2$, hence one also has to
include higher derivatives when coupling scalars to gravity.} as the basic ingredients for the construction of UV FP CFT algebras containing operators with higher derivatives
coupled to background geometry is an essential initial task in this
ambitious programme. These CFT's are obviously in strong interest to any high energy theorist.

The six-derivative scalar operator with conformal properties was originally
found for physical applications related to conformal anomaly in $d=6$ spacetime dimensions in 2001 by Hamada in \cite{Hamada:2000me}. In this work, Hamada analyzed
how the integrability conditions for conformal anomalies constrain
the possible form of the effective action for an even-dimensional
differential operators. His analysis first showed that these integrability
conditions are satisfied for an action that is constructed using a
four-derivative scalar conformal operator in $d=4$ spacetime dimensions (the Paneitz operator).
The general form of this operator makes manifest diffeomorphism invariance,
renormalization and regularization scheme-independence. Additionally
its coefficients are chosen such that the anomaly vanishes and the
action of the higher-derivative scalar field coupled to background gravitational field (background geometry) is fully conformally invariant. These results were crucial for further developments in \cite{Hamada:2002cm,Hamada:2003jc}. In \cite{Hamada:2000me} Hamada later generalized these
arguments to a new six-derivative scalar operator considered exclusively
in its critical dimension, namely in $d=6$. 
(As acknowledged by Hamada in his papers \cite{Hamada:2000me}, the first attempts to present such operators with six derivatives were also presented in \cite{Karakhanian:1994yd}.) This is the reason why in this paper we will call a six-derivative scalar conformal operator and acting on scalars by the name of Hamada operator.

\subsection{Scope and the plan of the article}
\label{s1a}

In this work we will study the conformal properties of scalar operators
with six derivatives in general dimensions $d$ of spacetime. Our
results will be applicable both in the Euclidean (space) and also
in the Minkowskian domain (spacetime) of the metric signature. The
first case besides the obvious applications in conformal Euclidean
differential geometry of curved manifolds will also be useful in condensed
matter theory to describe conformal behaviour of some higher-derivative
scalar quasi-particle excitations coupled to geometries of effectively
higher-dimensional spaces. We expect this situation to be present
in the close vicinity of the RG fixed points, so where the critical
phenomena occur. Our results can be applied, for example, to phonon
excitations propagating on a curved graphene 2-dimensional sheets
moving and deforming themselves in time \cite{Iorio:2013ifa} or to some analogue (Euclidean) gravity
models with the presence of scalars and of Weyl symmetry \cite{Iorio:2010pv,Iorio:2012xg}. In the Minkowskian case of the
signature, such a conformal scalar operator gives natural answers
to a problem of conformal coupling to gravity for a scalar field whose
kinetic operator on the flat spacetime contains higher derivatives, in our case precisely six derivatives.

Conformal invariants (conformally invariant tensors, conformally covariant differential operators, conformal holonomy groups, etc.) are of central significance in differential geometry and physics. Well-known examples of such operators are the Yamabe-, the Paneitz-, the Dirac- and the twistor operator. Our aim is to present the basic ideas and some of the recent developments around conformal operators and conformal holonomy. The part on Q-conformal curvature \cite{bookjuhl} has its origin and its relevance in geometry, spectral theory and high energy physics \cite{bookbaum}. Here the influence of ideas which have their origin in the AdS/CFT-correspondence becomes clearly visible. For example, the conformal holonomy describes recent classification results, its relation to Einstein metrics and to conformal Killing spinors, and related special geometries, which are still very important in high energy physics. Some interesting applications of this to theoretical high energy and discussions of Weyl and conformal symmetries are contained in the review paper \cite{Karananas:2015ioa}.

We will propose a general form of conformal scalar operators in general dimensions $d$ with arbitrary a priori undetermined coefficients. The tensorial structure of the terms in the expansion of the operators will be kept fixed, while we will search for values of these numerical coefficients by demanding full conformal covariance of these operators when acting on a scalar field with properly assigned conformal weight under Weyl transformations in the GR setting.
Of course, to check the conditions for conformality of operators, we
will need to investigate their transformations under conformal rescalings
(\ref{eq: rescalings}) since they will be built from metric tensors, covariant derivatives and gravitational curvature tensors, which are all functions of the metric as a basic atomic building element, which transforms as in (\ref{eq: rescalings}). We will especially make use of a method of
infinitesimal conformal variations because it is well known that one has to check the conformal invariance only to the first linearized order in the conformal parameters. In this procedure, we consider
conformal parameters (parameters $\Omega(x)$ of conformal transformations
in (\ref{eq: rescalings})) to have values very close to zero, so
the difference of $\exp(2\Omega(x))$ from $1$ is perturbative and very small. This lets us to consider
variations as linear operators and hence the problem of finding a
generalized form of conformal operators with higher derivatives in
$d$ dimensions will be reduced to an algebraic one of solving some linear system of equations for coefficients of expansion of operators in some basis of GR-invariant terms.

Although the second and fourth order in derivatives scalar conformal
operators in their generalized forms in any dimension  $d$ (\O{}rsted and Paneitz
operators respectively) were already reported in the physics literature \cite{oersted,paneitz,Fradkin:1982xc,deBarros:1997he},
we will analyze them again. This is because they have interesting
features that cannot be left behind, for example, their singular dependence
in coefficients of expansion in GR scalar terms on the number of dimensions
$d$ and the presence of some critical dimensions $d_{{\rm crit}}$
related to the order of derivatives. In particular, we will emphasize a known and clearly visible fact that in even dimension of space(time) smaller than $d_{{\rm crit}}$ such respective operators are not constructible.  Moreover, these conformal operators are useful building
blocks in order to construct a generalized form of the Hamada operator
(which we call an operator with six derivatives acting on scalars
and transforming covariantly under general conformal transformations)
in any dimension $d$.

This article is organized as follows: in section \ref{s2}, we present a brief
review of the main peculiarities and properties of conformal transformations as standardly used in GR. We also complement this section by a short review there of the \O{}rsted and Paneitz conformal operators analyzed in any dimension of spacetime.
In the next section (section \ref{s3}), we discuss the Hamada operator, first in $d=6$, following the original work of Hamada.  In the same section,
we also intend to clarify the need for the presence of some additional terms (higher in curvatures) in the construction of generalizations for both the Paneitz (away
from $d=4$) and Hamada operators (away from $d=6$).
This step is crucial in correctly understanding the listing of all basis elements for the construction of the Hamada
operator in general dimensions $d$, which we undertake in the next
sections.
Namely, in section \ref{s4}, we present a general method of derivation, based on the complete basis of GR-invariant terms but with undetermined coefficients, first concentrating on global
conformal transformations and their consequences and later on local infinitesimal transformations. Finally,
at the end of this section, we present all coefficients of the generalized Hamada operator $\Delta_6$ in arbitrary dimension
$d$ of spacetime. This gives us the main result of the paper. In the next main section \ref{s5}, we reconsider the original Hamada operator in $d=6$ and
we point out the singularities in its construction in dimensions $d=2$ and $d=4$. We also indicate there
what are the possible dimensions $d$ in which the operators like $\Delta_4$ and $\Delta_6 $ (generalized Paneitz and Hamada respectively) become
conformally covariant and in what dimensions the coefficients of expansion
of these operators have irremovable singularities. Furthermore, to strengthen these conclusions, we generalize these observations
and in the last part of section \ref{s5}
 we prove a theorem that a general operator $\Delta_n$  cannot be constructed in dimension $d=n-2$.
 In the second part of the paper, first we discuss thoroughly the construction of conformal covariant derivative and corresponding covariant box operator, this is in section \ref{s6}. We apply it to find the generalization of the Hamada operator written with conformal curvatures and derivatives and for this the expression due to Wunsch is much shorter. As a part with completely \emph{ new} results we also present section \ref{s7} when we use a new construction method with different orders of conformal derivatives.
 Finally in section \ref{s8} we compare all three methods for finding generalization of conformal Hamada operators.
In section \ref{s9} we draw our conclusions.
We also there give some historical remarks about Hamada operators.

\section{Conformal transformations}
\label{s2}

In this short section, we discuss our conventions about conformal transformations of the covariant metric tensor of spacetime and its derived quantities. We also postulate the conformally covariant transformations laws for scalar fields present in the theory. This finally lets us to show the simplest conformally covariant scalar operators when they act on corresponding scalar fields with proper conformal weights, namely we will show here the \O{}rsted and Paneitz operators. Here, we keep the dimension $d$ of spacetime and the signature of the metric in it completely arbitrary.

To set our convention we specify the conformal transformations on
the covariant metric tensor as
\begin{equation}
g_{\mu\nu}\to g'_{\mu\nu}=e^{2\Omega}g_{\mu\nu},
\label{conftrmetric}
\end{equation}
where $\Omega=\Omega(x)$ is a local (spacetime dependent) parameter
of conformal transformations \cite{Wald} in the GR framework for performing conformal (or Weyl) transformations. Under these transformations the infinitesimal
length element of the physical spacetime transforms as
\begin{equation}
ds^{2}\to ds'^{2}=e^{2\Omega}ds^{2},
\end{equation}
so the spacetime distances are not anymore invariant. However, the
angles remain conformally invariant. It is required that the conformal parameter function $\Omega(x)$ is everywhere real-valued, hence the exponential factor $\exp(2\Omega(x))$ in (\ref{conftrmetric}) is always positive-definite. This is important since then under general conformal transformations the signature of the metric tensor of space(time) remains invariant and also the metric never becomes degenerate. From this basic conformal transformation law in (\ref{conftrmetric}), one derives the corresponding transformations for the other objects like the contravariant (inverse metric), metric density, Levi-Civita covariant derivatives and finally various curvature tensors (Riemann tensor, Ricci tensor and Ricci scalar). We will not need explicit formulas for the conformally transformed versions of these objects, but one can consult, for example, the textbook \cite{Wald} as a useful reference for them. We just here for future convenience notice the conformal transformation law of the fully covariant Weyl tensor (tensor of conformal curvature) with all four spacetime indices covariant, which in arbitrary dimension $d$ of spacetime (where the Weyl tensor can be defined, so in $d\geqslant3$) reads,
\begin{equation}
C_{\mu\nu\!\rho\sigma}\to C_{\mu\nu\!\rho\sigma}'=e^{2\Omega}C_{\mu\nu\!\rho\sigma}.
\end{equation}
From the above formula, one explicitly sees that this tensor transforms covariantly (with a specific conformal weight that can be assigned to it) under general conformal transformations. This is a notable exception among other ordinary curvature tensors as considered in differential geometry. Hence the role of Weyl tensor for conformal transformations and for conformal methods in geometry is very significant. One can even consider a conformally invariant  GR-tensor $C_{\mu\nu\!\rho}{}^\sigma = g^{\sigma\kappa}C_{\mu\nu\!\rho\kappa}$ and this example proves that the position of indices (whether they are covariant or contravariant) on a general tensor in GR matters for an assignment of proper conformal weight, if such definition is possible. 

When the scalar field $\phi=\phi(x)$
has assigned conformal weight $w\in\mathbb{R}$, then it changes according
to
\begin{equation}
\phi\to\phi'=e^{w\Omega}\phi.
\end{equation}
The spacetime dependence of the scalar field $\phi(x)$ is not changed by conformal transformations, so for simplicity further we will omit writing it explicitly. The conformal weight $w$ of the scalar field is constant in a given model provided the dimensionality $d$ of the spacetime is fixed. Actually, as we will show by explicit computation from reasons of simple global scale invariance of the scalar models, this weight can only depend as a simple rational function on the general dimension $d$ of spacetime (it is a polynomial of the degree 1 in $d$ parameter). Moreover, we call the case when $w\neq 0$ as conformally covariant scalar field, while only in the special case of $w=0$ we can speak of conformally invariant (or inert) scalar field. As we will see in section \ref{s5}, this special case for conformally inert scalars is on the boundary of the regime where we cannot define conformal scalar operators acting on such scalars. Finally, here we also emphasize that for different scalar models (characterized by a different number of derivatives in the kinetic term on the flat spacetime background), but still considered within the spacetime of the same fixed dimensionality $d$, the conformal weights for the scalars are different since a priori these are different and unrelated scalar fields, and for example, their energy dimensions are also different.

One sees that the covariant metric can be assigned a weight $w=2$,
while its inverse (the contravariant metric) must then have $w=-2$. This must be in such a way that the Kronecker delta tensor (the metric with mixed position of indices) $\delta^\mu_\nu$ remains completely invariant, with $w=0$, and absolute object of any geometry.
One also notices the transformations of the relativistic spacetime
density measure
\begin{equation}
\sqrt{|\det g_{\mu\nu}|}\underset{{\rm df}}{=}\sqrt{|g|}\to\sqrt{|g|}'=e^{d\Omega}\sqrt{|g|}
\end{equation}
in spacetime of dimension $d$, while the integration volume element
in coordinates $d^{d}x$ remains conformally invariant. This means that the density measure (being the square root of the absolute value of the determinant of the covariant metric tensor $g_{\mu\nu}$) $\sqrt{|g|}$ may be assigned a conformal weight $w=d$.

In the infinitesimal form (for $\Omega\ll1$), these conformal transformations of the basic building blocks of geometrical GR-covariant tensors and covariant differential operators read,
\begin{equation}
g_{\mu\nu}\to g'_{\mu\nu}=(1+2\Omega)g_{\mu\nu},
\label{f8}
\end{equation}
\begin{equation}
g^{\mu\nu}\to g'^{\mu\nu}=(1-2\Omega)g^{\mu\nu},
\end{equation}
\begin{equation}
\sqrt{|g|}\to\sqrt{|g|}'=\left(1+d\Omega\right)\sqrt{|g|},
\label{densitytrlaw}
\end{equation}
\begin{equation}
C_{\mu\nu\!\rho\sigma}'=(1+2\Omega)C_{\mu\nu\!\rho\sigma},
\end{equation}
\begin{equation}
\phi\to\phi'=(1+w\Omega)\phi.
\label{f12}
\end{equation}
The transformation rules for curvature tensors or other covariant
differential operators are more complicated -- they might contain also derivatives
of the parameter $\Omega=\Omega(x)$ and we will not need to write
them explicitly here. The formulas in (\ref{f8}) to (\ref{f12}) express the forms of infinitesimal conformal variations of the objects on the left hand sides of these equations. In later sections, we will make an extensive use of the formalism of conformal variations of various tensors and differential operators of geometry since for conformal invariance or covariance we need to check only to the linearized level in the small parameter $\Omega(x)$ and hence only infinitesimal conformal variations settle the issues with conformal symmetry and good transformation properties under its corresponding transformations.

\subsection{Conformal operators and conformal actions}
\label{s20}

Below, we will present the final results for the construction of two- and four-derivative scalar conformal operators acting on scalar fields present on the spacetime manifold and coupled to it in the conformal way. We remark that conformal operators $\Delta$ when they are conformally covariant and when they act on conformally covariant scalar fields $\phi$ with properly assigned weights, give rise to another scalars $\Delta \phi$, which for conformality must also transform as conformally covariant objects. Here, the crucial thing is that the weight $w'$ for the new composed object $\Delta \phi$ will be almost always necessarily different than the original weight $w$ of the scalar field $\phi$. Therefore, the general action of conformally covariant differential operator on some conformal scalars is characterized and adjusted by specifying two constant numbers: $w$ and $w'$ for any type of conformally covariant differential operator. One could say that only in the rare case, when $w'=w$, we can tell that the differential operator is conformally invariant since then it does not add anything to the weight of the scalar field it acts upon on the right. In general, the weight of the differential operator $w_\Delta$ could be defined (following \cite{Erdmenger:1996yc,Erdmenger:1997wy}) as the difference
\begin{equation}
w_\Delta=w'-w.
\end{equation}
This definition works also for endomorphic operators, that is such ones that they do not have any differential character and they do not contain any derivatives, nor partial nor covariant ones in their definitions, and in that case, it coincides with the standard definition of the natural conformal weight of a conformally covariant tensor. One assumes judiciously here that in the case of a product of two conformally covariant objects, let them be tensors, scalars or differential operators, the product is also an object that transforms covariantly under the conformal rescalings and that its conformal weight is equal to the sum of weights. This additivity of conformal weights is a natural property like the additivity of charge for $U(1)$ gauge transformations since also conformal transformations have some formal similarities to the Abelian one-parameter gauge transformations. One can prove this additive relation between conformal weights of tensors in a product by exploiting the linearity and the Leibniz rule for the infinitesimal conformal variations of products. For the case of action of differential operators on some tensors, which could be viewed as a ``product'' only by going to suitable momentum space, we adopt this rule and actually based on this we define the weights of various conformally covariant differential operators as the differences like $w'-w$ as used above.

Having at our disposal a conformally covariant operator with the conformal weight $w_\Delta$ defined above and acting on precisely one tensorial field on the right, one can construct also a conformally invariant action functional. One then treats the conformal operator as the kinetic term for the tensorial field coupled non-trivially to a background geometry. This way of coupling to the external gravitational field is non-minimal, but it is also conformal from the definition of the conformally covariant operator. In order to finish the explicit construction of the action, one uses the conformally covariant object schematically denoted as $\Delta\phi$, which is linear in the scalar field $\phi$ and couple it on the left hand side with the conjugate tensor field. For the uncharged scalar field case, that we consider in this work, the conjugation means here no operations, so we just multiply by the same scalar field evaluated at the same spacetime point $\phi=\phi(x)$. In the more general case of various tensorial representations of various internal symmetry groups, one has to conjugate this tensorial field on the left in the charge conjugation sense, internal indices sense, and also use the tensors with the complementary positions of all spacetime and Lorentz indices present on the field. Or in other words, one has to exploit the form of the kinetic term for the field $\phi$ and naturally quadratic in this tensorial field, as it is done in a standard manner in any QFT model with these tensorial fields. The Lagrangian built in such a way is already a conformally covariant GR scalar. When it is properly densitized by adjoining the factor of $\sqrt{|g|}$, then one gets a conformal invariant  and a GR scalar density. Finally, after integrating over the whole spacetime points of the manifold (with the conformally invariant coordinate integration element $d^dx$) this yields an action functional which also enjoys the same properties of being fully conformally invariant and also of being a nonlocal scalar action functional from the point of view of GR. Such action functional written in the schematic form as
\begin{equation}
 S=\!\int\!d^dx\sqrt{|g|}\phi\Delta\phi
 \label{action}
\end{equation}
can be chosen as well as the basis for studies of conformal invariance properties of the field theory model and of the differential operator in question. This is what we are going to use in some further parts of this paper. Hence one can study various conformally covariant differential operators or related to them conformally invariant action functionals. The fact that the operator $\Delta$ is conformally covariant and that it gives rise to the action (\ref{action}) implies that it also possesses many nice and interesting features. We can already mention one of them here since this operator automatically has to be self-adjoint with respect to the conjugation and the operation of integrating by parts under the volume spacetime integral in (\ref{action}). This also means that its action on the right scalar field $\phi$ (canonically) must give the same results as its conjugate action on the left scalar field $\phi$ in (\ref{action}) for the case of scalar fields, when conjugation boils down to no operations at all. This operator $\Delta$ arises also naturally as the second variational derivative of the action $S$ in (\ref{action}) with respect to scalar fields $\phi$, here without the need of additional self-adjoining procedure (such latter operation is needed for ordinary operators not being conformally covariant, for example). Only the self-adjointness property in such a form allows us to have both conformally covariant operator $\Delta$ and conformally invariant action $S$, quadratic in scalar fields $\phi$ and with the same operator $\Delta$ in it (as in (\ref{action})), at the same time.

It is also possible, in principle, to consider operators that involve non-linear actions on conformally covariant tensorial fields as their arguments, they even do not have to be the same tensorial fields. Such complication is actually quite easy to tackle on the level of differential geometry considerations. On the level of the corresponding QFT models, this is related to the non-trivial interactions between various conformal fields of the model and not only with non-trivial interactions of these fields with gravitational background geometry. We will comment briefly on these possibilities in section \ref{s5} for the case of scalar fields. However, here one has to take into account that from the perspective of QFT the addition of such non-trivial interactions is possible but then the weights of the scalar fields participating in such interactions must be determined firstly and solely from the kinetic term.  The considerations of the conformal invariance of the action as constructed in (\ref{action}) with only the kinetic (quadratic) term for the scalar field $\phi$ are the most relevant here since the kinetic term is the most important one and this is  because it contains the smallest possible number of powers of fields, when we also neglect tadpole diagrams.

In the light of what we have said above, in a consistent conformally symmetric theory coupled to background gravitational field, it is possible to have only one unique even number of derivatives in the kinetic term on the flat spacetime background, for example, for the scalar field, at the same time. So, for example, it is forbidden to have simultaneously, in one model, \O{}rsted and Paneitz operators for one and the same scalar field $\phi$. This is already due to scale invariance requirement, so from the considerations of the energy dimensions of the scalar fields. The same restriction will apply to non-linearities  and basically only non-linearities containing less derivatives than in the leading kinetic term of the model will be allowed. One does not allow here cases with negative or rational, fractional powers of the covariant derivatives and also cases of the same pathologies of exponents for scalar fields -- this is due to requirement of locality of the Lagrangian. It is perhaps possible to add some non-linear interaction terms with derivatives, but we have not yet checked this option. Still, one can very easily add non-linearities containing no derivatives at all and corresponding to the very specific powers of the scalar fields or equivalently to the form of local conformal interaction vertices with no momentum dependence in the language of QFT. Then in the last case there is no issue with their generalization to the curved spacetime background since the latter is easily achieved by just supplying the $\sqrt{|g|}$ factor for conformality of the resulting Lagrangian density, because that part of the interaction Lagrangian does not contain derivatives at all and all the fields present in the interaction vertex transform conformally covariantly. One is sure that on the flat spacetime background the non-linear interaction terms with derivatives exist and are there scale-invariant, however the question is really whether there exists a suitable generalization of such terms to the case of non-trivial gravitational backgrounds. If this is possible, we can add such terms as well to the final expression for the conformally invariant action functional of the theory, which will now include also non-trivial but conformal self-interaction terms of the scalar field $\phi$, possibly also with some number of background covariant derivatives acting between these fields.

However, as it should be clear from the preceding analysis, this issue with non-linearities is in no relation to the main issue of this paper, namely of constructing the kinetic operators so such that they act precisely on one power of the scalar field $\phi$ on the right. This is because for conformal covariance of the latter operator one has to analyze only operatorial terms which act linearly and on one scalar field $\phi$, and non-linearities here do not play any role. Simply, the propagator of quantum modes for the field $\phi$ on a curved background, and in particular on flat Minkowskian background, depends only and exclusively on terms which are quadratic in the field $\phi$ on the level of Lagrangian and higher non-linearities do not matter for this issue. This is what also conformal coupling to background geometry repeats here, namely the structure of the kinetic term describing propagation of modes of the field $\phi$ and this is what needs to be only checked here for conformal covariance properties of the latter operator. When one couples the kinetic term describing the propagation of modes of the fields $\phi$ to background geometry, one can resort to the  construction of a new conformally covariant operator which only describes a propagation on this non-trivial background and consistently with the requirements of conformal symmetry and at the same time the issue of higher interactions or self-interactions of the scalar field $\phi$ may be for this purpose completely omitted. In other words, one does not need to include such non-linearities  for the main parts of this paper, where we show explicitly the construction for conformal kinetic operators acting between two scalar fields on the level of the action functional, like in (\ref{action}).

The examples of the kinetic scalar conformal operators when coupled to external geometry give us explicit results for the conformal coupling to background geometry procedure. This is not a minimal gravitational coupling, and so it may not exist for any operator that one can consider on the level of flat spacetime theory, without intervening gravity. The question whether such conformal covariantization from flat spacetime to a curved background exists and whether it is successful in all interesting us cases is the main scientific question that we want to address here. Having discussed general issues of the construction of conformal scalar operators in the form of suitable kinetic operators coupling the scalar fields in a conformal way to background geometry, now we can show the details of the construction of \O{}rsted and Paneitz operator. We achieve this in the two subsections below. We adopt the notation here that the operator with $n$ derivatives in the leading kinetic term, which is a term which comes with no gravitational curvatures (so equivalently as it is considered on the flat spacetime background), is denoted by the symbol $\Delta_n$. 

\subsection{\O{}rsted operator $\Delta_2$}
\label{s2a}

Two-derivative scalar conformal operator\footnote{Note that this operator is a two-derivative Klein-Gordon kinetic operator (acting on the scalar field) non-minimally
coupled to a Ricci scalar $R$ of the background geometry and without
the mass since a constant mass term would generally break the conformal symmetry. (But actually here one can treat effectively the Ricci scalar $R$ as a possibly spacetime-dependent mass term for the scalar field $\phi$). This constitutes the case of standard conformal coupling of a scalar field to background gravitational field.} $\Delta_{2}$
\begin{equation}
\Delta_{2}=\square-\frac{d-2}{4(d-1)}R
\label{orsted}
\end{equation}
was first studied by Penrose in $d=4$ and by \O{}rsted in general dimensions $d$ and gives the scalar action
\begin{equation}
S=\!\int\!d^{d}x\sqrt{|g|}\phi\Delta_{2}\phi
\label{orstedaction}
\end{equation}
conformally invariant if the conformal weight of the scalar field
is given by
\begin{equation}
w=\frac{2-d}{2}.
\end{equation}
Such a two-derivative operator in (\ref{orsted}) can be defined in any spacetime dimension
$d>1$. It can be easily found here that $w_{\Delta_2} = -2$, so in the result $w'=-\frac{d+2}{2}$, which is the weight of the scalar $\Delta_2\phi$. The action in (\ref{orstedaction}) is conformally invariant (has effectively $w=0$) because of the compensating total conformal weight of the metric density $\sqrt{|g|}$ as in (\ref{densitytrlaw}). The action in (\ref{orstedaction}) describes just a free standard scalar field without self-interactions and only coupled in a conformal way to the background geometry. This action consists only of the two-derivative kinetic term for scalars, which is the leading term in the  number of derivatives. Moreover, only this term survives in a limiting case of flat spacetime backgrounds giving rise to a standard scalar flat space propagator with two derivatives in momentum space.

The action in (\ref{orstedaction}) and in $d=4$ familiar spacetime dimensions contains the famous non-minimal coupling $\xi R\phi^2$ of the scalar field $\phi$ to the background geometry represented here by the Ricci scalar $R$. In $d=4$, this coupling still gives rise to a renormalizable quantum theory of scalar matter coupled to quantum gravity (since the coefficient $\xi$ is dimensionless in $d=4$). The coupling of this type is naturally generated when one puts massless scalar matter (with two-derivative kinetic operator) from flat spacetime background on a non-trivial gravitational configuration. This is an additional term of interactions between scalars and gravitation that has to be added by the token of conformal version of the DeWitt-Utiyama arguments for higher derivatives \cite{utiyama}. So this coupling term is still renormalizable, but non-minimal and required by the consistency of the coupled quantum matter and gravitational theory. In $d=4$, one finds a special value of the coefficient $\xi_{d=4} = -\frac16$. For other values  $\xi\neq-\frac16$ the theory is not conformally invariant. Actually, for non-conformal values of the $\xi$  coefficient in $d=4$, the theory is still scale-invariant (under global Weyl transformations and under restricted conformal transformations satisfying $\square \Omega(x)=0$). Moreover, the coupling of the form $\xi R\phi^2$ is the most general one, that one can add to the two-derivative scalar theory, but still being consistent on the quantum level with renormalizability of the coupled theory. Other types of higher couplings of the scalar field $\phi$  containing for example, other gravitational curvature tensors or other powers of $R$, necessarily come with higher-derivative character, so with the benefits and drawbacks of higher-derivative models of gravitational dynamics.

One notices that for $d=2$ the term proportional to the Ricci scalar curvature $R$ in (\ref{orsted})
drops down. This is again a special thing of the case of $d=2$ space(time) dimensions, which is related to special extended conformal symmetry on the 2-dimensional complex plane (the algebra of symmetry generators is in such a case infinite-dimensional), to the speciality of a 2-dimensional quantum gravity based on a two-derivative Einstein-Hilbert action, and also to the whole idea of construction of a conformal 2-dimensional Minkowskian worldsheet spacetimes as used in string theories. In such a case, one can say that the background GR-covariant box operator (the generalization from the flat spacetime of the d'Alembertian operator $\partial^2$, namely $\square=g^{\mu\nu}\nabla_\mu\nabla_\nu$) when acting on massless, dimensionless and conformally inert scalar field $\phi$ with $w=0$ in $d=2$, is perfectly conformally covariant with the conformal weight $w_\square=-2$.

Another interesting thing is that for the action of the \O{}rsted operator in \eqref{orsted} on a scalar field $\phi$, the conformal weight $w$ of the scalar field in interesting and physically sensible dimensions, namely for $d\geqslant2$, is fixed and completely determined by the dimension of spacetime and it is also always non-positive, reaching zero only in the case of $d=2$. As we will show later this conformal weight is closely related to the energy dimensionality of the scalar field with the same kinetic term on the flat spacetime, so in our case this is just the $\partial^2$  operator and this is canonically normalized standard massless scalar field on flat spacetime with two-derivative action. Another known speciality of the $d=2$ case (both for Euclidean and Minkowskian signature of the metric) is that there such a scalar field is dimensionless. From this we derive that its conformal weight vanishes too, so this is a case for a conformally invariant scalar field $\phi$.

\subsection{Paneitz operator $\Delta_4$}
\label{s2b}

Four-derivative scalar conformal operator $\Delta_{4}$ was originally
found independently -- first by Fradkin and Tseytlin and later
by Paneitz and Eastwood and Singer. Its form in $d=4$ is the following\footnote{Our notation regarding differential operators acting on non-operatorial
objects is the most compact one and does not use additional parentheses.
That is the sequence of differential operators (product of operators)
only acts on the first tensorial object it finds on its right and
finishes its derivative action there. For example, by $\nabla_{\mu}R\nabla^{\mu}$
we mean $\left(\nabla_{\mu}R\right)\nabla^{\mu}$. Such a form of
writing all expanded formulas in this paper can be achieved by extensive
use of Leibniz rule. If the operator acts on a product of tensors,
then we put them in one common parenthesis.}
\begin{equation}
\Delta_{4}=\square^{2}+\frac{1}{3}\nabla_{\mu}R\nabla^{\mu}+2R_{\mu\nu}\nabla^{\mu}\nabla^{\nu}-\frac{2}{3}R\square.
\label{paneitzd4}
\end{equation}
We first observe that in $d=4$, terms quadratic in curvatures do
not appear in the construction of this operator. Moreover, there the conformal
weight of the scalar field must be chosen as $w=0$. This coincides
with the fact known from simple dimensional analysis that the scalar
field in $d=4$ dimensions with four-derivative kinetic term must
be dimensionless (in energy units), when, for example, analyzed on the flat spacetime background.

The Paneitz operator was later generalized by Paneitz to arbitrary
number of dimensions of spacetime $d$ (except the case of $d=2$
where such simple generalization described below does not work). Its
form reads 
\[
\Delta_{4}=\square^{2}-\frac{d-6}{2(d-1)}\nabla_{\mu}R\nabla^{\mu}+\frac{4}{d-2}R_{\mu\nu}\nabla^{\mu}\nabla^{\nu}-\frac{(d-2)^{2}+4}{2(d-1)(d-2)}R\square+
\]
\begin{equation}
+\frac{(d-4)\left(d^{3}-4d^{2}+16d-16\right)}{16(d-1)^{2}(d-2)^{2}}R^{2}-\frac{d-4}{(d-2)^{2}}R_{\mu\nu}R^{\mu\nu}-\frac{d-4}{4(d-1)}\square R.
\label{paneitzgend}
\end{equation}

The conformal weight of the scalar field must then be given by 
\begin{equation}
w=\frac{4-d}{2},
\label{weightpaneitz}
\end{equation}
which is negative for $d>4$ and positive for $d<4$. One notices
immediately that terms quadratic in gravitational curvatures or containing
four energy dimensions in the background quantities (namely $R^{2}$,
$R_{\mu\nu}R^{\mu\nu}$ and $\square R$) come with coefficients which
are proportional to the factor $(d-4)$. This confirms that in $d=4$
they are not necessary in the construction of the $\Delta_{4}$ operator.
One also sees that various coefficients in front of various terms
contain the singularity $\frac{1}{d-2}$, hence they cannot be defined
as a simple limit in $d=2$ spacetime dimensions. We will come back
to this issue later. Some coefficients explode which means that Paneitz
operator cannot be constructed in this way in $d=2$. This happens
when the dimension $d$ of spacetime (being an even integer number)
is lower than the critical dimension given by the index on the $\Delta$
operator. Namely $\Delta_{n=4}=\Delta_{4}$ is a Paneitz operator,
$\Delta_{n=2}=\Delta_{2}$ is conformal two-derivative operator, etc.
This is also related to the negativity of the conformal weight $w$
of the scalar field and vanishing weight happens precisely in the
critical dimension. As we will see below, in the case when the weight
is positive and equals to one, we are able to prove nonexistence of
the dimensionally generalized $\Delta_{n}$ for $d=n-2$ as a mathematical
theorem.

One can, of course, freely change the basis for writing terms quadratic
in curvature in the Paneitz operator. For example, squares of the
Riemann tensor are possible there too $R_{\mu\nu\rho\sigma}R^{\mu\nu\rho\sigma}$.
However, we notice one peculiar fact, that in dimensions $d>3$, one
can add for free and with an arbitrary coefficient a conformally invariant
operator whose structure makes evident its conformal properties. This
operator is the square of the Weyl tensor $C_{\mu\nu\rho\sigma}C^{\mu\nu\rho\sigma}\underset{\rm df}{=}C^2$
and it acts as endomorphism on the scalar field (without derivatives).
This is why in the form of writing the operator $\Delta_{4}$ we need
to use only two quadratic in curvature invariants in $d\neq4$ (and
we have chosen them as $R^{2}$ and $R_{\mu\nu}R^{\mu\nu})$. Notice
that in critical dimensions the terms with no derivatives acting on
the scalar $\phi$ (so with two gravitational curvatures) are never needed for the construction of the minimal $\Delta_4$ operator in such a critical dimension $d=4$, but the additional term $C^2\phi$  can be added here for free and with arbitrary chosen value of the real front coefficient. We call the operator $\Delta_4$  without such addition $C^2\phi$ as minimal one, contrary to the non-minimal case. This addition can be realized in any dimension $d\geqslant4$.

Note that the general Paneitz operator in \eqref{paneitzgend} is a four-derivative kinetic operator (acting on the scalar field) non-minimally
coupled to various gravitational curvatures of the background geometry and without
the mass since a constant mass term again would generally break here the conformal symmetry of the model. (Actually here one can treat effectively the curvature terms, like $R^2$, $R_{\mu\nu}R^{\mu\nu}$ and $\square R$ as a possibly spacetime-dependent mass terms for the scalar field $\phi$).
One sees that the non-minimal coupling to the geometry is realized by more than one term  (one term with Ricci scalar curvature was the case for the $\Delta_2$ operator). Moreover, terms with lower number of derivatives than 4 still acting non-endomorphically on the scalar field are present in \eqref{paneitzgend}. They are generally possible and contain two or one background covariant derivative still acting on the scalar field $\phi$ on the right. The terms with three derivatives are not possible to construct here due to the requirement of general covariance symmetry of the form for the operator in \eqref{paneitzgend} since we cannot construct local GR-covariant terms not being derivatives with precisely one unit of energy dimension. The same type of argument excluded terms with one covariant derivative still acting on the scalar field in the case of the \O{}rsted conformal operator $\Delta_2$ - there we cannot have any additional differential operators constructed with covariant derivatives besides the leading one, that is a kinetic term $\partial^2$ with the highest number of derivatives.  The form of the operator in \eqref{paneitzgend} constitutes the case of a generalized conformal coupling of a scalar field with four-derivative action and four-derivative kinetic term (on flat spacetime) to background gravitational field. The conformally invariant action of the theory is given by \eqref{action} with the operator $\Delta=\Delta_4$  for this case.

Such a four-derivative operator in (\ref{paneitzgend}) can be defined in any spacetime dimension, except the case of
$d=2$. It can be easily found here that $w_{\Delta_2} = -4$, so in the result with \eqref{weightpaneitz}, we find that $w'=-\frac{d+4}{2}$, which is the weight of the scalar $\Delta_2\phi$. The action in (\ref{action}) for the case of Paneitz operator $\Delta=\Delta_4$ is conformally invariant (has effectively $w=0$) because of the compensating total conformal weight of the metric density $\sqrt{|g|}$ as in (\ref{densitytrlaw}). The action based on the kinetic operator as in \eqref{paneitzgend} describes just a free 
scalar field with four-derivative kinetic term and without self-interactions and only coupled in a conformal way to the background geometry. This action consists of the four-derivative kinetic term for scalars, which is the leading term in the  number of derivatives and also some other subleading terms in the number of derivatives which therefore must be accompanied in their construction by some curvature terms. Moreover, only the leading in derivatives four-derivative term $\square^2$ survives in a limiting case of flat spacetime backgrounds giving rise to a scalar flat space propagator with four derivatives in momentum space.

In $d=4$, the action of the Paneitz operator $\Delta_4$  from \eqref{paneitzd4} still contains various terms  expressing non-trivial and non-minimal couplings to the background geometry. These couplings still give rise to a renormalizable quantum theory of scalar matter coupled to quantum gravity, provided that the scalar field is taken as dimensionless, so then it is not canonically normalized scalar field. The coupling of this type is naturally generated when one puts massless dimensionless scalar matter (with a four-derivative kinetic operator) from flat spacetime background on a non-trivial gravitational configuration. These are some additional terms of interactions between scalars and gravitation that have to be added by the token of conformal version of the DeWitt-Utiyama arguments for higher derivatives \cite{utiyama}. So these coupling terms are still renormalizable, but non-minimal and required by the consistency of the coupled quantum matter and gravitational theory. Moreover, the couplings of the forms as in \eqref{paneitzd4} are the most general ones, that one can add to the four-derivative scalar theory, but still being consistent on the quantum level with renormalizability of the coupled theory.

One notices that for the critical dimension of $d=4$ all the endomorphic terms, so not containing derivatives when acting on the scalar field $\phi$, so terms proportional to the Ricci scalar curvature squared $R^2$, Ricci tensor squared $R_{\mu\nu}R^{\mu\nu}$ and the GR-covariant box operator acting on a scalar curvature $\square R$ in (\ref{paneitzgend})
drop down. 

Another interesting thing is that for the action of the Paneitz operator in \eqref{paneitzgend} on a scalar field $\phi$, the conformal weight $w$ of the scalar field in interesting and physically sensible dimensions, namely for $d\geqslant4$, is fixed and completely determined by the dimension of spacetime and it is also always non-positive, reaching zero only in the special case of $d=4$. As we will show later this conformal weight is closely related to the energy dimensionality of the scalar field with the same kinetic term on the flat spacetime, so in our case this is just the $(\partial^2)^2$  operator and this is canonically normalized massless scalar field on flat spacetime with four-derivative action. Another known speciality of the $d=4$ case (both for Euclidean and Minkowskian signature of the metric) is that there such a scalar field is dimensionless. From this we derive that its conformal weight vanishes too, so this is a case for a conformally invariant scalar field $\phi$.

\section{Hamada operator in $d=6$}
\label{s3}

The operator with six-derivatives $\Delta_{6}$ was first found by
Hamada in the critical number of dimensions, so for the dimensionless
scalar field in $d=6$ with the vanishing conformal weight $w=0$.
As explained above, terms with energy dimensions $E^{6}$ in background
quantities are not needed in its construction. Still, the structure
of the operator contains 20 terms (in the basis chosen below). Its
form is
\[
\Delta_{6}=\square^{3}+4R_{\mu\nu}\nabla^{\mu}\nabla^{\nu}\square-R\square^{2}+4\nabla_{\rho}R_{\mu\nu}\nabla^{\rho}\nabla^{\mu}\nabla^{\nu}+4\square R_{\mu\nu}\nabla^{\mu}\nabla^{\nu}-\frac{3}{5}\square R\square+
\]
\[
+\zeta_{1}R_{\mu\rho\sigma\tau}R_{\nu}{}^{\rho\sigma\tau}\nabla_{\mu}\nabla_{\nu}+\zeta_{2}R_{\mu\nu\rho\sigma}R^{\mu\nu\rho\sigma}\square+\zeta_{1}R_{\rho\sigma}R_{\mu}{}^{\rho\sigma}{}_{\nu}\nabla^{\mu}\nabla^{\nu}+\left(6-\frac{3}{4}\zeta_{1}\right)R_{\mu\rho}R_{\nu}{}^{\rho}\nabla^{\mu}\nabla^{\nu}+
\]
\[
+\left(-1+\frac{1}{8}\zeta_{1}-\zeta_{2}\right)R_{\mu\nu}R^{\mu\nu}\square+\left(-2+\frac{1}{4}\zeta_{1}\right)RR_{\mu\nu}\nabla^{\mu}\nabla^{\nu}+\left(\frac{9}{25}-\frac{1}{40}\zeta_{1}+\frac{1}{10}\zeta_{2}\right)R^{2}\square+
\]
\[
+\frac{2}{5}\nabla_{\mu}\square R\nabla^{\mu}+\left(\zeta_{1}+4\zeta_{2}\right)R_{\mu\nu\rho\sigma}\nabla^{\sigma}R^{\mu\nu\rho}{}_{\tau}\nabla^{\tau}-\zeta_{1}R_{\mu\nu\rho\sigma}\nabla^{\mu}R^{\nu\rho}\nabla^{\sigma}+\left(6+\frac{1}{4}\zeta_{1}\right)R_{\mu\nu}\nabla^{\mu}R^{\nu}{}_{\rho}\nabla^{\rho}+
\]
\begin{equation}
+\left(-2-\frac{3}{4}\zeta_{1}-2\zeta_{2}\right)R_{\mu\nu}\nabla_{\rho}R^{\mu\nu}\nabla^{\rho}+\left(1-\frac{1}{8}\zeta_{1}\right)R_{\mu\nu}\nabla^{\mu}R\nabla^{\nu}+\left(-\frac{7}{25}+\frac{3}{40}\zeta_{1}+\frac{1}{5}\zeta_{2}\right)R\nabla_{\mu}R\nabla^{\mu}.\label{eq: hamada}
\end{equation}
One observes that the operator $\Delta_{6}$ is not unique in $d=6$
since it depends on two arbitrary real parameters $\zeta_{1}$ and
$\zeta_{2}$ which can take values, which may simplify in some special
cases the form of the operator. This two-parameter family of conformal
operators is a common feature since in higher dimensions we have more
ways to construct gravitational conformal invariants (we have more
ways to contract various Weyl tensors)\footnote{For example, in $d=4$, there is only one way of contracting two Weyl
tensors $C_{\mu\nu\rho\sigma}C^{\mu\nu\rho\sigma}$. Using the identity
$C_{\mu\nu\rho\sigma}C^{\mu\rho\nu\sigma}=\frac{1}{2}C_{\mu\nu\rho\sigma}C^{\mu\nu\rho\sigma}$
due to cyclicity properties of the Weyl tensor, the second possible
contraction $C_{\mu\nu\rho\sigma}C^{\mu\rho\nu\sigma}$ is eliminated.} In dimensions $d=6$, we have three independent ways to construct
gravitational conformal invariants \cite{Bastianelli:2000hi}.
In the sector of non-differentiated Weyl tensors only, they reduce
to two different contractions of the cube of the Weyl tensor, namely
$C_{\mu\nu\rho\sigma}C^{\rho\sigma}{}_{\kappa\lambda}C^{\kappa\lambda\mu\nu}$
and the other $C_{\mu\rho\sigma\nu}C^{\rho\kappa\lambda\sigma}C_{\kappa\mu\nu\lambda}$
(the other possible contractions again reduce to the two above by
exploiting the cyclicity). Only in lower dimensions, like in $d=4$,
there exists an identity relating $C_{\mu\rho\sigma\nu}C^{\rho\kappa\lambda\sigma}C_{\kappa\mu\nu\lambda}$
and $C_{\mu\nu\rho\sigma}C^{\rho\sigma}{}_{\kappa\lambda}C^{\kappa\lambda\mu\nu}$.
Surely the number of free parameters in the construction of the
$\Delta_{6}$ operator is related to the number of invariants that
can be constructed by contractions of three Weyl tensors.\\
\\

\subsection{Terms in the generalization of Hamada operator (away from $d=6$)}
\label{s3a}

The generalization of the Hamada operator to arbitrary dimensions
$d$ was not known before. First, one has to complete the basis of
invariants as given above in the expansion of the $\Delta_{6}$ operator
in the critical dimension $d=6$. Two terms are additionally possible
which contain free derivatives acting on the scalar field (which is
always understood to be on the right of the operator), namely 
\begin{equation}
\nabla_{\mu}R\nabla^{\mu}\square\quad{\rm and}\quad\nabla_{\mu}\nabla_{\nu}R\nabla^{\mu}\nabla^{\nu}.\label{eq: new_elements}
\end{equation}
The basis of all together 22 terms containing derivatives is now complete.
The other possible terms that one would think should be added are
reduced by the following identities
\begin{equation}
R_{\nu\rho\sigma\tau}\nabla_{\mu}R^{\nu\sigma\rho\tau}\nabla^{\mu}\phi=\frac{1}{2}R_{\nu\rho\sigma\tau}\nabla_{\mu}R^{\nu\rho\sigma\tau}\nabla^{\mu}\phi,
\end{equation}
\begin{equation}
R_{\rho\sigma\tau\mu}\nabla^{\mu}R^{\rho\tau\sigma}{}_{\nu}\nabla^{\nu}\phi=\frac{1}{2}R_{\rho\sigma\tau\mu}\nabla^{\mu}R^{\rho\sigma\tau}{}_{\nu}\nabla^{\nu}\phi,
\end{equation}
\begin{equation}
R_{\nu\rho\sigma\tau}\nabla_{\mu}R^{\nu\sigma\rho\tau}\nabla^{\mu}\phi=\frac{1}{2}R_{\nu\rho\sigma\tau}\nabla_{\mu}R^{\nu\rho\sigma\tau}\nabla^{\mu}\phi
\end{equation}
arising from the usage of cyclicity of Riemann tensor ($R_{\mu\nu\rho\sigma}+R_{\mu\rho\sigma\nu}+R_{\mu\sigma\nu\rho}=0$)
and
\begin{equation}
R_{\nu\rho\sigma\tau}\nabla_{\mu}R^{\nu\rho\sigma\tau}\nabla^{\mu}\phi=2R_{\rho\sigma\tau\mu}\nabla^{\mu}R^{\rho\sigma\tau}{}_{\nu}\nabla^{\nu}\phi,
\end{equation}
\begin{equation}
R_{\nu\rho\sigma\tau}\nabla^{\rho}R^{\mu\nu\sigma\tau}=-\frac{1}{2}R_{\nu\rho\sigma\tau}\nabla^{\mu}R^{\nu\rho\sigma\tau}
\end{equation}
arising from use of Bianchi identity ($\nabla_{\mu}R_{\nu\rho\sigma\tau}+\nabla_{\nu}R_{\rho\mu\sigma\tau}+\nabla_{\rho}R_{\mu\nu\sigma\tau}=0$)
satisfied by Riemann tensor. 

Therefore, following Hamada, the basis we choose to write the full
expansion of the derivative part of the $\Delta_{6}$ operator reads
\[
\Delta_{6,{\rm \,der}}=\square^{3}+z_{1}R_{\mu\nu}\nabla^{\mu}\nabla^{\nu}\square+z_{2}R\square^{2}+z_{3}\nabla_{\rho}R_{\mu\nu}\nabla^{\rho}\nabla^{\mu}\nabla^{\nu}+z_{4}\nabla_{\mu}R\nabla^{\mu}\square+z_{5}\nabla_{\mu}\nabla_{\nu}R\nabla^{\mu}\nabla^{\nu}+
\]
\[
+z_{6}\square R_{\mu\nu}\nabla^{\mu}\nabla^{\nu}+z_{7}\square R\square+z_{8}R_{\mu\rho\sigma\tau}R_{\nu}{}^{\rho\sigma\tau}\nabla_{\mu}\nabla_{\nu}+z_{9}R_{\mu\nu\rho\sigma}R^{\mu\nu\rho\sigma}\square+
\]
\[
+z_{10}R_{\rho\sigma}R_{\mu}{}^{\rho\sigma}{}_{\nu}\nabla^{\mu}\nabla^{\nu}+z_{11}R_{\mu\rho}R_{\nu}{}^{\rho}\nabla^{\mu}\nabla^{\nu}+z_{12}R_{\mu\nu}R^{\mu\nu}\square+z_{13}RR_{\mu\nu}\nabla^{\mu}\nabla^{\nu}+z_{14}R^{2}\square+
\]
\[
+z_{15}\nabla_{\mu}\square R\nabla^{\mu}+z_{16}R_{\mu\nu\rho\sigma}\nabla^{\sigma}R^{\mu\nu\rho}{}_{\tau}\nabla^{\tau}+z_{17}R_{\mu\nu\rho\sigma}\nabla^{\mu}R^{\nu\rho}\nabla^{\sigma}+z_{18}R_{\mu\nu}\nabla^{\mu}R^{\nu}{}_{\rho}\nabla^{\rho}+
\]
\begin{equation}
+z_{19}R_{\mu\nu}\nabla_{\rho}R^{\mu\nu}\nabla^{\rho}+z_{20}R_{\mu\nu}\nabla^{\mu}R\nabla^{\nu}+z_{21}R\nabla_{\mu}R\nabla^{\mu},\label{eq: delta_der}
\end{equation}
where we already set the coefficient of the term leading in the number
of derivatives (so of the $\square^{3}$ term) to unity. One sees
that the choice of the basis made by Hamada with two arbitrary parameters
$\zeta_{1}$ and $\zeta_{2}$ corresponds to allowing for completely
arbitrary coefficients in front of the terms
\begin{equation}
R_{\mu\rho\sigma\tau}R_{\nu}{}^{\rho\sigma\tau}\nabla_{\mu}\nabla_{\nu}\quad{\rm and}\quad R_{\mu\nu\rho\sigma}R^{\mu\nu\rho\sigma}\square
\end{equation}
and these two terms can be traded to two ones containing only contractions
of the Weyl tensors and no ordinary covariant derivatives acting on
Weyl tensors (though they still will act on the scalar field $\phi$
on the right). Actually, the change of a basis to the one dominated
by Weyl tensors is possible here as well. However, for further computation
we have to stick to one basis. Of course, the results for the existence
of the operator, etc. do not depend on the choice of such a basis.
And we here adopt the same choice as of Hamada. One sees a posteriori
that the two new elements (\ref{eq: new_elements}) of the basis (\ref{eq: delta_der})
are generated in general dimension $d\neq6$.

However, in general dimension $d$ one also expects terms in $\Delta_{6}$
which are without any derivative acting on the scalar (so they are
endomorphisms). The construction of the basis for these terms is another
tedious task, which was solved in Gilkey. Before we list all the terms
present in this additional non-derivative part of the operator $\Delta_{6}$,
we discuss the issue related to the general construction of the action
for scalars, which supposed to be conformally invariant. The action
we consider is quadratic in the scalar fields $\phi$ and therefore
has the general form
\begin{equation}
S=\!\int\!d^{d}x\sqrt{|g|}\phi\Delta_{6}\phi\label{eq: general_action}
\end{equation}
in general dimension $d$. We can decompose the operator $\Delta_{6}$
as follows
\begin{equation}
\Delta_{6}=\Delta_{6,\,{\rm der}}+\Delta'_{6}.
\end{equation}
Of course, all terms in $\Delta_{6,\,{\rm der}}$ contain derivatives
(from one to six) acting on the right scalar field. For such terms
we cannot perform integration by parts of covariant derivatives, if
we want to preserve the structure of the general action (\ref{eq: general_action}),
that is that the left scalar field is not differentiated at all. This
fixes the possible form of the $\Delta_{6,\,{\rm der}}$ operator.
However, in $\Delta'_{6}$ there are no derivatives and one may think
that a complete basis of terms consists of all six-dimensional gravitational
invariants, which can give rise to conformally invariant gravitational
action in $d=6$ dimensions. Or to globally scale-invariant gravitational
actions in $d=6$ dimensions. It is known that there are precisely
10 such independent terms. However, the basis for $\Delta'_{6}$ operator
has to be extended by inclusion of 7 new terms, which on the level
of gravitational action in $d=6$ would be total derivatives (or related
to them by subtracting terms from the original basis with 10 elements).
On the level of the scalar action (conformally coupled to gravitation)
in (\ref{eq: general_action}) these terms are not total derivatives
because of the presence of two scalar fields (both left and right).
Again if we require the conservation of the structure as in (\ref{eq: general_action}),
we cannot allow for doing of integration by parts and this 10-element
basis must be extended to the general 17-element one. This last basis
we can call as Gilkey basis.

Having discussed this issue, we give below the explicit form of terms
in this basis,
\[
\Delta'_{6}=x_{1}R\square R+x_{2}R_{\mu\nu}\square R^{\mu\nu}+x_{3}R^{3}+x_{4}RR_{\mu\nu}R^{\mu\nu}+x_{5}RR_{\mu\nu\rho\sigma}R^{\mu\nu\rho\sigma}+x_{6}R_{\mu\nu}R^{\mu}{}_{\rho}R^{\nu\rho}+
\]
\[
+x_{7}R_{\mu\nu}R_{\rho\sigma}R^{\mu\rho\nu\sigma}+x_{8}R_{\mu\nu}R^{\mu}{}_{\rho\sigma\tau}R^{\nu\rho\sigma\tau}+x_{9}R_{\mu\nu\rho\sigma}R^{\rho\sigma}{}_{\tau\omega}R^{\tau\omega\mu\nu}+x_{10}R_{\mu\nu\rho\sigma}R^{\mu\tau\rho\omega}R^{\nu}{}_{\tau}{}^{\sigma}{}_{\omega}+
\]
\[
+x_{11}\nabla_{\mu}\nabla_{\nu}RR^{\mu\nu}+x_{12}\nabla_{\mu}R\nabla^{\mu}R+x_{13}\nabla_{\mu}R_{\nu\rho}\nabla^{\mu}R^{\nu\rho}+x_{14}\square^{2}R+x_{15}\nabla_{\mu}R_{\nu\rho}\nabla^{\nu}R^{\mu\rho}+
\]
\begin{equation}
+x_{16}\nabla_{\mu}\nabla_{\nu}R_{\rho\sigma}R^{\mu\rho\nu\sigma}+x_{17}\nabla_{\mu}R_{\nu\rho\sigma\tau}\nabla^{\mu}R^{\nu\rho\sigma\tau}.\label{eq: delta_endo}
\end{equation}
The other possible terms one could think about like
\begin{equation}
R_{\mu\nu\rho\sigma}\square R^{\mu\nu\rho\sigma},\,\nabla_{\mu}R_{\nu\rho\sigma\tau}\nabla^{\mu}R^{\nu\sigma\rho\tau},\,\nabla_{\mu}R_{\nu\rho\sigma\tau}\nabla^{\nu}R^{\mu\rho\sigma\tau},\,\nabla_{\mu}R_{\nu\rho\sigma\tau}\nabla^{\nu}R^{\mu\sigma\rho\tau},\,R_{\mu\nu}R^{\mu}{}_{\rho\sigma\tau}R^{\nu\sigma\rho\tau}
\end{equation}
are reduced by performing commutation of covariant derivatives, using
cyclicity or Bianchi identities (sometimes also in the contracted
forms) of the Riemann tensor. Explicitly, we use the following formulas
which are valid at any point of the spacetime, where we do not have
to permit integration by parts:
\[
R_{\mu\nu\rho\sigma}\square R^{\mu\nu\rho\sigma}=
\]
\begin{equation}
=2R_{\mu\nu}R^{\mu}{}_{\rho\sigma\tau}R^{\nu\rho\sigma\tau}+4\nabla_{\mu}\nabla_{\nu}R_{\rho\sigma}R^{\mu\rho\nu\sigma}-R_{\mu\nu\rho\sigma}R^{\rho\sigma}{}_{\tau\omega}R^{\tau\omega\mu\nu}-4R_{\mu\nu\rho\sigma}R^{\mu\tau\rho\omega}R^{\nu}{}_{\tau}{}^{\sigma}{}_{\omega},
\end{equation}
\begin{equation}
\nabla_{\mu}R_{\nu\rho\sigma\tau}\nabla^{\mu}R^{\nu\sigma\rho\tau}=\frac{1}{2}\nabla_{\mu}R_{\nu\rho\sigma\tau}\nabla^{\mu}R^{\nu\rho\sigma\tau},
\end{equation}
\begin{equation}
\nabla_{\mu}R_{\nu\rho\sigma\tau}\nabla^{\nu}R^{\mu\rho\sigma\tau}=\frac{1}{2}\nabla_{\mu}R_{\nu\rho\sigma\tau}\nabla^{\mu}R^{\nu\rho\sigma\tau},
\end{equation}
\begin{equation}
\nabla_{\mu}R_{\nu\rho\sigma\tau}\nabla^{\nu}R^{\mu\sigma\rho\tau}=\frac{1}{2}\nabla_{\mu}R_{\nu\rho\sigma\tau}\nabla^{\nu}R^{\mu\rho\sigma\tau}=\frac{1}{4}\nabla_{\mu}R_{\nu\rho\sigma\tau}\nabla^{\mu}R^{\nu\rho\sigma\tau},
\end{equation}
\begin{equation}
{\rm and}\quad R_{\mu\nu}R^{\mu}{}_{\rho\sigma\tau}R^{\nu\sigma\rho\tau}=\frac{1}{2}R_{\mu\nu}R^{\mu}{}_{\rho\sigma\tau}R^{\nu\rho\sigma\tau}.
\end{equation}
\\
\\

\section{Derivation of $\Delta_6$ in arbitrary $d$}
\label{s4}
\subsection{Method of derivation of $\Delta_{6}$ in general $d$}
\label{s4a}

Since now we have selected and described the basis of terms (both
in $\Delta_{6,\,{\rm der}}$ and in $\Delta'_{6}$) which will be
present in the generalized form of the Hamada operator in an arbitrary
dimension $d$, we can describe shortly our method of derivation of
the conformal scalar operator $\Delta_{6}$. The operator will be
determined (up to possible ambiguities like in two parameters $\zeta_{1}$
and $\zeta_{2}$ in (\ref{eq: hamada})) completely by specifying
the values of all coefficients $z_{1},\ldots,z_{21}$ and $x_{1},\ldots,x_{17}$
where the normalization is kept by fixing the coefficient in front
of $\square^{3}$ to unity in the expansion of the $\Delta_{6}$ operator.
We will solve the system of linear equations needed to determine the
$21+17=38$ coefficients $z_{1},\ldots,z_{21}$ and $x_{1},\ldots,x_{17}$
or relations between them, if there will be some free parameters here.

The condition of conformal invariance of the action (\ref{eq: general_action})
implies that
\begin{equation}
\delta_{c}S=\frac{\delta S}{\delta g_{\mu\nu}}\delta_{c}g_{\mu\nu}+\frac{\delta S}{\delta\phi}\delta_{c}\phi=0.\label{eq: conf_condition}
\end{equation}
This last condition for conformally invariant theories is also known
as Noether identity for local conformal symmetry. Remembering that
under infinitesimal local conformal transformations we have that
\begin{equation}
\delta_{c}g_{\mu\nu}=2\Omega g_{\mu\nu}
\end{equation}
and
\begin{equation}
\delta_{c}\phi=w\Omega\phi,
\end{equation}
this condition is rewritten as
\begin{equation}
0=\delta_{c}S=2\frac{\delta S}{\delta g_{\mu\nu}}\Omega g_{\mu\nu}+\frac{\delta S}{\delta\phi}w\Omega\phi
\end{equation}
or defining the densitized trace of the energy-momentum tensor of
the system and the scalar equation of motion (EOM)
\begin{equation}
\sqrt{|g|}T=\sqrt{|g|}g_{\mu\nu}T^{\mu\nu}=2\frac{\delta S}{\delta g_{\mu\nu}}
\end{equation}
and
\begin{equation}
E=\frac{\delta S}{\delta\phi},
\end{equation}
this condition is written as
\begin{equation}
0=\delta_{c}S=\sqrt{|g|}T\Omega+Ew\Omega\phi=\Omega\left(\sqrt{|g|}T+Ew\phi\right).
\end{equation}
One sees explicitly that the trace of the EMT of the matter action
defined in (\ref{eq: general_action}) is not vanishing, but is balanced
by the term proportional to the scalar classical equation of motion.
When $w=0$, so we consider the situation in critical spacetime dimensions
$d=n$ (as we also explain in the next section), then the trace $T$
of EMT of the scalar matter system must vanish identically. If we
are not in a critical dimension, then the trace $T$ of the EMT in
matter models conformally coupled to gravitation, vanishes but only
on-shell (that is using EOM from the matter sector). In general off-shell
situation one finds the following expression for the densitized EMT
in conformally coupled matter models
\begin{equation}
\sqrt{|g|}T=-wE\phi.
\end{equation}

However, in our derivation for the algebraic system of equations,
we decide not to use integration by parts after performing the conformal
transformations on terms and this implies that the equation in (\ref{eq: conf_condition})
has only a formal meaning. The conformal variations of metric and
of the scalar fields will be still under the derivatives in our variation
of the action $\delta_{c}S$. Instead, we remark that such operations
of integration by parts under spacetime volume integral of the action
of the model are typically performed to obtain the expressions for
the EMT and for the matter EOM as considered above.\\
\\

\subsection{Consequences of global scale-invariance}
\label{s4b}

First, it is instructive to analyze the situation for global scale
invariance, so the case in which $\Omega$ being the parameter of
infinitesimal conformal transformations is constant ($\Omega={\rm const}$)
and hence all derivatives acting on it vanish identically. Then conformal
invariance agrees with the results of dimensional analysis and this
is only an algebraic constraint on the possible value of the conformal
weight $w$ of the scalar field. Here, we can analyze the situation
for the first leading term $\square^{\frac{n}{2}}$ (in number of
derivatives) of the $\Delta_{n}$ operator in $d$ dimensions. This
part of the action now reads
\begin{equation}
\int d^{d}x\sqrt{|g|}\phi\square^{\frac{n}{2}}\phi.
\end{equation}
Requiring for this to be dimensionless quantity, we find the energy
dimension of the scalar field
\begin{equation}
[\phi]=E^{\frac{d-n}{2}}
\end{equation}
or in terms of the conformal weight
\begin{equation}
w=-\frac{d-n}{2}=\frac{n-d}{2}.
\end{equation}
This counting of energy dimensions agrees precisely with the counting
of coefficients in front of $\Omega$ for the infinitesimal transformations.
And similarly for conformal invariance we want to have that $\delta_{c}S\propto\Omega$
with the proportionality coefficient equal to zero. For this one has
to recall that for global infinitesimal conformal transformation 
\begin{equation}
\delta_{c}\left(d^{d}x\sqrt{|g|}\right)=d\Omega d^{d}x\sqrt{|g|},
\end{equation}
\begin{equation}
\delta_{c}\left(\square\right)=\delta_{c}\left(g^{\mu\nu}\nabla_{\mu}\nabla_{\nu}\right)=-2\Omega g^{\mu\nu}\nabla_{\mu}\nabla_{\nu}=-2\Omega\square,
\end{equation}
\begin{equation}
{\rm and}\quad\delta_{c}\phi=w\Omega\phi
\end{equation}
in precise opposite analogy (the minus sign is only different) to
the energy dimension assignments of the same elements of the action,
namely
\begin{equation}
\left[d^{d}x\sqrt{|g|}\right]=E^{-d},
\end{equation}
\begin{equation}
\left[\square\right]=\left[g^{\mu\nu}\nabla_{\mu}\nabla_{\nu}\right]=\left[\nabla\right]^{2}=E^{2},
\end{equation}
\begin{equation}
{\rm and}\quad[\phi]=E^{\frac{d-n}{2}}.
\end{equation}
The conformal weight $w=\frac{n-d}{2}$ of the scalar field $\phi$
derived this way must be uniformly valid for all terms in the operator
$\Delta_{n}$, hence this is already a first result for the issue
of determination of $\Delta_{6}$ in general dimensions $d$. This
is of course in agreement with the general methodology of our construction
of the scalar conformal operators $\Delta_{n}$ in general spacetime
of dimensionality $d$. First, we need to assign the conformal weight
$w$ to the scalar field $\phi$ and only after this we can search
for coefficients of all terms in the expansion of the operator $\Delta_{n}.$
As emphasized above the first task is easily solved by exploiting
global scale-invariance (or dimensional analysis).\\
\\

\subsection{Linearized infinitesimal conformal transformations}
\label{s4c}

For other terms in the conformal variation of the action $\delta_{c}S$,
where the derivatives act on spacetime-dependent parameter of local
conformal transformation $\Omega=\Omega(x)$, it is enough to resort
to considering invariance of the action under infinitesimal transformations,
so to the linear level in $\Omega$, where we neglect terms with higher
powers of it. This corresponds to the analysis of the algebra of conformal
transformations. Finite conformal transformations are obtained by
exponentiation of infinitesimal ones and we expect no problem with
invariance with respect to them, if the first is established provided
that there are no any topological obstacles to finite conformal transformations.
The detail computation shows that the scalar actions with $\Delta_{6}$
as we find them below are without any problems invariant also under
finite conformal rescalings.

After performing the conformal transformation, we need to order terms
resulting from the variation and keep only these linear in $\Omega$.
For this we can only keep terms linear in the derivatives of the scalar
$\Omega$. The global part was analyzed above with the results for
$w$ only. Next, we need to use Bianchi identities, cyclicity of Riemann
tensor and commutation of derivatives to reduce the terms and find
only linearly independent combination of them in the final result
for $\delta_{c}S$ written in an irreducible basis. This step is very
important to get a correct solution for the unknown coefficients $z_{1},\ldots,z_{21}$
and $x_{1},\ldots,x_{17}$. We remark that the number of equations
we get is typically bigger than $21+17=38$, because the number of
independent terms in the results for $\delta_{c}S$ is bigger than
the total number of invariants in (\ref{eq: delta_der}) and in (\ref{eq: delta_endo}).
For example, in $\delta_{c}S$ we have scalars with derivatives on
$\Omega$, which are neither present in the basis in (\ref{eq: delta_der}),
nor in (\ref{eq: delta_endo}). Hence the basis of all possible terms
in $\delta_{c}S$ is a completely different and unrelated to the ones
in (\ref{eq: delta_der}), (\ref{eq: delta_endo}). 

Only for the global case, when $\Omega={\rm const}$ we find in $\delta_{c}S$
the same terms as in $\Delta_{6}$ just multiplied by $\Omega$. Therefore,
we must have precisely $22+17=39$ terms in the results for $\delta_{c}S$
with $\Omega={\rm const}$\footnote{One more term without associated a numerical $z$ coefficient in the
basis (\ref{eq: delta_der}) is simply $\square^{3}$.}. All of the conditions for vanishing of these terms hold, when $w=\frac{n-d}{2}$
is plugged in.

For terms with derivatives of $\Omega$ we find in $\delta_{c}S$
precisely $37+22=59$ independent terms (this decomposition will be
explained below). Hence, at the end we get a system of $59$ equations
for $38$ unknown coefficients. The check that this is not a contradictory
system is a very important verification step for our method of computation
(there are $21$ excess equations). At the end, we find that this
system still has some free parameters (like $\zeta_{1}$ and $\zeta_{2}$
in the case of the original Hamada operator). We will not write explicitly
the independent terms in the basis of these $59$ elements of expansion
of $\delta_{c}S$, since this would be quite lengthy and moreover
this basis will not be used by us further. We will need it just to
specify the system of linear equations for $38$ coefficients.

Actually, here one can easily determine and distinguish the form of
terms in $\delta_{c}S$, which come from conformal variation of the
part $\Delta'_{6}$ of the $\Delta_{6}$ operator. These terms will
of course carry with them the linear dependence on the $x_{1},\ldots,x_{17}$
coefficients. It is clear from their character that they will result
exclusively in terms, which will be proportional to $\phi^{2}$, so
no derivatives will act on the scalar field, while they may act on
$\Omega$ or on other gravitational curvatures. We can single out
these basis terms, which are proportional to $\phi^{2}$. We find
precisely $22$ of them. The other $39$ terms are built with derivatives
on one of the scalar field (right field) and on $\Omega$. We find
the ordering of terms there in number of derivatives acting on scalars
quite useful. To deal with these terms and with the big systems of
linear equations we find help from special Mathematica packages.

As a matter of fact one sees that the system of coefficients $x_{1},\ldots,x_{17}$
can be solved (or relations between them found) based on mentioned
above $22$ equations arising from $22$ independent terms in $\delta_{c}S$
proportional to $\phi^{2}$. On the other hand, for the coefficients
$z_{1},\ldots,z_{21}$, we need to solve or reduce the bigger system
of $37$ equations. At the end, two systems of equations are coupled,
so the solutions for $x_{1},\ldots,x_{17}$ depend also on solutions
found for $z_{1},\ldots,z_{21}$.\\
\\

\subsection{Solutions for coefficients of $\Delta_{6}$}
\label{s4d}

The results for the solutions of these systems are as follows. The
system of $z_{1},\ldots,z_{21}$ coefficients has the form
\[
z_{1}=\frac{16}{d-2}
\]
\[
z_{2}=-\frac{3d^{2}-12d+44}{4(d-1)(d-2)}=-\frac{3(d-6)^{2}+24(d-6)+80}{4(d-1)(d-2)}
\]
\[
z_{3}=\frac{16}{d-2}
\]
\[
z_{4}=-\frac{(d-6)(3d-10)}{2(d-1)(d-2)}=-\frac{(d-6)[3(d-6)+8]}{2(d-1)(d-2)}
\]
\[
z_{5}=-\frac{(d-6)(d-8)}{(d-1)(d-4)}=-\frac{(d-6)[(d-6)-2]}{(d-1)(d-4)}
\]
\[
z_{6}=\frac{8}{d-4}
\]
\[
z_{7}=-\frac{3d^{2}-26d+72}{4(d-1)(d-4)}=-\frac{3(d-6)^{2}+10(d-6)+24}{4(d-1)(d-4)}
\]
\[
z_{10}=\frac{4\left[\left(z_{8}+4\right)d-4\left(z_{8}+6\right)\right]}{(d-2)(d-4)}=\frac{4\left[\left(z_{8}+4\right)(d-6)+2z_{8}\right]}{(d-2)(d-4)}
\]
\[
z_{11}=-\frac{2\left(z_{8}-8\right)d}{(d-2)^{2}}=-\frac{2\left(z_{8}-8\right)[(d-6)+6]}{(d-2)^{2}}
\]
\[
z_{12}=-\frac{2\left[2\left(z_{9}+1\right)d^{2}-\left(z_{8}+12z_{9}+16\right)d+4\left(z_{8}+4z_{9}+10\right)\right]}{(d-2)^{2}(d-4)}=
\]
\[
=-\frac{2\left[2\left(z_{9}+1\right)(d-6)^{2}+\left(-z_{8}+12z_{9}+8\right)(d-6)-2\left(z_{8}-8z_{9}-8\right)\right]}{(d-2)^{2}(d-4)}
\]
\[
z_{13}=-\frac{4\left[d^{3}-\left(z_{8}+8\right)d^{2}+\left(5z_{8}+36\right)d-4\left(z_{8}+16\right)\right]}{(d-1)(d-2)^{2}(d-4)}=
\]
\[
=-\frac{4\left[(d-6)^{3}-\left(z_{8}-10\right)(d-6)^{2}+\left(-7z_{8}+48\right)(d-6)-10\left(z_{8}-8\right)\right]}{(d-1)(d-2)^{2}(d-4)}
\]
\[
z_{14}=\frac{1}{16(d-1)^{2}(d-2)^{2}(d-4)}\left\{ 3d^{5}-36d^{4}+8\left(4z_{9}+29\right)d^{3}-32\left(z_{8}+7z_{9}+30\right)d^{2}+\right.
\]
\[
\left.+16\left(10z_{8}+28z_{9}+155\right)d-64\left(2z_{8}+4z_{9}+39\right)\right\} =
\]
\[
=\frac{1}{16(d-1)^{2}(d-2)^{2}(d-4)}\left\{ 3(d-6)^{5}+54(d-6)^{4}+32\left(z_{9}+14\right)(d-6)^{3}-\right.
\]
\[
\left.-32\left(z_{8}-11z_{9}-60\right)(d-6)^{2}-32\left(7z_{8}-38z_{9}-136\right)(d-6)-64\left(5z_{8}-20z_{9}-72\right)\right\} 
\]
\[
z_{15}=-\frac{d-8}{d-1}=-\frac{(d-6)-2}{d-1}
\]
\[
z_{16}=z_{8}+4z_{9}
\]
\[
z_{17}=-\frac{2(d-4)z_{8}}{d-2}=-\frac{2[(d-6)+2]z_{8}}{d-2}
\]
\[
z_{18}=\frac{2\left[\left(z_{8}+8\right)d-4z_{8}\right]}{(d-2)^{2}}=\frac{2\left[\left(z_{8}+8\right)(d-6)+2\left(z_{8}+24\right)\right]}{(d-2)^{2}}
\]
\[
z_{19}=-\frac{4\left[\left(z_{8}+2z_{9}+4\right)d-3z_{8}-4z_{9}-16\right]}{(d-2)^{2}}=-\frac{4\left[\left(z_{8}+2z_{9}+4\right)(d-6)+3z_{8}+8z_{9}+8\right]}{(d-2)^{2}}
\]
\[
z_{20}=-\frac{d^{3}+\left(z_{8}-10\right)d^{2}+\left(-5z_{8}+4\right)d+4\left(z_{8}+10\right)}{(d-1)(d-2)^{2}}=
\]
\[
=-\frac{(d-6)^{3}+\left(z_{8}+8\right)(d-6)^{2}+\left(7z_{8}-8\right)(d-6)+10\left(z_{8}-8\right)}{(d-1)(d-2)^{2}}
\]
\[
z_{21}=\frac{3d^{4}-40d^{3}+8\left(2z_{8}+4z_{9}+27\right)d^{2}-32\left(2z_{8}+3z_{9}+25\right)d+16\left(3z_{8}+4z_{9}+55\right)}{8(d-1)^{2}(d-2)^{2}}=
\]
\begin{equation}
=\frac{3(d-6)^{4}+32(d-6)^{3}+16\left(z_{8}+2z_{9}+9\right)(d-6)^{2}+32\left(4z_{8}+9z_{9}+2\right)(d-6)+16\left(15z_{8}+40z_{9}-56\right)}{8(d-1)^{2}(d-2)^{2}},
\end{equation}
while the system of solutions for $x_{1},\ldots,x_{17}$ is
\[
x_{1}=\frac{1}{16(d-1)^{2}(d-2)^{2}(d-4)}\left\{ 3d^{5}-44d^{4}+8\left(z_{8}+4z_{9}+37\right)d^{3}-8\left(11z_{8}+44z_{9}-32x_{17}+152\right)d^{2}+\right.
\]
\[
\left.+16\left(17z_{8}+68z_{9}-80x_{17}+163\right)d-64\left(3z_{8}+12z_{9}-16x_{17}+33\right)\right\} =
\]
\[
=\frac{1}{16(d-1)^{2}(d-2)^{2}(d-4)}\left\{ 3(d-6)^{5}+46(d-6)^{4}+8\left(z_{8}+4z_{9}+40\right)(d-6)^{3}+\right.
\]
\[
\left.+8\left(7z_{8}+28z_{9}+32x_{17}+136\right)(d-6)^{2}+16\left(5z_{8}+20z_{9}+112z_{17}+88\right)(d-6)+2560x_{17}\right\} 
\]
\[
x_{2}=\frac{-\left(z_{8}+4z_{9}+4\right)d^{2}+2\left(5z_{8}+20z_{9}-16x_{17}+16\right)d-8\left(3z_{8}+12z_{9}-16x_{17}+6\right)}{(d-2)^{2}(d-4)}=
\]
\[
=-\frac{\left(z_{8}+4z_{9}+4\right)(d-6)^{2}+2\left(z_{8}+4z_{9}+16x_{17}+6\right)(d-6)+64x_{17}}{(d-2)^{2}(d-4)}
\]
\[
x_{3}=-\frac{1}{64(d-1)^{3}(d-2)^{3}(d-4)}\left\{ d^{7}-16d^{6}+4\left(8z_{9}+35\right)d^{5}-16\left(6z_{9}-4x_{10}+53\right)d^{4}-\right.
\]
\[
-16\left(132z_{9}+64x_{9}+16x_{10}-192x_{17}-231\right)d^{3}+64\left(190z_{9}+104x_{9}-5x_{10}-304x_{17}-186\right)d^{2}-
\]
\[
\left.-64\left(300z_{9}+184x_{9}-24x_{10}-512x_{17}-309\right)d+256\left(36z_{9}+24x_{9}-4x_{10}-64x_{17}-45\right)\right\} =
\]
\[
=-\frac{1}{64(d-1)^{3}(d-2)^{3}(d-4)}\left\{ (d-6)^{7}+26(d-6)^{6}+32\left(z_{9}+10\right)(d-6)^{5}+\right.
\]
\[
+32\left(27z_{9}+2x_{10}+71\right)(d-6)^{4}+64\left(111z_{9}-16x_{9}+20x_{10}+48x_{17}+156\right)(d-6)^{3}+
\]
\[
+64\left(352z_{9}-184x_{9}+139x_{10}+560x_{17}+408\right)(d-6)^{2}+256\left(90z_{9}-166x_{9}+99x_{10}+512x_{17}+120\right)(d-6)-
\]
\[
\left.-1280\left(36x_{9}-19x_{10}-112x_{17}\right)\right\} 
\]
\[
x_{4}=\frac{1}{(d-1)(d-2)^{3}(d-4)}\left\{ \left(z_{9}+1\right)d^{4}+\left(2z_{9}+3x_{10}-12\right)d^{3}+\right.
\]
\[
+\left(-112z_{9}-48x_{9}-9x_{10}+128x_{17}+72\right)d^{2}+8\left(56z_{9}+33x_{9}-3x_{10}-84x_{17}-34\right)d+
\]
\[
\left.+16\left(-24z_{9}-18x_{9}+3x_{10}+40x_{17}+21\right)\right\} =
\]
\[
=\frac{1}{(d-1)(d-2)^{3}(d-4)}\left\{ \left(z_{9}+1\right)(d-6)^{4}+\left(26z_{9}+3x_{10}+12\right)(d-6)^{3}+\right.
\]
\[
+\left(140z_{9}-48x_{9}+45x_{10}+128x_{17}+72\right)(d-6)^{2}+8\left(23z_{9}-39x_{9}+24x_{10}+108x_{17}+20\right)(d-6)+
\]
\[
\left.+4\left(-108x_{9}+57x_{10}+304x_{17}\right)\right\} 
\]
\[
x_{5}=-\frac{z_{9}d(d-4)-12z_{9}-24x_{9}+6x_{10}+32x_{17}}{4(d-1)(d-2)}=-\frac{z_{9}(d-6)^{2}+8z_{9}(d-6)+2\left(-12x_{9}+3x_{10}+16x_{17}\right)}{4(d-1)(d-2)}
\]
\[
x_{6}=\frac{z_{8}d^{2}+2\left(-4z_{8}-4z_{9}+8x_{9}-3x_{10}\right)d+4\left(3z_{8}+12z_{9}-4x_{9}+2x_{10}-16x_{17}\right)}{(d-2)^{3}}=
\]
\[
=\frac{z_{8}(d-6)^{2}+2\left(2z_{8}-4z_{9}+8x_{9}-3x_{10}\right)(d-6)+4\left(20x_{9}-7x_{10}-16x_{17}\right)}{(d-2)^{3}}
\]
\[
x_{7}=-\frac{1}{(d-2)^{2}(d-4)}\left\{ \left(z_{8}+12z_{9}+3x_{10}\right)d^{2}-2\left(5z_{8}+60z_{9}+12x_{9}+6x_{10}-48x_{17}-8\right)d+\right.
\]
\[
\left.+24\left(z_{8}+12z_{9}+4x_{9}-16x_{17}-4\right)\right\} =
\]
\[
=-\frac{1}{(d-2)^{2}(d-4)}\left\{ \left(z_{8}+12z_{9}+3x_{10}\right)(d-6)^{2}-2\left(-z_{8}-12z_{9}+12x_{9}-12x_{10}-48x_{17}-8\right)(d-6)-\right.
\]
\[
\left.-12\left(4x_{9}-3x_{10}-16x_{17}\right)\right\} 
\]
\[
x_{8}=\frac{2z_{9}(d-6)-12x_{9}+3x_{10}+16x_{17}}{d-2}=\frac{2z_{9}(d-6)-12x_{9}+3x_{10}+16x_{17}}{d-2}
\]
\[
x_{11}=-\frac{2(d-6)^{2}}{(d-1)(d-2)(d-4)}
\]
\[
x_{12}=\frac{(d-6)\left(d^{3}-\left(z_{8}+10\right)d^{2}+\left(z_{8}-16z_{9}+32x_{17}+52\right)d+8\left(2z_{9}-4x_{17}-7\right)\right)}{8(d-1)^{2}(d-2)^{2}}=
\]
\[
=\frac{(d-6)}{8(d-1)^{2}(d-2)^{2}}\left\{ (d-6)^{3}+\left(-z_{8}+8\right)(d-6)^{2}+\right.
\]
\[
\left.+\left(-11z_{8}-16z_{9}+32x_{17}+40\right)(d-6)+2\left(-15z_{8}-40z_{9}+80x_{17}+56\right)\right\} 
\]
\[
x_{13}=\frac{z_{8}d^{2}-2\left(3z_{8}-8z_{9}+12x_{17}+4\right)d-16\left(6z_{9}-11x_{17}-3\right)}{2(d-2)^{2}}=
\]
\[
=\frac{z_{8}(d-6)^{2}+2\left(3z_{8}+8z_{9}-4-12x_{17}\right)(d-6)+32x_{17}}{2(d-2)^{2}}
\]
\[
x_{14}=-\frac{d-6}{4(d-1)}
\]
\[
x_{15}=-\frac{z_{8}d^{2}-2\left(3z_{8}-8z_{9}+8x_{17}\right)d-32\left(3z_{9}-5x_{17}\right)}{2(d-2)^{2}}=
\]
\[
=-\frac{z_{8}(d-6)^{2}+2\left(3z_{8}+8z_{9}-8x_{17}\right)(d-6)+64x_{17}}{2(d-2)^{2}}
\]
\begin{equation}
x_{16}=\frac{\left(z_{8}+4z_{9}\right)d-2\left(3z_{8}-12z_{9}+16x_{17}\right)}{d-2}=\frac{\left(z_{8}+4z_{9}\right)(d-6)+32x_{17}}{d-2}.
\end{equation}
One has to make several observations pertaining to the new results
obtained above. First, in solving for the system of $z_{1},\ldots,z_{21}$
we find again two-parameter freedom. We have chosen $z_{9}$ and $z_{10}$
as free parameters following the choice made by Hamada motivated firstly
in $d=6$ dimensions. All solutions for $z$ coefficients: from $z_{1}$
to $z_{8}$ and from $z_{11}$ to $z_{21}$ depend only on these two
parameters and the dimension of spacetime $d$. The system is solved
for $19$ coefficients in terms of $2$ parameters $z_{9}$, $z_{10}$
and the dimension $d$.

For the system of $17$ a priori unknown coefficients $x_{1},\ldots,x_{17}$
we find that we were able to determine $14$ of them in terms of $3$
free parameters from the set $x_{1},\ldots,x_{17}$, with addition
of $z_{9}$ and $z_{10}$ from the previous set and the dimension
of spacetime $d$. As free parameters here we decide to choose $x_{9}$,
$x_{10}$ and $x_{\ensuremath{17}}$. The reason for such a choice
will be explained later. Therefore we solved for $x$ coefficients
from $x_{1}$ to $x_{8}$ and from $x_{11}$ to $x_{16}$.

Now, we comment on the dependence on dimension in the results above.
First, one notices that some coefficients explode in dimensions $d=4$
and $d=2$ and $d=1$ due to denominators, where we find only decomposable
polynomials of the dimension $d$. The first two problematic dimensions
are even dimensions lower than the critical dimension $d=6$ for the
$\Delta_{6}$ conformal operator. One sees that generally one is unable
to define the conformal operator in dimensions $d=2$ and $d=4$ by
any limiting procedure. Below we can prove as a theorem that for dimensions
smaller by two than the critical one for a given operator this is
impossible and one cannot have the operator $\Delta_{n}$ in dimensions
$d=n-2$ obtained using this method. In numerators of the expressions
above we find polynomials of $d$ and of parameters $z_{9}$ and $z_{10}$
(and also of $x_{9}$, $x_{10}$ and $x_{17}$ for the solved $x$
coefficients). The degree in $d$ variable of these polynomials is
up to fifth in solved $z$ coefficients (and up to seventh in $x$
coefficients), while in other variables $z_{9}$ and $z_{10}$ (and
also $x_{9}$, $x_{10}$ and $x_{17}$) the polynomials are linear.\\
\\

\section{Singularities in construction of conformal operators}
\label{s5}

\subsection{Hamada operator in $d=6$ reconsidered}
\label{s5a}

Next, we consider the limiting situation in critical dimension $d=6,$
which is special for the operator $\Delta_{6}$. To facilitate this
we wrote above when this was possible the expansion of the polynomials
in the numerator in the shifted variable $(d-6)$. One notices that
the constant term of this last expansion is what really matters for
the situation in $d=6$. We therefore reproduce exactly the coefficients
as found by Hamada. In the basis chosen by him they take the form
\[
z_{1}=4,\quad z_{2}=-1,\quad z_{3}=4,\quad z_{4}=0,\quad z_{5}=0,\quad z_{6}=4,\quad z_{7}=-\frac{3}{5},
\]
\[
z_{8}=\zeta_{1},\quad z_{9}=\zeta_{2},\quad z_{10}=z_{8},\quad z_{11}=-\frac{3}{4}\left(z_{8}-8\right),\quad z_{12}=\frac{1}{8}\left(z_{8}-8z_{9}-8\right),\quad z_{13}=\frac{1}{4}\left(z_{8}-8\right),
\]
\[
z_{14}=-\frac{1}{200}\left(5z_{8}-20z_{9}-72\right),\quad z_{15}=\frac{2}{5},\quad z_{16}=z_{8}+4z_{9},\quad z_{17}=-z_{8},\quad z_{18}=\frac{1}{4}\left(z_{8}+24\right),
\]
\begin{equation}
z_{19}=-\frac{1}{4}\left(3z_{8}+8z_{9}+8\right),\quad z_{20}=-\frac{1}{8}\left(z_{8}-8\right),\quad z_{21}=\frac{1}{200}\left(15z_{8}+40z_{9}-56\right).
\end{equation}
Notice that the coefficients $z_{4}$ and $z_{5}$ vanish in $d=6$
since their general dimension expressions are both proportional to
the $(d-6)$ factor. Also almost all coefficients after $z_{10}$
(except $z_{15}$) carry some dependence on $z_{8}$ and $z_{9}$
coefficients, so when we set them both to zero a few of coefficients
after $z_{10}$ get vanish. This may have some explanation below.
Finally, one also notices that the particular combination $\left(z_{8}-8\right)$
appears above quite frequently, so again for the special choice $z_{8}=8$
more coefficients vanish.

When one analyzes the situation with $x_{1},\ldots,x_{17}$ coefficients
again in the limit $d=6$, one finds that a few terms there drop out
(namely $x_{11}$, $x_{12}$ and $x_{14}$), namely we have
\[
x_{1}=\frac{x_{17}}{5},\quad x_{2}=-2x_{17},\quad x_{3}=\frac{36x_{9}-19x_{10}-112x_{17}}{800},\quad x_{4}=\frac{-108x_{9}+57x_{10}+304x_{17}}{160},
\]
\[
x_{5}=\frac{12x_{9}-3x_{10}-16x_{17}}{40},\quad x_{6}=\frac{20x_{9}-7x_{10}-16x_{17}}{16},\quad x_{7}=\frac{3\left(4x_{9}-3x_{10}-16x_{17}\right)}{8},
\]
\begin{equation}
x_{8}=\frac{-12x_{9}+3x_{10}+16x_{17}}{4},\quad x_{11}=0,\quad x_{12}=0,\quad x_{13}=x_{17},\quad x_{14}=0,\quad x_{15}=-2x_{17},\quad x_{16}=8x_{17},
\end{equation}
while the other are linear functions of only the coefficients from
the $x$ set, that is of $x_{9}$, $x_{10}$ and $x_{17}$. The constant
terms in the numerator (independent of the above three variables)
do not appear and there is no any dependence on the two free parameters
$z_{8}$ and $z_{9}$. This proves that it is consistent to set all
three $x_{9}$, $x_{10}$ and $x_{17}$ to zero and end up with Hamada
operator as a conformal scalar operator with six derivatives in $d=6$
without a need to include terms in $\Delta'_{6}$ there. Here, instead
we provide the generalization of the Hamada operator still in $d=6$,
when we include additional three-parameter family of endomorphism
terms acting on scalar $\phi$ described by these $x_{9}$, $x_{10}$
and $x_{17}$ free coefficients. We will show below that when we change
the basis this extension of the Hamada operator has a natural interpretation. In a sense the size of this freedom has to do with the number of Weyl co-cycles that one can create with cubes of Weyl tensors \cite{Bonora:1985cq,Arakelian:1995ye}.

Although the system of equations as analyzed before was derived from the invariance under infinitesimal conformal transformations, we also checked that generalized Hamada operator is invariant under finite transformations too. Therefore this proves that there is not any topological obstacle in the group space of the full conformal group and we can easily extend the results obtained in the vicinity of identity (or origin in the conformal algebra) to the full group.

\subsection{Theorem about the nonexistence of $\Delta_{n}$ in
$d=n-2$}
\label{s5b}

We already have seen that this theorem works for $n=4$ and $d=2$
(singular case of the generalized Paneitz operator) and for $n=6$
and $d=4=n-2$ and $d=2=n-4$ from the generalized Hamada operator.

To prove this theorem in a bigger generality, one has to recall a
small number of facts about the construction of the operator $\Delta_{n}$
in general dimension $d$. The leading (in derivatives) term of this
operator must necessarily have the form $\square^{\frac{n}{2}}$.
Moreover, following the listing of terms as written by Hamada we know
that the following term with exactly two derivatives acting on the
scalar field $\phi$ can be present
\begin{equation}
z_{*}\square^{\frac{n-4}{2}}R_{\mu\nu}\nabla^{\mu}\nabla^{\nu}\phi,
\end{equation}
of course, among many others. But for the sake of the proof it is
necessary to concentrate on these two terms in $\Delta_{n}$. Next,
one recalls the infinitesimal conformal transformation laws for the
scalar field $\phi$ (with for the moment arbitrary conformal weight
$w$) and for the covariant metric tensor of spacetime, which are
\begin{equation}
\delta_{c}\phi=w\Omega\phi,
\end{equation}
\begin{equation}
\delta_{c}g_{\mu\nu}=2\Omega g_{\mu\nu}.
\end{equation}
The main idea of the proof concentrates on the following term
\begin{equation}
X=\square^{\frac{n-4}{2}}R_{\mu\nu}\nabla^{\mu}\Omega\nabla^{\nu}\phi\label{eq: Xterm}
\end{equation}
in the long expression for the infinitesimal local conformal variation
of the action of the operator on the scalar field, so on the quantity
$\delta_{c}\left(\Delta_{n}\phi\right)$. We will not write other
terms in $\delta_{c}\left(\Delta_{n}\phi\right)$ since they are irrelevant
and, of course, for increasing $n$ the number of these terms grows
very fast.

The first pertinent observation is that only two terms in the general
expansion of the operator $\Delta_{n}$ contribute anything to the
$X$ term in $\delta_{c}\left(\Delta_{n}\phi\right)$. They are precisely:
the leading term (with fixed overall coefficient)
\begin{equation}
\square^{\frac{n}{2}}\phi
\end{equation}
and the one we have singled out already
\begin{equation}
z_{*}\square^{\frac{n-4}{2}}R_{\mu\nu}\nabla^{\mu}\nabla^{\nu}\phi.
\end{equation}

That all other possible terms in $\Delta_{n}\phi$ do not contribute
to $X$ one can easily convince oneself by considering the structure
of the background of the $X$ term. The choice $\square^{\frac{n-4}{2}}R_{\mu\nu}$
as the background quantity in $X$ is very peculiar since the number
of derivatives acting on Ricci tensor is here maximal (if in the $X$
term we have to have derivatives of conformal parameter $\Omega$)
and hence it is very difficult to obtain it as a result of conformal
variation from other terms in $\Delta_{6}$. 

In order to arrive to such results of the expansion of the conformal
variation of the action of the operator on the scalar field $\Delta_{n}\phi$,
one must pick up some convention how to order derivatives in various
terms. We decide to pick up one which seems to us a very natural (and
it is also motivated by the work of Barvinsky, Vilkovisky and others).
First, derivative operators should act on a single quantity (not on
a product of them), like on a scalar field $\phi$, parameter $\Omega$
or on some gravitational curvatures. To achieve this one has to sufficiently
many times perform expansion of the action of derivatives using the
Leibniz rules. When derivatives acting on the same quantity have indices
contracted, these two derivatives should be collected in powers of
the GR-covariant box operator ($\square=g^{\mu\nu}\nabla_{\mu}\nabla_{\nu}$).
When derivatives and box differential operators act on the same quantity,
then the box operators should act with priority. To achieve this order
of covariant derivatives and boxes acting before covariant derivatives,
one has to sufficiently many times use commutation of covariant derivatives,
producing of course terms with gravitational curvatures. This is a
convention of writing all terms in $\delta_{c}\left(\Delta_{n}\phi\right)$.
This convention still leaves some ambiguity (of how to order free
uncontracted derivatives), however for the term $X$ this is completely
unambiguous (since there are no more than one free covariant derivative
acting on scalars $\phi$ and $\Omega$ there). Moreover, to continue
with the proof some convention has to be chosen, but of course, the
final result of nonexistence does not depend at all on the choice
of convention of how the terms are written in $\delta_{c}\left(\Delta_{n}\phi\right)$,
because different ways of writing all of them are perfectly equivalent
and they related by identities from differential geometry.

Having discussed the convention of writing terms in $\delta_{c}\left(\Delta_{n}\phi\right)$,
we can now give final results. We find that the conformal variation
of the leading term $\square^{\frac{n}{2}}\phi$ contains
\begin{equation}
\delta_{c}\left(\square^{\frac{n}{2}}\phi\right)=\delta_{c}\left(\square^{\frac{n}{2}-2}\square^{2}\phi\right)\supset\square^{\frac{n}{2}-2}\left(\delta_{c}\left(\square^{2}\phi\right)\right)\supset2(2w+d-2)\square^{\frac{n}{2}-2}R_{\mu\nu}\nabla^{\mu}\Omega\nabla^{\nu}\phi,
\end{equation}
where the inclusion symbol $\supset$ denotes that of course there
are also other terms in the conformal variation, but they do not interest
us here. To obtain this formula we use first the general formula for
the conformal variation of the scalar box acting on some scalar field
$Y$
\begin{equation}
\delta_{c}\left(\square Y\right)=\square\delta_{c}Y-2\Omega\square Y+(d-2)\nabla_{a}\Omega\nabla_{a}Y,
\end{equation}
where $Y$ is a scalar from the point of view of GR, but it does not
even have to have assigned a conformal weight\footnote{It is only important everything which is encoded in the transformation
$\delta_{c}Y$.}. Of course, when $Y=\phi$, we know that conformal weight can be
assigned, and then $w(Y)=w(\phi)$, but for example such a weight
cannot be defined when $Y=\square\phi$, since the last does not transform
co-covariantly. Here we treat the operator $\square^{\frac{n}{2}-2}$
as a spectator acting on the variation of $\square^{2}\phi$ and that
is why nothing in the coefficient of the result depend on $n$. We
order terms with derivatives in this conformal variation and we use
the commutation rules
\begin{equation}
\left[\square,\nabla_{\mu}\right]\phi=R_{\mu\nu}\nabla^{\nu}\phi
\end{equation}
and
\begin{equation}
\left[\square,\nabla_{\mu}\right]\Omega=R_{\mu\nu}\nabla^{\nu}\Omega.
\end{equation}

Similarly, we find that
\begin{equation}
\delta_{c}\left(\square^{\frac{n-4}{2}}R_{\mu\nu}\nabla^{\mu}\nabla^{\nu}\phi\right)\supset\square^{\frac{n-4}{2}}R_{\mu\nu}\delta_{c}\left(\nabla^{\mu}\nabla^{\nu}\phi\right)\supset2(w-1)\square^{\frac{n}{2}-2}R_{\mu\nu}\nabla^{\mu}\Omega\nabla^{\nu}\phi.
\end{equation}
It is important for the proof to notice the exact coefficients in
front of the $X$ term in the conformal variation $\delta_{c}\left(\Delta_{n}\phi\right)$.
In the second expression we see that the final result does not completely
depend on $n$ since $\square^{\frac{n}{2}-2}R_{\mu\nu}$ is treated
here like a spectator for this derivation. Similarly, there is no
dependence on the dimension $d$ of the spacetime, since the dimension
is not needed in the linearized conformal variation
\begin{equation}
\delta_{c}\left(\nabla_{\mu}\nabla_{\nu}\phi\right)=\nabla_{\mu}\nabla_{\nu}\delta_{c}\phi-2\nabla_{(\mu}\Omega\nabla{}_{\nu)}\phi+g_{\mu\nu}\nabla_{\rho}\Omega\nabla^{\rho}\phi.
\end{equation}

Now, the key point for the proof is that the conformal weight of the
scalar field $\phi$ is in $d=n-2$ equal precisely to $w=1$. Then
this implies that the contribution of the leading term is non-vanishing
in $\delta_{c}\left(\Delta_{n}\phi\right)$
\begin{equation}
\delta_{c}\left(\square^{\frac{n}{2}}\phi\right)\supset2d\square^{\frac{n}{2}-2}R_{\mu\nu}\nabla^{\mu}\Omega\nabla^{\nu}\phi,
\end{equation}
while from the other term we find
\begin{equation}
\delta_{c}\left(\square^{\frac{n-4}{2}}R_{\mu\nu}\nabla^{\mu}\nabla^{\nu}\phi\right)\supset2\left.(w-1)\right|_{w=1}\square^{\frac{n}{2}-2}R_{\mu\nu}\nabla^{\mu}\Omega\nabla^{\nu}\phi=0.
\end{equation}
So no contribution from the second term! Hence it is impossible by
playing with the front coefficient $z_{*}$ (of the second term) to
cancel the contribution 
\begin{equation}
2d\square^{\frac{n}{2}-2}R_{\mu\nu}\nabla^{\mu}\Omega\nabla^{\nu}\phi
\end{equation}
of the leading term to the conformal variation $\delta_{c}\left(\Delta\phi\right)$
on a general non-Ricci-flat background. The other possible terms in
$\Delta_{n}$ cannot help with the cancellation of this contribution
either. Therefore by any choice of the arbitrary coefficients (like
$z_{*}$) in the expansion of the operator $\Delta_{n}$ we are unable
to make the whole local conformal variation vanish. In conclusion,
such a conformal operator $\Delta_{n}$ acting on scalar $\phi$ does
not exist in $d=n-2$ (so two below the critical dimension $d=n$
where the scalar field is dimensionless). This completes the proof
of the absence of $\Delta_{n}\phi$ conformal kinetic operator for
scalars in spacetime of dimensionality $d=n-2$. Probably the operator
also does not exist for any other even dimension $d<n-2$. We therefore
may conjecture that for even dimensionalities $d<n$ smaller than
critical $d_{{\rm crit}}=n$, the operator $\Delta_{n}$ is not constructable
in this way.

The particular examples of application of this theorem with $n=6$
(in $d=4$) and with $n=4$ (in $d=2$) we have seen on explicit forms
of the generalized Paneitz and Hamada operators respectively.

We also comment that a slightly more generalized theorem about the inexistence of conformally covariant operators was presented in \cite{gover0}, where however the more complicated arguments were used involving Q-curvature, tractor calculus and conformal holonomy. We believe that our simple proof using only the methods of  infinitesimal conformal variations is more accessible to any high energy physicist.\\

\section{Wunsch conformal covariant derivatives}
\label{s6}

In this section we discuss various other ways how the general Hamada
operator can be presented. In the sense we will employ different bases
than this one used in section \ref{s4} and \ref{s5}.

\subsection{Review of the construction of conformal covariant derivative due
to Wunsch}

Firstly, we comment on the results obtained by Wunsch \cite{Wunsch1,Wunsch2}.
The main element of his construction is the usage of the specially
constructed conformal covariant derivative, denoted by $\overset{c}{\nabla}_{\mu}$
and differential operators built with it. In order to introduce the
mathematical description of this new covariant derivative, we need
a few additional technical details. We define the Schouten tensor
$P_{\mu\nu}$ by
\begin{equation}
P_{\mu\nu}=\frac{1}{d-2}\left(R_{\mu\nu}-\frac{1}{2(d-1)}g_{\mu\nu}R\right)\label{eq: schouten}
\end{equation}
in any dimension $d\neq1,2$, where $R_{\mu\nu}$ and $R$ denote
above standard Ricci tensor and Ricci (curvature) scalar. The derived
property of the Schouten tensor is its trace, which equals to $P=g^{\mu\nu}P_{\mu\nu}=\frac{R}{2(d-1)}$
in any dimension $d\neq1$. In particular, the last formula, holds
also for the case $d=2$, despite the formula in (\ref{eq: schouten})
being apparently singular in $d=2$. Knowing that in $d=2$ the Ricci
tensor is not independent since it is completely expressed via the
Ricci scalar with the formula $R_{\mu\nu}=\frac{1}{2}g_{\mu\nu}R$
and the same for other curvature tensors, symmetric and with two indices
(so also $P_{\mu\nu}=\frac{1}{2}g_{\mu\nu}P$), one can reexpress
the Schouten tensor there. We have that in $d=2$, $P=\frac{1}{2}R$,
hence $P_{\mu\nu}=\frac{R}{4}g_{\mu\nu}$ and this proves that the
Schouten tensor is not singular in $d=2$ and its expression is finite
and does not vanish since for 2-dimensional manifolds Ricci scalar
$R$ can take arbitrary values. However, we have to be very careful
with the expressions for Schouten tensor and its trace in $d=2,$
as we also show later (with the discussion of the impossibility of
the construction of the Paneitz operator in $d=2$).

An important property of the Schouten tensor is that under infinitesimal
conformal transformation its variation takes quite simple form, namely
$\delta_{c}P_{\mu\nu}=-\nabla_{\mu}\nabla_{\nu}\Omega$, where $\Omega$
is the parameter of transformation. Below we will work with differential
operators acting only on scalar fields $\phi$, which have already
good transformation law under conformal symmetry. That is we assume,
as earlier, that
\begin{equation}
\delta_{c}\phi=w_{0}\Omega\phi.
\end{equation}
The idea of the construction with conformal covariant derivatives
acting in sequence and eventually on a scalar field $\phi$ is the
following. We treat a sequence of conformal covariant derivatives
with various Lorentz indices as one differential operator, which always
acts on a scalar $\phi$ and in this way we do not have to consider
conformal derivatives acting on different more complicated representations
of the Poincare group. For example, the expression $\overset{c}{\nabla}_{\nu}\overset{c}{\nabla}_{\mu}\phi$
(to be defined shortly) we treat as one differential operator acting
on $\phi$ and not as a Wunsch conformal covariant derivative $\overset{c}{\nabla}_{\nu}$
acting on a GR vector $v_{\mu}=\overset{c}{\nabla}_{\mu}\phi$.

We initiate the explicit construction of Wunsch conformal covariant
derivatives by the simple case of one derivative. Then we have the
definition that
\begin{equation}
\overset{c}{\nabla}_{\mu}\phi=\nabla_{\mu}\phi.
\end{equation}

Below we summarize the important conformal transformation properties
of the first conformal covariant derivative. We have first that
\begin{equation}
\delta_{c}\left(\overset{c}{\nabla}_{\mu}\phi\right)=w_{0}\nabla_{\mu}\Omega\phi+w_{0}\Omega\nabla_{\mu}\phi.
\end{equation}
The derivative on the scalar field $\phi$ can be for free promoted
to the conformal one here. We can decompose this expression of the
infinitesimal conformal transformation (always linear in the parameter
$\Omega$) for the part without and with only first derivatives of
the parameter $\Omega$ according to
\begin{equation}
\delta_{c}\left(\overset{c}{\nabla}_{\mu}\phi\right)=\left.\left(\delta_{c}\left(\overset{c}{\nabla}_{\mu}\phi\right)\right)\right|_{\Omega}+\left.\left(\delta_{c}\left(\overset{c}{\nabla}_{\mu}\phi\right)\right)\right|_{\nabla\Omega},
\end{equation}
where we also find now the identifications that
\begin{equation}
\left.\left(\delta_{c}\left(\overset{c}{\nabla}_{\mu}\phi\right)\right)\right|_{\Omega}=w_{0}\Omega\nabla_{\mu}\phi=w_{0}\Omega\overset{c}{\nabla}_{\mu}\phi
\end{equation}
(the last equality is written to be consistent with the formalism
of all conformal derivatives) and
\begin{equation}
\left.\left(\delta_{c}\left(\overset{c}{\nabla}_{\mu}\phi\right)\right)\right|_{\nabla\Omega}=w_{0}\nabla_{\mu}\Omega\phi=\nabla_{\gamma}\Omega X^{\gamma}{}_{\mu}\phi,
\end{equation}
where in the last equality we also have defined the new tensor $X$,
here with two indices in the mixed position $X^{\gamma}{}_{\mu}$,
so the first GR-covariant derivative of the parameter $\Omega$ can
be with arbitrary index $\gamma$ and there is a linear matrix multiplication
involved to get $\left.\left(\delta_{c}\left(\overset{c}{\nabla}_{\mu}\phi\right)\right)\right|_{\nabla\Omega}$
from the first derivatives of $\Omega$. Explicitly, we can define
formally the tensor $X$ as the result of the following linear operation
of double variational derivative:
\begin{equation}
X^{\gamma}{}_{\mu}=\frac{\delta^{2}\left(\left.\left(\delta_{c}\left(\overset{c}{\nabla}_{\mu}\phi\right)\right)\right|_{\nabla\Omega}\right)}{\delta\left(\nabla_{\gamma}\Omega\right)\delta\phi}\label{eq: Xtensorfirstdef}
\end{equation}
and its expression reads $X^{\gamma}{}_{\mu}=w_{0}\delta_{\mu}^{\gamma}$
in this simple case.

Next, we proceed recursively, definining a sequence of two conformal
derivative as
\begin{equation}
\overset{c}{\nabla}_{\nu}\overset{c}{\nabla}_{\mu}\phi=\nabla_{\nu}\overset{c}{\nabla}_{\mu}\phi+P_{\nu\gamma}X^{\gamma}{}_{\mu}\phi=\nabla_{\nu}\nabla_{\mu}\phi+P_{\nu\gamma}X^{\gamma}{}_{\mu}\phi,\label{eq: twoderconfWunsch}
\end{equation}
where the tensor $X^{\gamma}{}_{\mu}=w_{0}\delta_{\mu}^{\gamma}$
we use from the previous definitions for the first conformal derivative.
One can pause here for a moment to contemplate the properties of the
just defined new differential operators with respect to conformal
transformations. We have them explicitly. The infinitesimal conformal
variation of $\overset{c}{\nabla}_{\mu}\phi$ contains only terms
with no derivatives on the parameter $\Omega$ or with precisely first
derivative on $\Omega$. Regarding the transformation of $\overset{c}{\nabla}_{\nu}\overset{c}{\nabla}_{\mu}\phi$
one can convince oneself that the same is true, mainly because of
the properties of the Schouten tensor and the fact that the $X^{\gamma}{}_{\mu}$
tensor is not touched by conformal transformation since it is built
with Kronecker delta(s). For the last case, the speciality is that
there are no terms like $\nabla_{\alpha}\nabla_{\beta}\Omega$ generated
in the expression for $\delta_{c}\left(\overset{c}{\nabla}_{\nu}\overset{c}{\nabla}_{\mu}\phi\right)$.
This is the defining property and feature of Wunsch construction.
His conformal covariant derivative operators $\overset{c}{\nabla}_{\mu}$
transform infinitesimally only up to the first derivative of the parameter
$\Omega$ and higher derivatives are not allowed. (For the specification
and counting of derivatives on $\Omega$ we can just use standard
covariant derivatives from GR, that is $\nabla_{\mu}$.) 

One can see here that this notion of ``conformal'' covariant
derivative is generalized since in standard gauge theory (like in
QED, QCD) the respective covariant derivatives transform without terms
with derivatives of the parameters of gauge transformations; they
are constructed in such a way to transform precisely as the matter
field they are acting upon. This is there obtained by introducing
special gauge connections which balance the first and higher derivatives
of the parameter of gauge transformations in the law for gauge transformations
of total gauge covariant derivatives acting on some matter field charged
with respect to a gauge group. Here, with conformal symmetry, we are
with a different set of tricks. Firstly, we do not employ additional
conformal connection, special for the conformal transformations, although
this is possible and it would lead to the conformal gauge theory.
Secondly, since we accept that the first (in a sequence, that is the
most internal or the most closed to the scalar field $\phi)$ conformal
covariant derivative is identical to the standard GR-covariant derivative
(and also identical here to the partial derivative since on a scalar
we get $\nabla_{\mu}\phi=\partial_{\mu}\phi$), then we must also
accept that this derivative has the conformal transformation law with
precisely first derivative on $\Omega$. And there is no a theoretical
way to cancel these first derivatives keeping general covariance of
the results (this is why here we cannot use for compensation the Christoffel
symbols of the first kind, that is $\Gamma^{\mu}=g^{\nu\rho}\Gamma^{\mu}{}_{\nu\rho}$
or $\Gamma_{\rho}=\Gamma^{\mu}{}_{\mu\rho}$). Had we also accepted
that the sequence of two derivatives is identical with two normal
GR-derivatives, then we would have to accept inevitability of two
derivatives on $\Omega$ in the conformal transformation law and so
on for higher terms. We decided, following Wunsch, not to use a special
conformal connection, which would be a GR vector field $A_{\mu}$
quite similar to the electromagnetic potential in QED, which is alien
to the GR framework; however, we could still use the geometric elements:
tensors and connections as all used in framework of differential geometry.
As a matter of fact, we cannot compensate and modify the first conformal
covariant derivative $\overset{c}{\nabla}_{\mu}$ and we must accept
the fact of the presence of derivatives of $\Omega$ in the transformation
laws. However, for the definition of the sequence of two conformal
derivatives we may employ some addition of a special tensor, as in
(\ref{eq: twoderconfWunsch}), chosen in such a way to precisely cancel
the terms with two derivatives of the conformal parameter $\Omega$.
In a sense, the Schouten tensor plays here the ``role''
of conformal compensator for two derivatives (so it acts like a conformal
``connection'' but it is built entirely out of the elements
available in GR, does not require new additional connection, and it
behaves as a tensor with respect to the diffeomorphism transformations
of GR, differently than for example the Levi-Civita connection in
GR). The Schouten tensor is the first object that we could use here
from GR, which is built with tensors and has a useful infinitesimal
conformal transformation law. The fact that it is a tensor helps because
then also the new differential operator $\overset{c}{\nabla}_{\mu}$
behaves like a good tensorial (or precisely vectorial) operatorial
expression from the point of view of GR.

Consistently with the previous discussion, here we also list the properties
under infinitesimal conformal transformations of the just defined
sequence of two Wunsch conformal derivatives. We have first that
\begin{equation}
\delta_{c}\left(\overset{c}{\nabla}_{\nu}\overset{c}{\nabla}_{\mu}\phi\right)=w_{0}\Omega\overset{c}{\nabla}_{\nu}\overset{c}{\nabla}_{\mu}\phi+2\left(w_{0}-1\right)\nabla_{(\nu}\Omega\overset{c}{\nabla}_{\mu)}\phi+g_{\nu\mu}\nabla_{\gamma}\Omega\overset{c}{\nabla}{}^{\gamma}\phi.
\end{equation}
The derivatives on the scalar field $\phi$ are all written as conformal
ones here. We can decompose this expression of the infinitesimal conformal
transformation (always linear in the parameter $\Omega$) for the
part without and with only first derivatives of the parameter $\Omega$
according to
\begin{equation}
\delta_{c}\left(\overset{c}{\nabla}_{\nu}\overset{c}{\nabla}_{\mu}\phi\right)=\left.\left(\delta_{c}\left(\overset{c}{\nabla}_{\nu}\overset{c}{\nabla}_{\mu}\phi\right)\right)\right|_{\Omega}+\left.\left(\delta_{c}\left(\overset{c}{\nabla}_{\nu}\overset{c}{\nabla}_{\mu}\phi\right)\right)\right|_{\nabla\Omega},
\end{equation}
where we also find now the following identifications that
\begin{equation}
\left.\left(\delta_{c}\left(\overset{c}{\nabla}_{\nu}\overset{c}{\nabla}_{\mu}\phi\right)\right)\right|_{\Omega}=w_{0}\Omega\overset{c}{\nabla}_{\nu}\overset{c}{\nabla}_{\mu}\phi
\end{equation}
and
\begin{equation}
\left.\left(\delta_{c}\left(\overset{c}{\nabla}_{\nu}\overset{c}{\nabla}_{\mu}\phi\right)\right)\right|_{\nabla\Omega}=\left(w_{0}-1\right)\nabla_{\nu}\Omega\overset{c}{\nabla}_{\mu}\phi+\left(w_{0}-1\right)\nabla_{\mu}\Omega\overset{c}{\nabla}_{\nu}\phi+g_{\nu\mu}\nabla_{\gamma}\Omega\overset{c}{\nabla}{}^{\gamma}\phi\label{eq: var2ndder}
\end{equation}
(the last equality is written to be consistent with the formalism
of all conformal derivatives on $\phi$). The last equation can be
also written in the formal way as
\begin{equation}
\left.\left(\delta_{c}\left(\overset{c}{\nabla}_{\nu}\overset{c}{\nabla}_{\mu}\phi\right)\right)\right|_{\nabla\Omega}=\nabla_{\gamma}\Omega X^{\gamma}{}_{\nu\mu}{}^{\alpha}\overset{c}{\nabla}_{\alpha}\phi,\label{eq: twoderXtensor}
\end{equation}
where in the last equality we also have defined the new tensor $X$,
here with four indices in the mixed position $X^{\gamma}{}_{\nu\mu}{}^{\alpha}$,
here two indices are covariant and two others are contravariant, but
these are all not at all equivalent indices and their positions matter.
Therefore, in the expression (\ref{eq: twoderXtensor}) the first
GR-covariant derivative of the parameter $\Omega$ can be with arbitrary
index $\gamma$ and there is a linear matrix multiplication involved
to get $\left.\left(\delta_{c}\left(\overset{c}{\nabla}_{\nu}\overset{c}{\nabla}_{\mu}\phi\right)\right)\right|_{\nabla\Omega}$
from the first derivatives of $\Omega$. Moreover, we have also a
linear matrix multiplication to produce $\left.\left(\delta_{c}\left(\overset{c}{\nabla}_{\nu}\overset{c}{\nabla}_{\mu}\phi\right)\right)\right|_{\nabla\Omega}$
from the first conformal covariant derivatives of the scalar field
$\phi$ here, i.e. $\overset{c}{\nabla}_{\alpha}\phi$ with an arbitrary
index $\alpha$. This form of the equation as in (\ref{eq: twoderXtensor})
containing two first derivatives of $\Omega$ and $\phi$ respectively
is also required based on dimensional arguments. Explicitly, here
we can define formally the tensor $X$ as the result of the following
linear operation of double variational derivative:
\begin{equation}
X^{\gamma}{}_{\nu\mu}{}^{\alpha}=\frac{\delta^{2}\left(\left.\left(\delta_{c}\left(\overset{c}{\nabla}_{\nu}\overset{c}{\nabla}_{\mu}\phi\right)\right)\right|_{\nabla\Omega}\right)}{\delta\left(\nabla_{\gamma}\Omega\right)\delta\left(\overset{c}{\nabla}_{\alpha}\phi\right)}\label{eq: Xtensorseconddef}
\end{equation}
and its expression based on (\ref{eq: var2ndder}) reads 
\begin{equation}
X^{\gamma}{}_{\nu\mu}{}^{\alpha}=\left(w_{0}-1\right)\delta_{\nu}^{\gamma}\delta_{\mu}^{\alpha}+\left(w_{0}-1\right)\delta_{\mu}^{\gamma}\delta_{\nu}^{\alpha}+g_{\nu\mu}g^{\gamma\alpha}\label{eq: X4ind}
\end{equation}
in this case. By inspection of the above formula, we see that this
tensor is accidentally symmetric in $\mu$ and $\nu$ pair of covariant
indices, together with explicit symmetry in contravariant indices
$\gamma$ and $\alpha$. However, the role of the two contravariant
indices $\gamma$ and $\alpha$ is different, hence we view this last
symmetry as without deeper meaning which is also proved by considerations
for higher number of derivatives, where it is absent.

Now, we discuss the next step in the recurrence procedure of defining
the sequence of conformal covariant derivatives. For three consecutive
derivatives, we postulate that
\begin{equation}
\overset{c}{\nabla}_{\rho}\overset{c}{\nabla}_{\nu}\overset{c}{\nabla}_{\mu}\phi=\nabla_{\rho}\overset{c}{\nabla}_{\nu}\overset{c}{\nabla}_{\mu}\phi+P_{\rho\gamma}X^{\gamma}{}_{\nu\mu}{}^{\alpha}\overset{c}{\nabla}_{\alpha}\phi,\label{eq: threedersW}
\end{equation}
where this time more complicated version of the $X$ tensor with four
indices we used from the formula (\ref{eq: X4ind}). We see that in
the definition (\ref{eq: threedersW}) we invoke the expression with
two conformally covariant derivatives (\ref{eq: twoderconfWunsch}),
when the most external (most distant from the scalar field $\phi$)
derivative, i.e. here $\overset{c}{\nabla}_{\rho}$, is opened. At
the same time, in (\ref{eq: threedersW}), the form of the $X$ tensor
is more complicated because of addition of two new indices and finally
due to dimensional arguments, since the $X$ tensor is always constructed
as dimensionless, at the end the expression must be proportional to
the first conformal covariant derivative on the scalar field $\phi$,
i.e. $\overset{c}{\nabla}_{\alpha}\phi$. Here this last in the expression,
the first covariant derivative comes with a Lorentz index $\alpha$
that is a dummy index and contracted with the last contravariant index
$\alpha$ on the $X$ tensor. The more complicated form of the $X$
tensor is required because of new indices, however its form is unique
to establish the fact that the sequence of three Wunsch conformal
covariant derivatives on the scalar field $\overset{c}{\nabla}_{\rho}\overset{c}{\nabla}_{\nu}\overset{c}{\nabla}_{\mu}\phi$
under conformal transformation changes maximally only up to the first
derivatives of the parameter $\Omega$. One notices that the $X$
tensor here is covariantly conserved that is $\nabla_{\beta}X^{\gamma}{}_{\nu\mu}{}^{\alpha}=0$
(because it is constructed out of Kronecker deltas and metric tensors)
and that it does not at all transform under conformal transformations,
i.e. $\delta_{c}X^{\gamma}{}_{\nu\mu}{}^{\alpha}=0$ (because it has
the same number of covariant and contravariant indices -- all on
deltas or on metrics).

We also have the following properties of three conformal covariant
derivatives under the transformations of conformal symmetry. We have
first the known decomposition
\begin{equation}
\delta_{c}\left(\overset{c}{\nabla}_{\rho}\overset{c}{\nabla}_{\nu}\overset{c}{\nabla}_{\mu}\phi\right)=\left.\left(\delta_{c}\left(\overset{c}{\nabla}_{\rho}\overset{c}{\nabla}_{\nu}\overset{c}{\nabla}_{\mu}\phi\right)\right)\right|_{\Omega}+\left.\left(\delta_{c}\left(\overset{c}{\nabla}_{\rho}\overset{c}{\nabla}_{\nu}\overset{c}{\nabla}_{\mu}\phi\right)\right)\right|_{\nabla\Omega},
\end{equation}
where we obviously and naturally find that
\begin{equation}
\left.\left(\delta_{c}\left(\overset{c}{\nabla}_{\rho}\overset{c}{\nabla}_{\nu}\overset{c}{\nabla}_{\mu}\phi\right)\right)\right|_{\Omega}=w_{0}\Omega\overset{c}{\nabla}_{\rho}\overset{c}{\nabla}_{\nu}\overset{c}{\nabla}_{\mu}\phi,
\end{equation}
since this is the part without derivatives on $\Omega$, so effectively
when the parameter of conformal transformation $\Omega$ is constant,
so we consider here only global conformal transformations. Obviously
$w_{0}$ is the conformal weight of the scalar $\phi$ and of any
sequence of conformal (or GR-covariant, or partial) derivatives acting
on it, because for the definition of the weight only global transformations
matter. For the part with precisely one derivative on $\Omega$, the
last equation can be also written in the formal way as
\begin{equation}
\left.\left(\delta_{c}\left(\overset{c}{\nabla}_{\rho}\overset{c}{\nabla}_{\nu}\overset{c}{\nabla}_{\mu}\phi\right)\right)\right|_{\nabla\Omega}=\nabla_{\gamma}\Omega X^{\gamma}{}_{\rho\nu\mu}{}^{\alpha\beta}\overset{c}{\nabla}_{\alpha}\overset{c}{\nabla}_{\beta}\phi,\label{eq: threederXtensor}
\end{equation}
where in the last equality we also have defined the new tensor $X$,
here with six indices (twice the number of conformal derivatives in
(\ref{eq: threederXtensor})) in the mixed position $X^{\gamma}{}_{\rho\nu\mu}{}^{\alpha\beta}$,
here three indices are covariant (coming from the indices on conformal
derivatives on the LHS and we preserved their order) and three others
are contravariant, but these are all not at all equivalent indices
and their positions matter. Therefore, in the expression (\ref{eq: threederXtensor})
the first GR-covariant derivative of the parameter $\Omega$ can be
with arbitrary index $\gamma$ and there is a linear matrix multiplication
involved to get $\left.\left(\delta_{c}\left(\overset{c}{\nabla}_{\rho}\overset{c}{\nabla}_{\nu}\overset{c}{\nabla}_{\mu}\phi\right)\right)\right|_{\nabla\Omega}$
from the first derivatives of $\Omega$. Moreover, we have also a
linear matrix multiplication to produce $\left.\left(\delta_{c}\left(\overset{c}{\nabla}_{\nu}\overset{c}{\nabla}_{\nu}\overset{c}{\nabla}_{\mu}\phi\right)\right)\right|_{\nabla\Omega}$
from the two conformal covariant derivatives of the scalar field $\phi$
here, they are in a sequence$\overset{c}{\nabla}_{\alpha}\overset{c}{\nabla}_{\beta}\phi$
with arbitrary indices names $\alpha$ and $\beta$. This form of
the equation as in (\ref{eq: threederXtensor}) containing exactly
one first derivative of $\Omega$ and one second order conformal covariant
derivative of $\phi$ respectively is also required based on dimensional
arguments. We remind the reader that the tensor $X$ is not assumed
to be symmetric in the group of all covariant indices, here $\rho$,
$\nu$, $\mu$ since this is the order of conformal covariant derivatives,
that also do not commute. But as an accident we see that this tensor
is symmetric in $\alpha$ and $\beta$ indices. This is the result
originating from the definition in (\ref{eq: twoderconfWunsch}),
symmetry of two GR-covariant derivatives on a scalar and explicit
symmetry of Schouten tensor in its indices. Of course, as denoted
by horizontal spacing of group of these indices on the tensor $X$,
here on $X^{\gamma}{}_{\rho\nu\mu}{}^{\alpha\beta}$, the last group
of two contravariant indices plays a completely different role than
the first contravariant index $\gamma$, which is contracted with
first derivative on the parameter $\Omega$, while the former ones
are contracted with conformal derivatives acting on scalar $\phi$.
Hence there cannot be any general symmetry between all three contravariant
indices here. Explicitly, here we can define formally the tensor $X$
as the result of the following linear operation of double variational
derivative:
\begin{equation}
X^{\gamma}{}_{\rho\nu\mu}{}^{\alpha\beta}=\frac{\delta^{2}\left(\left.\left(\delta_{c}\left(\overset{c}{\nabla}_{\rho}\overset{c}{\nabla}_{\nu}\overset{c}{\nabla}_{\mu}\phi\right)\right)\right|_{\nabla\Omega}\right)}{\delta\left(\nabla_{\gamma}\Omega\right)\delta\left(\overset{c}{\nabla}_{\alpha}\overset{c}{\nabla}_{\beta}\phi\right)}\label{eq: Xtensorthird}
\end{equation}
and its explicit expression will be given below in a case with general
number $N$ of conformal covariant derivatives acting on a scalar
$\phi$.

Equipped with the knowledge of the case of three conformal covariant
derivatives acting on the scalar field $\phi$ we can now present
the general construction of a sequence of $N$ consecutive derivatives
acting there. This construction is based inductively on the knowledge
of $N-1$ conformal derivatives and the fact that these $N-1$ conformal
derivatives transform under conformal transformations in a decent
way as explained above (that is that only the terms with maximum up
to the first derivative of the parameter $\Omega$ are present). We
define that
\begin{equation}
\overset{c}{\nabla}_{\rho_{1}}\overset{c}{\nabla}_{\rho_{2}}\cdots\overset{c}{\nabla}_{\rho_{N}}\phi=\nabla_{\rho_{1}}\overset{c}{\nabla}_{\rho_{2}}\cdots\overset{c}{\nabla}_{\rho_{N}}\phi+P_{\rho_{1}\gamma}X^{\gamma}{}_{\rho_{2}\ldots\rho_{N}}{}^{\beta_{1}\ldots\beta_{N-2}}\overset{c}{\nabla}_{\beta_{1}}\cdots\overset{c}{\nabla}_{\beta_{N-2}}\phi\label{eq: Ndersconf}
\end{equation}
in a way very similarly to the previous definitions. This definition
of a sequence of $N$ conformally covariant derivatives works for
$N\geqslant2$. The general expression for the $X$ tensor, here with
total of $2(N-1)$ indices with half of them covariant and other half
contravariant, reads
\[
X^{\gamma}{}_{\rho_{2}\ldots\rho_{N}}{}^{\beta_{1}\ldots\beta_{N-2}}=\sum_{k=1}^{N-1}\left(w_{0}-N+k+1\right)\delta_{\rho_{k+1}}^{\gamma}\delta_{\rho_{2}}^{\beta_{1}}\cdots\delta_{\rho_{k}}^{\beta_{k-1}}\delta_{\rho_{k+2}}^{\beta_{k}}\cdots\delta_{\rho_{N}}^{\beta_{N-2}}+
\]
\begin{equation}
+\sum_{k=2}^{N-1}\sum_{l=1}^{k-1}\left(g_{\rho_{k+1}\rho_{l+1}}g^{\gamma\beta_{k-1}}-\delta_{\rho_{k+1}}^{\gamma}\delta_{\rho_{l+1}}^{\beta_{k-1}}\right)\delta_{\rho_{2}}^{\beta_{1}}\cdots\delta_{\rho_{l}}^{\beta_{l-1}}\delta_{\rho_{l+2}}^{\beta_{l}}\cdots\delta_{\rho_{k}}^{\beta_{k-2}}\delta_{\rho_{k+2}}^{\beta_{k}}\cdots\delta_{\rho_{N}}^{\beta_{N-2}}\,.\label{eq: Xgenexplicit}
\end{equation}
This expression corectly reproduces cases for $N=2$ and $N=3$ as
reported above in text and in (\ref{eq: X4ind}). In the expression
(\ref{eq: Ndersconf}), one finds the skeleton of the same construction
as originally in (\ref{eq: twoderconfWunsch}), but this time the
tensor $X$ possesses $2(N-1)$ indices grouped into three sets. And
it is contracted with $(N-2)$-nd order conformal derivative on the
same scalar $\phi$. The first special index on the tensor $X$ is
$\gamma$ to be contracted with precisely the first derivative on
$\Omega$ in the conformal transformation law. The other group of
indices from $\rho_{2}$ to $\rho_{N}$ are remaining free $(N-1)$
indices on the tensor of $N$-th order conformal derivative. In the
final group we find $(N-1)$ indices $\beta_{1}$ to $\beta_{N-1}$
which are all contracted with indices on the $(N-1)$-st order conformal
derivative standing to the most right in (\ref{eq: Ndersconf}). The
equivalent form of the formula (\ref{eq: Ndersconf}) can be written
slightly shorter as
\begin{equation}
\overset{c}{\nabla}_{\rho_{1}}\overset{c}{\nabla}_{\rho_{2}}\cdots\overset{c}{\nabla}_{\rho_{N}}\phi=\nabla_{\rho_{1}}\overset{c}{\nabla}_{\rho_{2}}\cdots\overset{c}{\nabla}_{\rho_{N}}\phi+P_{\rho_{1}\gamma}\frac{\delta\left(\left.\left(\delta_{c}\left(\overset{c}{\nabla}_{\rho_{2}}\cdots\overset{c}{\nabla}_{\rho_{N}}\phi\right)\right)\right|_{\nabla\Omega}\right)}{\delta\left(\nabla_{\gamma}\Omega\right)}\,,
\end{equation}
where we use in recursion the transformation properties of a sequence
of precisely $(N-1)$ conformal derivatives. In this formula we can
call the sequence of the derivatives $\overset{c}{\nabla}_{\rho_{2}}\cdots\overset{c}{\nabla}_{\rho_{N}}\phi$
as an object $Y$ (and forget its indices). It is important that the
object $Y$ transforms up to terms with the first derivatives of the
$\Omega$ parameter. And then we write a general formula for adding
new conformal derivative $\overset{c}{\nabla}_{\rho_{1}}$ on an arbitrary
object $Y$
\begin{equation}
\overset{c}{\nabla}_{\rho_{1}}Y=\nabla_{\rho_{1}}Y+P_{\rho_{1}\gamma}\frac{\delta\left(\left.\left(\delta_{c}Y\right)\right|_{\nabla\Omega}\right)}{\delta\left(\nabla_{\gamma}\Omega\right)}\,.\label{eq: gendefconfder}
\end{equation}

In general, we have for the infinitesimal conformal variation of the
arbitrary long sequence of conformal derivatives
\begin{equation}
\delta_{c}\left(\overset{c}{\nabla}_{\rho_{1}}\overset{c}{\nabla}_{\rho_{2}}\cdots\overset{c}{\nabla}_{\rho_{N}}\phi\right)=\left.\left(\delta_{c}\left(\overset{c}{\nabla}_{\rho_{1}}\overset{c}{\nabla}_{\rho_{2}}\cdots\overset{c}{\nabla}_{\rho_{N}}\phi\right)\right)\right|_{\Omega}+\left.\left(\delta_{c}\left(\overset{c}{\nabla}_{\rho_{1}}\overset{c}{\nabla}_{\rho_{2}}\cdots\overset{c}{\nabla}_{\rho_{N}}\phi\right)\right)\right|_{\nabla\Omega}\label{eq: decompgenN}
\end{equation}
and we find as a consequence of global conformal symmetry that
\begin{equation}
\left.\left(\delta_{c}\left(\overset{c}{\nabla}_{\rho_{1}}\overset{c}{\nabla}_{\rho_{2}}\cdots\overset{c}{\nabla}_{\rho_{N}}\phi\right)\right)\right|_{\Omega}=w_{0}\Omega\overset{c}{\nabla}_{\rho_{1}}\overset{c}{\nabla}_{\rho_{2}}\cdots\overset{c}{\nabla}_{\rho_{N}}\phi.
\end{equation}
For the part linear in derivatives of $\Omega$, we have a general
definition that
\begin{equation}
\left.\left(\delta_{c}\left(\overset{c}{\nabla}_{\rho_{1}}\overset{c}{\nabla}_{\rho_{2}}\cdots\overset{c}{\nabla}_{\rho_{N}}\phi\right)\right)\right|_{\nabla\Omega}=\nabla_{\gamma}\Omega X^{\gamma}{}_{\rho_{1}\ldots\rho_{N}}{}^{\beta_{1}\ldots\beta_{N-1}}\overset{c}{\nabla}_{\beta_{1}}\overset{c}{\nabla}_{\beta_{2}}\cdots\overset{c}{\nabla}_{\beta_{N-1}}\phi,\label{eq: dersigmagenN}
\end{equation}
where here we refer to the tensor $X$ of the one level higher, that
is it must have above precisely $2N$ indices. The general conclusion
is that the $X$ tensor with $2N$ indices appears in the conformal
transformation law of a sequence of exactly $N$ conformal derivatives,
but is heavily used in the definition of conformal derivative for
the sequence of $N+1$ derivatives, like this is done in (\ref{eq: Ndersconf})
for the case of $N$ derivatives. In a formal way we obtain a general
expression for the tensor $X$ as
\begin{equation}
X^{\gamma}{}_{\rho_{1}\ldots\rho_{N}}{}^{\beta_{1}\ldots\beta_{N-1}}=\frac{\delta^{2}\left(\left.\left(\delta_{c}\left(\overset{c}{\nabla}_{\rho_{1}}\overset{c}{\nabla}_{\rho_{2}}\cdots\overset{c}{\nabla}_{\rho_{N}}\phi\right)\right)\right|_{\nabla\Omega}\right)}{\delta\left(\nabla_{\gamma}\Omega\right)\delta\left(\overset{c}{\nabla}_{\beta_{1}}\overset{c}{\nabla}_{\beta_{2}}\cdots\overset{c}{\nabla}_{\beta_{N-1}}\phi\right)}\,.\label{eq: XgenN}
\end{equation}
With the explicit formula for the tensor $X$ in (\ref{eq: Xgenexplicit})
valid for any $N\geqslant1$, one can check that indeed the sequence
of conformal derivatives constructed in a way as in (\ref{eq: Ndersconf})
at the level with $(N+1)$ derivatives changes under conformal transformation
only up to maximal first derivatives of the parameter $\Omega$. For
this one needs to establish a certain quadratic (so non-linear) relation
for the $X$ tensors that can be verified tediously and explicitly
knowing the expression in (\ref{eq: Xgenexplicit}). It reads explicitly
\begin{equation}
X^{[\gamma}{}_{\rho_{1}\ldots\rho_{N}}{}^{\beta_{1}\ldots\beta_{N-1}}X^{\gamma']}{}_{\beta_{1}\ldots\beta_{N-1}}{}^{\alpha_{1}\ldots\alpha_{N-2}}=0,
\end{equation}
where the antisymmetrization is only between two indices $\gamma$
and $\gamma'$, while all indices $\beta_{1}$ to $\beta_{N-1}$ are
in contraction. Moreover, in the above condition the indices which
are additionally free are $\rho_{1}$ to $\rho_{N}$ and in a different
set $\alpha_{1}$ to $\alpha_{N-2}$. This condition can be solved
having the initial condition for $N=1$ that says $X_{\mu}^{\gamma}=w_{0}\delta_{\mu}^{\gamma}$
and an useful recurrence relation satisfied by these $X$ tensors
relating the $X$ tensor with $2N$ indices to the one with $2N-2$.
The recurrence relation satisfied by the $X$ tensor is
\[
X^{\gamma}{}_{\rho_{1}\ldots\rho_{N}}{}^{\beta_{1}\ldots\beta_{N-1}}=\left(w_{0}-N\right)\delta_{\rho_{1}}^{\gamma}\delta_{\rho_{2}}^{\beta_{1}}\cdots\delta_{\rho_{N}}^{\beta_{N-1}}-\sum_{k=2}^{N}\delta_{\rho_{k}}^{\gamma}\delta_{\rho_{2}}^{\beta_{1}}\cdots\delta_{\rho_{k-1}}^{\beta_{k-2}}\delta_{\rho_{1}}^{\beta_{k-1}}\delta_{\rho_{k+1}}^{\beta_{k}}\cdots\delta_{\rho_{N}}^{\beta_{N-1}}+
\]
\begin{equation}
+\sum_{k=2}^{N}g_{\rho_{1}\rho_{k}}g^{\gamma\beta_{k-1}}\delta_{\rho_{2}}^{\beta_{1}}\cdots\delta_{\rho_{k-1}}^{\beta_{k-2}}\delta_{\rho_{k+1}}^{\beta_{k}}\cdots\delta_{\rho_{N}}^{\beta_{N-1}}+\delta_{\rho_{1}}^{\beta_{1}}X^{\gamma}{}_{\rho_{2}\ldots\rho_{N}}{}^{\beta_{2}\ldots\beta_{N-1}}.\label{eq: recrel}
\end{equation}

The general tensor $X$ possesses the same properties as discussed
above in the case $N=3$. It is completely covariantly conserved and
it does not change under conformal transformations. There is no any
assumed symmetry between all indices covariant $\rho_{1}$ to $\rho_{N}$,
neither between contravariant ones since $\gamma$ is a special type
of index, neither between the group of $(N-1)$ indices $\beta_{1}$
to $\beta_{N-1}$. With its explicit form one can write in full detail
the result for the infinitesimal conformal variation of the sequence
of $N$ conformal derivatives on the scalar $\phi$. This formula
(part linear in the derivative of the $\Omega$ parameter) is given
by
\[
\left.\left(\delta_{c}\left(\overset{c}{\nabla}_{\rho_{1}}\overset{c}{\nabla}_{\rho_{2}}\cdots\overset{c}{\nabla}_{\rho_{N}}\phi\right)\right)\right|_{\nabla\Omega}=\nabla_{\gamma}\Omega X^{\gamma}{}_{\rho_{1}\ldots\rho_{N}}{}^{\beta_{1}\ldots\beta_{N-1}}\overset{c}{\nabla}_{\beta_{1}}\overset{c}{\nabla}_{\beta_{2}}\cdots\overset{c}{\nabla}_{\beta_{N-1}}\phi=
\]
\[
=\sum_{k=1}^{N}\left(w_{0}+k-N\right)\nabla_{\rho_{k}}\Omega\overset{c}{\nabla}_{\rho_{1}}\cdots\overset{c}{\nabla}_{\rho_{k-1}}\overset{c}{\nabla}_{\rho_{k+1}}\cdots\overset{c}{\nabla}_{\rho_{N}}\phi-
\]
\[
-\sum_{k=2}^{N}\sum_{l=1}^{k-1}\nabla_{\rho_{k}}\Omega\overset{c}{\nabla}_{\rho_{1}}\cdots\overset{c}{\nabla}_{\rho_{l-1}}\overset{c}{\nabla}_{\rho_{l+1}}\cdots\overset{c}{\nabla}_{\rho_{k-1}}\overset{c}{\nabla}_{\rho_{l}}\overset{c}{\nabla}_{\rho_{k+1}}\cdots\overset{c}{\nabla}_{\rho_{N}}\phi+
\]
\begin{equation}
+\sum_{k=2}^{N}\sum_{l=1}^{k-1}g_{\rho_{k}\rho_{l}}\nabla_{\gamma}\Omega\overset{c}{\nabla}_{\rho_{1}}\cdots\overset{c}{\nabla}_{\rho_{l-1}}\overset{c}{\nabla}_{\rho_{l+1}}\cdots\overset{c}{\nabla}_{\rho_{k-1}}\overset{c}{\nabla}{}^{\gamma}\overset{c}{\nabla}_{\rho_{k+1}}\cdots\overset{c}{\nabla}_{\rho_{N}}\phi.
\end{equation}

With the usage of conformally covariant derivatives $\overset{c}{\nabla}_{\mu}$
and the tensor of conformal curvature (Weyl tensor $C_{\mu\nu\rho\sigma}$),
metric tensor of spacetime $g_{\mu\nu}$ and basic conformally covariant
matter fields, like the field $\phi$, we can in principle construct
any invariant and conformally covariant expressions (that is not transforming
or transforming only with the $\Omega$ parameter not differentiated
respectively). The theorem here is that any conformal invariant or
conformally covariant expression (or a differential operator) can
be written in a form that uses only these basic building blocks as
mentioned above, and nothing else. The algebra of conformally covariant
and invariant terms is spanned by metric tensors, matter fields $\phi$
and any completely symmetrized sequences of conformal derivatives
acting on basic fields or on Weyl tensors in various combinations
of products. This is the same like in GR where all invariant and GR-covariant
expressions and differential operators are generated by metric tensors,
matter fields $\phi$ and any completely symmetrized sequences of
covariant derivatives acting on basic fields or on curvature tensors
(here Riemann tensors) and all this in various combinations in tensor
products. Here, for conformal symmetry, we must be precise since the
above building elements generate algebra of all expressions and differential
operators that may transform up to the first derivative of the $\Omega$
parameter. Eventually, to assess whether the expression, tensor or
the operator is conformally covariant or invariant we must check whether
the first derivative of $\Omega$ cancel out in the final conformal
transformation law.

We also remark on the range of applicability of our just given definition
of arbitrary Wunsch conformal covariant derivatives. Above, we have
defined it as acting on a sequence of previously defined and constructed
conformal derivatives (with one less derivative in the sequence).
In this way of proceeding we see a clear path from basic conformally
covariant expressions, like the scalar field $\phi$ leading to arbitrary
long expressions with a sequence of conformal derivatives. Finally
such different expressions can be multiplied tensorially to form a
product expression or a differential operator with the desired conformal
transformation law. But one can ask for a definition of conformal
derivative on a general object $Y$ (all Lorentz indices hidden for
writing here), be it a tensorial or operatorial expression and abstracting
from its previous form as a product of expressions with a sequence
of conformal derivatives. Here there is an important restriction for
the possible form of the expression-{}-object $Y$ on which our definition
of $\overset{c}{\nabla}_{\mu}$ is valid. The conformal transformation
of the object $Y$ must be up to terms with first derivatives of the
parameter $\Omega$. For example, when the object $Y$ transforms
with $\square\Omega$ or $\nabla_{\mu}\nabla_{\nu}\Omega$ terms,
then we cannot define the result of acting on $Y$ with the conformal
derivative $\overset{c}{\nabla}_{\mu}$. Because in a sense we cannot
rectify the fact there are already higher than the first derivative
of $\Omega$ present in the transformation law of $Y$ and then they
would also be present in a tentative expression for $\overset{c}{\nabla}_{\mu}Y$
and we cannot make it transforming only up to the first derivatives
of $\Omega$. Then we have two cases: or the infinitesimal transformation
law is 
\begin{equation}
\delta_{c}Y=w(Y)\Omega Y
\end{equation}
or
\begin{equation}
\delta_{c}Y=w(Y)\Omega Y+\left.\left(\delta_{c}Y\right)\right|_{\nabla\Omega},\label{eq: case2}
\end{equation}
where the last term in the second formula is written very schematically.
Obviously, in the first case $Y$ is a conformally covariant tensor
expression or differential operator. In the second case, it is a an
expression transforming with up to the first derivative of the $\Omega$
parameter. In these two cases we can very easily define the action
of conformal covariant derivative. We have, with the conformal object
$Y$ (first case) $\overset{c}{\nabla}_{\mu}Y=\nabla_{\mu}Y$ and
its conformal transformation law is simply
\begin{equation}
\delta_{c}\left(\overset{c}{\nabla}_{\mu}Y\right)=w(Y)\Omega\overset{c}{\nabla}_{\mu}Y+w(Y)\nabla_{\mu}\Omega Y.
\end{equation}
In that case, $Y$ is simply a basic conformally covariant tensor
or matter field (e.g. it could be the metric $g_{\mu\nu}$, Weyl tensor
$C_{\mu\nu\rho\sigma}$ or a scalar field $\phi$ with assigned conformal
weight). In the second case, $Y$ is a more complicated object and
it arises only as a result of conformal derivatives acting in a sequence
on some basic conformally covariant object or a tensorial product
of such expressions. But we can forget about this detail of the structure
of the $Y$ object in this case and consider general and abstract
case here. Then the first conformal derivative of the object $Y$
is defined as in formula (\ref{eq: gendefconfder}) and that is why
in the transformation of the second case in (\ref{eq: case2}) we
explicitly singled out the term with first derivatives of the $\Omega$
parameter. Then the conformal transformation law is simply
\begin{equation}
\delta_{c}\left(\overset{c}{\nabla}_{\mu}Y\right)=w(Y)\Omega\overset{c}{\nabla}_{\mu}Y+\left.\left(\delta_{c}\overset{c}{\nabla}_{\mu}Y\right)\right|_{\nabla\Omega},
\end{equation}
where the last term in the above equation is written schematically.
However, using the recurrence relations (\ref{eq: recrel}) as satisfied
by the $X$ tensors, this $\left.\left(\delta_{c}\overset{c}{\nabla}_{\mu}Y\right)\right|_{\nabla\Omega}$
term can be related to $w(Y)$ and $\left.\left(\delta_{c}Y\right)\right|_{\nabla\Omega}$.
Actually, for this last operation, as one can see from solving (\ref{eq: Xgenexplicit})
of the recurrence relations in (\ref{eq: recrel}), one needs to count
the number of conformal covariant derivatives acting on basic conformal
tensors/scalars as hidden in the general object $Y$. In the other
case, when $Y$ is a product of terms, each one possibly transforming
up to the first derivatives of the $\Omega$ parameter, we first should
apply the Leibniz rule for the conformal derivative and then the new
to be defined conformal derivative $\overset{c}{\nabla}_{\mu}$ in
each term in the resulting sum is adding only to the one existing
sequence of conformal derivatives acting on a basic conformal object.
So then the above argumentation with counting the number of these
already existing conformal derivatives and using recurrence for $X$
tensors applies again.

Here we also mention some important properties of the conformal covariant
derivative mimicking the discussion of them as in the standard GR
case. The conformal covariant derivative is a differential operator
mapping general objects transforming with up to the first derivative
of $\Omega$ to another class of objects conformally transforming
with up to the first derivative of $\Omega$. As such differential
operator it is a linear operator satisfying also the standard Leibniz
rule for the derivative on the product. This conformal covariant derivatives
preserves the metric tensor, hence we can define in an unambiguous
way the conformal covariant box operator as it is done in the next
subsection. In general, the conformal derivatives do not commute,
similarly as GR-covariant derivatives. However, the law for conformal
commutation of derivatives is slightly more complicated as we also
show below since it is important for clarifying any ambiguities regarding
the order of conformal derivatives in some cases. Here from the simple
commutation of two covariant derivatives on a vector field with some
assigned conformal weight, we define the Weyl tensor as the tensor
of conformal curvature, the result of such a commutator. 

Finally, some properties do not hold contrary to the case of standard
differential calculus with covariant derivatives in GR. Firstly, the
conformal covariant derivative as an operation on some terms in general
does not commute with the infinitesimal conformal transformations
of the same object. (In GR we have that the covariant derivative obviously
commutes with the infinitesimal diffeomorphism symmetry transformations
since the covariant derivative is constructed in such a way to be
symmetric there.) For example, we do not get a simplified expression
for any conformal analogue version of the Bianchi identity, with conformally
covariant derivatives and on Weyl tensor. This was the case even with
standard GR-covariant derivatives and Weyl tensor, where the differential
Bianchi identity hold only when derivatives are applied on Riemann
tensor. We remind the reader that here on Weyl tensors the conformal
covariant derivative is identical to the GR-covariant one. As the
last property which does not hold (or holds in a restricted and modified
sense) we mention integration by parts. One has to be very careful
with integration by parts of some conformal derivatives under the
action integral. Due to the presence of the Schouten tensor in the
definition in (\ref{eq: gendefconfder}), in general the result of
integration by parts does not produce this conformal derivative acting
on the rest of tensorial expression together with a minus sign. A
term with Schouten tensor constitutes here a correction term. For
this last property under the spacetime volume integral we do not have
to require that the integral is conformally invariant, only properties
under conformal transformation of each of the terms matter and not
of the full spacetime integral with the proper measure. For example,
when all the terms in the integral are conformally covariant objects,
then one use of conformal derivative is identical to the standard
GR one and this conformal derivative can be integrated by parts without
any correction terms. But in general the Schouten tensor appears after
the process of integration by parts under volume spacetime integral
of conformal covariant derivatives.

The fact that the first conformal derivative on any conformally covariant
tensor is equal from the definition to the ordinary GR-covariant derivative
has important consequences as we wrote also earlier. In simple words
we start compensating for the conformal transformations in the definition
of a new conformally adjusted conformal derivative only from the level
of second conformal derivative. Moreover, we do not use a special
independent connection to compensate for transformation in the definition
of these conformal derivatives. From the second level up we use only
the Schouten tensor to compensate since its infinitesimal conformal
transformation law is in a simple form which contains precisely two
derivatives of the $\Omega$ parameter. Because only the second and
higher conformal derivatives are properly adjusted to compensate (but
only partially and not completely) for conformal transformations,
we have that all the objects defined with conformal derivatives transform
up to the first derivative of $\Omega$ parameter. There was no a
modification for the first conformal derivative and this is the mere
reason why the objects with proper conformal derivatives are not fully
conformally covariant. They in most cases do have also the terms in
the transformation law, which are linear in the first derivative of
the $\Omega$ parameter. This is precisely what we mean by only partial
compensation for the conformal transformations. Another implication
of this fact is that as explained above we must use the complicated
construction of arbitrary order of conformal derivatives employing
the definition of $X$ tensors, where the special care is given to
terms linear in the first derivative of $\Omega$. We remind the reader
that the same procedure is exploited for standard compensation of
gauge transformations in the definition of gauge covariant derivatives,
where the gauge connection, which is added on the level of the first
(and also all higher) covariant derivative uses the infinitesimal
transformations with the level of linearity in the $\Omega$ parameter
(but not in its derivatives). In this simple case, one could also
speak about the expression for the corresponding $X$ tensor and its
infinitesimal gauge transformation law, but obviously they are much
simpler. The need for special care on terms linear in $\nabla\Omega$
also gives a small drawback of the whole formalism of such constructed
conformal derivatives, where the terms with first derivative of $\Omega$
must be properly accounted for and finally the cancellation of them
must be sought for the fully conformal operators, tensors and expressions.

We comment that the procedure of constructing covariant derivatives
as presented by Wunsch, can be generalized to any other symmetry group
that acts non-linearly on the curvature tensors constructed out of
the metric, while on the metric as we know the conformal group in
the infinitesimal form does act linearly. We think that there was
nothing special here about conformal symmetry, although of course,
this is a paradigmatic case. If we do not want to use additional connection
related to this non-linear symmetry, and then if we are able to find
some curvature tensor that can play the role of the compensator for
some number of derivatives, then we could employ the same procedure
as proposed by Wunsch. But then generally within this procedure we
must remember that the constructed general symmetry covariant new
derivatives are not exactly ``covariant'' with respect
to this symmetry, but transform with some terms containing up to fixed
number of derivatives on the parameter of the symmetry transformation.
(This number is by one smaller than the number of derivatives on the
level, when we first time used the compensator tensor for the symmetry.
For example, for conformal symmetry as resolved by Wunsch, we used
Schouten tensor, hence this level was with two derivatives.) Therefore,
by using only conformally covariant derivatives as pioneered by Wunsch,
we do not have a certainty that the objects constructed with them
are always totally conformally covariant (transforming infinitesimally
only linearly in $\Omega$ and without derivatives of it) and we must
always eventually verify that the terms with first derivatives of
$\Omega$ in the conformal transformation law do cancel out. On the
other hand, in a search for conformally invariant and covariant expressions,
we might take an approach in which all objects are built using only
Wunsch conformal covariant derivatives and other conformally covariant
tensors (like Weyl or Bach tensor and generalization thereof). In
a sense, if an invariant or conformally covariant expression (or operator)
exists, then it is always possible to write it using the conformal
derivatives and other conformal tensors, but as we saw before this
way of presentation of it is not mandatory.

Above, we discussed in detail the construction of conformal covariant
derivative, where the ``good'' covariant objects transform
up to first derivatives of the parameter of the transformation. For
a general symmetry and along the lines presented above we can still
construct a sequence of $N$ covariant derivatives with respect to
some symmetry. If we add a compensating tensor on the level with $N_{0}$
derivatives, then it means that objects in general transform up to
$N_{0}-1$ derivatives of the parameter. Then we think that the formulas
like (\ref{eq: Ndersconf}), (\ref{eq: decompgenN}) and (\ref{eq: XgenN})
can still work, just that we need more $\gamma$ type of indices (precisely
$N_{0}-1$ of them) and more indices on the tensor $X$, and more
indices on the compensating tensor $P$. The decomposition of the
general infinitesimal transformation law can be still as in (\ref{eq: decompgenN}),
but we need to write there all the terms up to $(N_{0}-1)$ GR-covariant
derivatives of the parameter $\Omega$ of symmetry. All previous terms
in such decomposition are defined and determined by previous order.
However, only the last term (with $(N_{0}-1)$ derivatives of $\Omega$
could be expressed via the newly defined and more general $X$ tensor,
similarly to what was written above in (\ref{eq: dersigmagenN}) for
the case of $N_{0}=2$, now being contracted with precisely $(N_{0}-1)$
derivatives on $\Omega$. In reverse, the general definition of the
$X$ tensor via variational derivative as in (\ref{eq: XgenN}) can
be always applied. Here we only need to derive with respect to $(N_{0}-1)$-st
order of derivatives on the $\Omega$ parameter. We believe that the
analog version of the recurrence relation and its solutions also exist
in that more general case. As it is clearly obvious, above we discussed
explicitly the case of conformal transformations as they act on the
metric in GR and we choose $N_{0}=2$ with the help of Schouten tensor
and this is why we were able to write all formulas explicitly, but
the procedure is more general for any non-linearly acting symmetries
on curvatures and for any $N_{0}$.

\subsection{Conformal box operator and its powers}

Naturally, with the conformal covariant derivative at hand we can
construct a scalar differential operator, which would play the role
of the analogue of the GR-covariant d'Alembertian operator. We use
formula that
\begin{equation}
\overset{c}{\square}\phi=g^{\nu\mu}\overset{c}{\nabla}_{\nu}\overset{c}{\nabla}_{\mu}\phi.\label{eq: boxcdef}
\end{equation}
However, due to the remark made earlier about the character of all
operators and tensors made out of $\overset{c}{\nabla}_{\mu}$ under
conformal transformation law, we must say that the result of the action
of this operator is in general not a conformally covariant object
even if the scalar field was such object. From the point of view of
GR the operator $\overset{c}{\square}\phi=\phi'$ is clearly a new
scalar. However, this scalar with the conformal weight $w_{0}-2$
(because of the contraction with the contravariant metric $g^{\nu\mu}$
in (\ref{eq: boxcdef})), does not only transform conformally covariantly.
In general, here there is also a part with the first derivative of
the parameter $\Omega$ of the transformation. We have explicitly
that
\begin{equation}
\delta_{c}\left(\overset{c}{\square}\phi\right)=\left.\left(\delta_{c}\left(\overset{c}{\square}\phi\right)\right)\right|_{\Omega}+\left.\left(\delta_{c}\left(\overset{c}{\square}\phi\right)\right)\right|_{\nabla\Omega},
\end{equation}
with
\begin{equation}
\left.\left(\delta_{c}\left(\overset{c}{\square}\phi\right)\right)\right|_{\Omega}=\left(w_{0}-2\right)\Omega\overset{c}{\square}\phi
\end{equation}
and
\[
\left.\left(\delta_{c}\left(\overset{c}{\square}\phi\right)\right)\right|_{\nabla\Omega}=\nabla_{\gamma}\Omega g^{\nu\mu}X^{\gamma}{}_{\nu\mu}{}^{\alpha}\overset{c}{\nabla}_{\alpha}\phi=\nabla_{\gamma}\Omega\left\{ 2\left(w_{0}-1\right)+d\right\} g^{\gamma\alpha}\overset{c}{\nabla}_{\alpha}\phi=
\]
\begin{equation}
=\left\{ 2\left(w_{0}-1\right)+d\right\} \nabla^{\alpha}\Omega\overset{c}{\nabla}_{\alpha}\phi\neq0
\end{equation}
based on (\ref{eq: twoderXtensor}) and (\ref{eq: X4ind}). Still,
as with all other Wunsch conformal derivatives, there are no terms
with two (or more) derivatives on $\Omega$ in the conformal transformation
law for $\overset{c}{\square}\phi$. This means that one has to be
careful in defining powers of the $\overset{c}{\square}$ conformal
differential operator as acting on some conformally covariant scalars
or tensors, since strictly speaking they are not powers since the
result of action, that is $\phi'=\overset{c}{\square}\phi$ has different
transformation properties than the original field $\phi$ which is
conformally covariant with the weight $w_{0}$. (This is different
than in the case of normal GR-covariant d'Alembertian operator, where
$\phi'=\square\phi=g^{\mu\nu}\nabla_{\mu}\nabla_{\nu}\phi$ is also
a scalar and we can simply take powers of the same operator $\square$,
acting in the same representation of diffeomorphism group.) Here not
only the weight changes, namely $w(\phi')=w_{0}-2\neq w_{0}$, but
also we have these first derivatives $\nabla\Omega$ in the transformation
law for $\phi'$. Still formally we can define the powers of the $\overset{c}{\square}$
operator acting on the scalar field $\phi$ as
\begin{equation}
\overset{c}{\square}{}^{N}\underset{{\rm df}}{=}g^{\rho_{1}\rho_{2}}\cdots g^{\rho_{2N-1}\rho_{2N}}\overset{c}{\nabla}_{\rho_{1}}\overset{c}{\nabla}_{\rho_{2}}\cdots\overset{c}{\nabla}_{\rho_{2N}}\phi,
\end{equation}
where the order of contractions here matters a lot since in general
the conformal covariant derivatives do not commute (the same like
ordinary GR-covariant derivatives). One here can immediately notice
that the transformation law of the $\overset{c}{\square}{}^{N}$ operator
is determined from some (partial) trace of the corresponding $X$
tensor. Namely, we have
\[
\delta_{c}\left(\overset{c}{\square}{}^{N}\phi\right)=\left.\left(\delta_{c}\left(\overset{c}{\square}{}^{N}\phi\right)\right)\right|_{\Omega}+\left.\left(\delta_{c}\left(\overset{c}{\square}{}^{N}\phi\right)\right)\right|_{\nabla\Omega}=
\]
\begin{equation}
=\left(w_{0}-2N\right)\Omega\overset{c}{\square}{}^{N}\phi+\nabla_{\gamma}\Omega X^{\gamma\beta_{1}\ldots\beta_{2N-1}}\overset{c}{\nabla}_{\beta_{1}}\overset{c}{\nabla}_{\beta_{2}}\cdots\overset{c}{\nabla}_{\beta_{2N-1}}\phi,
\end{equation}
where we used the new definition of the partially contracted $X$
tensor as
\[
X^{\gamma\beta_{1}\ldots\beta_{2N-1}}=g^{\rho_{1}\rho_{2}}\cdots g^{\rho_{2N-1}\rho_{2N}}X^{\gamma}{}_{\rho_{1}\ldots\rho_{2N}}{}^{\beta_{1}\ldots\beta_{2N-1}}=
\]
\begin{equation}
=\sum_{k=1}^{N}\left(2w_{0}+4k-2-4N+d\right)g^{\beta_{1}\beta_{2}}\cdots g^{\beta_{2k-3}\beta_{2k-2}}g^{\gamma\beta_{2k-1}}g^{\beta_{2k}\beta_{2k+1}}\cdots g^{\beta_{2N-2}\beta_{2N-1}},
\end{equation}
so we find that
\[
\nabla_{\gamma}\Omega X^{\gamma\beta_{1}\ldots\beta_{2N-1}}\overset{c}{\nabla}_{\beta_{1}}\overset{c}{\nabla}_{\beta_{2}}\cdots\overset{c}{\nabla}_{\beta_{2N-1}}\phi=
\]
\begin{equation}
=\sum_{k=1}^{N}\left(2w_{0}+4k-2-4N+d\right)\nabla_{\gamma}\Omega\overset{c}{\square}{}^{k-1}\overset{c}{\nabla}{}^{\gamma}\overset{c}{\square}{}^{N-k}\phi.
\end{equation}

Hence finally we can write
\begin{equation}
\delta_{c}\left(\overset{c}{\square}{}^{N}\phi\right)=\left(w_{0}-2N\right)\Omega\overset{c}{\square}{}^{N}\phi+\sum_{k=1}^{N}\left(2w_{0}+4k-2-4N+d\right)\nabla_{\gamma}\Omega\overset{c}{\square}{}^{k-1}\overset{c}{\nabla}{}^{\gamma}\overset{c}{\square}{}^{N-k}\phi.\label{eq: boxnlaw}
\end{equation}
However, in the above formula we cannot simplify further since in
general the conformal covariant derivative does not commute with the
conformal box operator. This issue here is even more complicated than
in the case of standard GR-covariant box operator and GR-covariant
derivatives, because the conformal weight of the result is different
than the original of the scalar field $\phi$. (In GR we have that
the result of $\square^{N}$ on any tensorial representation of the
diffeomorphism group stays in the same representation.) Moreover,
we have also other various terms in the commutation rule for two Wunsch
conformal derivatives, which are related to the form of $X$ tensor.
For definiteness and exact formulas about this issue, we can refer
the interested reader to the original literature \cite{Wunsch1}.
In general, the law of commutation of conformal derivatives $\overset{c}{\nabla}_{\mu}$
(conformal Ricci identity) states that
\begin{equation}
\left[\overset{c}{\nabla}_{\mu},\overset{c}{\nabla}_{\nu}\right]=\left.\left(\vphantom{\overset{c}{\nabla}_{\nu}}\left[\nabla_{\mu},\nabla_{\nu}\right]\right)\right|_{R_{\alpha\beta\rho\sigma}\to C_{\alpha\beta\rho\sigma}}+\frac{1}{d-3}\nabla_{\alpha}C_{\mu\nu\gamma}{}^{\alpha}\frac{\delta\left(\left.\left(\delta_{c}\cdot\right)\right|_{\nabla\Omega}\right)}{\delta\left(\nabla_{\gamma}\Omega\right)},\label{eq: commders}
\end{equation}
where it is understood that it is applied on any conformal tensorial
object standing to the right. About this object it is assumed that
under infinitesimal conformal transformations it changes only with
terms up to the first derivative of the $\Omega$ parameter. The first
term in the above formula describes the same commutation as between
ordinary GR-covariant derivatives, but only with a substitution of
Riemann tensor by the proper tensor of conformal curvature, that is
by Weyl tensor. All terms with any contractions of Riemann tensor
here are substituted by zero since Weyl tensor is completely traceless.
In the second term, we find a modification proportional to the first
contracted derivative of Weyl tensor and to the conformal variation
of the original tensorial object from which the part with exactly
the first derivatives of $\Omega$ is considered and stripped of.
By an explicit derivation, one can convince oneself that a coefficient
of this term is singular in $d=3$ spacetime dimensions. However,
this does not lead to any problem in $d=3$, since there the Weyl
tensor explicitly vanishes and in the formula (\ref{eq: commders})
we are temporarily left only with the first term on the right giving
standard commutation as in GR, which finally in turn also vanishes
since the subsequent substitution of Riemann by Weyl tensor. Hence
our conclusion that in $d=3$ all Wunsch conformally covariant derivatives
completely commute. They also commute when the spacetime background is conformally flat, that is when Weyl tensor vanishes, $C_{\mu\nu\rho\sigma}=0$. This proves that on conformal background the sufficient conformal operator is just $\overset{c}{\square}{}^{N}$.

Another feature of the formula (\ref{eq: commders}) is that it shows
that it is more difficult to define the generalized tensor of curvature
for conformal symmetry as acting on some tensor representation on
the right. For the first term on the right hand side of (\ref{eq: commders})
we see this without a problem since this is in general a matrix which
mixes Lorentz indices of the tensor representation, but this is always
the same tensor representation (scalars, vectors, tensors, etc.) with
the same number of indices before the commutator on the right of $\left[\nabla_{\mu},\nabla_{\nu}\right]$
and in the result. However, the last term in (\ref{eq: commders})
is different since it mixes indices on the tensor representation with
one Lorentz index less (in a strict sense with one less conformal
covariant derivative) than the original tensor representation on which
action of the commutator $\left[\overset{c}{\nabla}_{\mu},\overset{c}{\nabla}_{\nu}\right]$
should be expressed. This is because of the presence of one derivative
on the Weyl tensor in the last term in (\ref{eq: commders}) and also
consistently due to the fact that in the infinitesimal transformation
law we extract the part with precisely one derivative acting on the
$\Omega$ parameter. Then this means that here the matrix in Lorentz
indices operates on a smaller tensorial representation than the original
(one less conformal derivative on the main tensor, while at the same
time one more derivative on the parameter $\Omega$ of the transformation).
This difference between the two representations (original and the
smaller ones used in the last term in (\ref{eq: commders})) is the
main problem in an attempt of describing the uniform matrix of the
tensor of generalized conformal curvature here.

From the analysis of the formula (\ref{eq: boxnlaw}), we see that
the $\overset{c}{\square}{}^{N}\phi$ operator does not in general
behaves like a conformally covariant object (except the case of $d=3$
spacetime dimensions). Although it is a fact that this is a scalar
operator with precisely $2N$ derivatives as considered on the flat
spacetime background. Because of its naturalness and simplicity of
the transformation law in (\ref{eq: boxnlaw}), we could use it, following
Wunsch, as a starting point for a successful construction of completely
conformally covariant differential operators on scalars with many
derivatives. For this purpose, we first can look at the case of small
$N$ of its power exponents.

It is easily seen that for $N=1$, the operator $\overset{c}{\square}\phi$
reproduces exactly the \O{}rsted-Penrose operator as discussed earlier.
From the technical side, one can just consider the equation for the
vanishing of the second coefficient in (\ref{eq: boxnlaw}), namely
\begin{equation}
2w_{0}+4k-2-4N+d=0
\end{equation}
for $N,\,k=1$ and find as a solution for the weigth $w_{0}=\frac{2-d}{2}$,
the same result as previously. The expansion of $\overset{c}{\square}\phi$
using (\ref{eq: schouten}), (\ref{eq: twoderconfWunsch}) and (\ref{eq: boxcdef})
produces exactly the operator in a GR-explicit form
\begin{equation}
\overset{c}{\square}\phi=\left(\square-\frac{d-2}{4(d-1)}R\right)\phi\label{eq: oersted}
\end{equation}
and this also proves that this is the only conformally covariant operator
possible here, with two derivatives, that is $\left.\left(\delta_{c}\left(\overset{c}{\square}\phi\right)\right)\right|_{\nabla\Omega}=0$
(and also $\left.\left(\delta_{c}\left(\overset{c}{\square}\phi\right)\right)\right|_{\Omega}=-\frac{d+2}{2}\Omega\overset{c}{\square}\phi$).
Moreover, as it is well known in $d=2$ the conformal operator coincides
with the standard GR-covariant box operator, namely we have $\left.\overset{c}{\square}\phi\right|_{d=2}=\square\phi$.
This is also explained by the fact that in $d=2$ we do not see any
singularity in (\ref{eq: oersted}) and the trace of the Schouten
tensor in (\ref{eq: schouten}) is not singular, because the interior
of the parenthesis in (\ref{eq: schouten}) vanishes after contracting
with the contravariant metric $g^{\mu\nu}$ and hence the Schouten
tensor and also its trace overall are not singular in $d=2$. This
is in agreement with the previous discussion of the form of the Schouten
tensor in the case of two dimensions. In an explicit expansion of
the formula (\ref{eq: oersted}) after (\ref{eq: twoderconfWunsch}),
we see that the trace of the Schouten tensor, which is non-vanishing
in $d=2$, is multiplied by the factor $(d-2)$ and this is why in
this dimension the Ricci scalar completely does not participate in
the construction of the conformally covariant box operator. In $d=2$,
the Schouten tensor and its trace are regular, but there is no such
contribution multiplying $R$, because the conformal weight $w_{0}$
must vanish in this dimension since it is a critical dimension for
a two-derivative scalar theory.

For the case $N=2$ we start seeing the advantage of using the formula
with conformal covariant derivatives. We derive that the square of
the conformal box operator $\overset{c}{\square}{}^{2}\phi$ behaves
correctly under conformal transformations. For this case, the law
of transformation reads
\[
\delta_{c}\left(\overset{c}{\square}{}^{2}\phi\right)=\left(w_{0}-4\right)\Omega\overset{c}{\square}{}^{N}\phi+\sum_{k=1}^{2}\left(2w_{0}+4k-2-8+d\right)\nabla_{\gamma}\Omega\overset{c}{\square}{}^{k-1}\overset{c}{\nabla}{}^{\gamma}\overset{c}{\square}{}^{2-k}\phi=
\]
\begin{equation}
=\left(w_{0}-4\right)\Omega\overset{c}{\square}{}^{N}\phi+\left(2w_{0}+d-6\right)\nabla_{\gamma}\Omega\overset{c}{\nabla}{}^{\gamma}\overset{c}{\square}\phi+\left(2w_{0}+d-2\right)\nabla_{\gamma}\Omega\overset{c}{\square}\overset{c}{\nabla}{}^{\gamma}\phi\label{eq: box2tr}.
\end{equation}
However, one can see by explicit computation that for three conformal
consecutive derivatives $\overset{c}{\nabla}_{\mu}$ contracted in
a form of conformal box operator $\overset{c}{\square}$ and one derivative
free, the order does not matter and we have as a consequence that
$\overset{c}{\square}\overset{c}{\nabla}{}^{\gamma}\phi=\overset{c}{\nabla}{}^{\gamma}\overset{c}{\square}\phi$.
(This result about the commutation can be understood since in GR-covariant
case the result is proportional to the Ricci tensor only and does
not involve at all usage of the Riemann tensor. With conformal derivatives
Ricci tensor cannot appear, since it is a trace of Riemann, but we
know that the tensor of conformal curvature -- commutator of two
conformal covariant derivatives -- Weyl tensor is completely traceless.
Moreover, from dimensional reasons and by counting indices we cannot
have any derivative of the Weyl tensor and we cannot contract indices
on Weyl tensor, so we must have that $\left[\overset{c}{\square},\overset{c}{\nabla}{}^{\gamma}\right]\phi=0$.)
This means that the law of transformation we simplify to the case
\begin{equation}
\delta_{c}\left(\overset{c}{\square}{}^{2}\phi\right)=\left(w_{0}-4\right)\Omega\overset{c}{\square}{}^{N}\phi+\left(4w_{0}+2d-8\right)\nabla_{\gamma}\Omega\overset{c}{\nabla}{}^{\gamma}\overset{c}{\square}\phi.
\end{equation}
Then the condition for conformal covariance again determines the weight
in this case. We find that $4w_{0}+2d-8=0$ implies that $w_{0}=\frac{4-d}{2}$.
If the conformal weight is assigned, then we have obviously that the
operator has
\begin{equation}
\delta_{c}\left(\overset{c}{\square}{}^{2}\phi\right)=-\frac{d+4}{2}\Omega\overset{c}{\square}{}^{2}\phi,
\end{equation}
so this is a conformally covariantly transforming expression. A bit
longer and tedious exercise shows that the operator $\overset{c}{\square}{}^{2}$
on scalar fields with the weight $w_{0}=\frac{4-d}{2}$ reproduces
completely the Paneitz operator. In this case, the operator is not
completely unique since we have also an ambiguity in adding a conformally
covariant scalar endomorphism. As explained earlier we can add here
$C^{2}\phi=C^{\mu\nu\rho\sigma}C_{\mu\nu\rho\sigma}\phi$ term with
arbitrary front coefficient. This is all consistent because the weight
of the expression $C^{\mu\nu\rho\sigma}C_{\mu\nu\rho\sigma}=C^{\mu\nu\rho}{}_{\sigma}C_{\mu\nu\rho}{}^{\sigma}$
equals $-4$ the same as in the overall expression for $\overset{c}{\square}{}^{2}$
(because of two contravariant metric tensors $g^{\mu\nu}$ used in
its construction). Therefore, the most general linear operator acting
on a scalar field $\phi$ possessing the conformal weight $w_{0}=\frac{4-d}{2}$,
containing up to 4 derivatives in the leading term takes the form
\begin{equation}
\overset{c}{\square}{}^{2}+\alpha C^{\mu\nu\rho\sigma}C_{\mu\nu\rho\sigma},
\end{equation}
where $\alpha$ is arbitrary real dimensionless and numerical constant.
In that case we acknowledge the fact that there is a one-parameter
freedom in defining the conformal operator. One can see that the singularity
$(d-2)$ in the denominator in the definition of the Schouten tensor
in (\ref{eq: schouten}) is here responsible for the fact that the
Paneitz operator $\overset{c}{\square}{}^{2}$ is not possible to
be defined in the limit $d\to2$, as noticed also earlier. Moreover,
here we cannot construct any differential operator with four derivatives,
where all indices on these conformal derivatives are not contracted
with themselves, so we must consider only scalar operators acting
on scalars. Due to the commutation rules of conformal covariant derivatives
mentioned above, we have that when acting on a scalar field
\begin{equation}
\overset{c}{\nabla_{\gamma}}\overset{c}{\nabla}{}^{\gamma}\overset{c}{\nabla_{\beta}}\overset{c}{\nabla}{}^{\beta}\phi=\overset{c}{\nabla_{\gamma}}\overset{c}{\nabla_{\beta}}\overset{c}{\nabla}{}^{\gamma}\overset{c}{\nabla}{}^{\beta}\phi=\overset{c}{\nabla_{\gamma}}\overset{c}{\nabla_{\beta}}\overset{c}{\nabla}{}^{\beta}\overset{c}{\nabla}{}^{\gamma}\phi=\overset{c}{\square}{}^{2}\phi,\label{eq: fourderpos}
\end{equation}
hence there is only one unique order of conformal derivatives here.
For the cases of $N=1$ and $N=2$, we see that the situation is quite
simple and obvious. Due to the algebraic properties we do not have
any other possibility of constructing the conformal operator and in
the transformation law (\ref{eq: boxnlaw}) the simplest operators
constructed must already have good conformal transformation laws (without
derivatives of the parameter $\Omega$). This is caused by the small
size of the algebraic space of independent terms that we could write
here, like in (\ref{eq: box2tr}) (where there is only one independent
term $\overset{c}{\square}\overset{c}{\nabla}{}^{\gamma}\phi$ within
the last two terms) and in (\ref{eq: fourderpos}). As we shall see
for higher $N$ the spaces are bigger, there is more freedom and therefore
the construction of conformally covariant operators is also more complicated.

Starting with $N=3$ the construction becomes more intricate and the
size of the algebraic spaces of terms grows. We find that the operator
$\overset{c}{\square}{}^{3}$ in action on the scalar field $\phi$
with the proper conformal weight $w_{0}$ is not conformally covariant.
It transforms with non-zero terms up to the first derivative of the
$\Omega$ parameter. According to the general formula in (\ref{eq: boxnlaw}),
for the case of $N=3$, we find
\[
\delta_{c}\left(\overset{c}{\square}{}^{3}\phi\right)=\left(w_{0}-6\right)\Omega\overset{c}{\square}{}^{3}\phi+\sum_{k=1}^{3}\left(2w_{0}+4k-14+d\right)\nabla_{\gamma}\Omega\overset{c}{\square}{}^{k-1}\overset{c}{\nabla}{}^{\gamma}\overset{c}{\square}{}^{3-k}\phi=
\]
\[
=\left(w_{0}-6\right)\Omega\overset{c}{\square}{}^{3}\phi+\left(2w_{0}-10+d\right)\nabla_{\gamma}\Omega\overset{c}{\nabla}{}^{\gamma}\overset{c}{\square}{}^{2}\phi+
\]
\begin{equation}
+\left(2w_{0}-6+d\right)\nabla_{\gamma}\Omega\overset{c}{\square}\overset{c}{\nabla}{}^{\gamma}\overset{c}{\square}\phi+\left(2w_{0}-2+d\right)\nabla_{\gamma}\Omega\overset{c}{\square}{}^{2}\overset{c}{\nabla}{}^{\gamma}\phi.
\end{equation}
In the formula above we could use the commutation of three last derivatives
acting on $\phi$ using the previous laws which also apply here, namely
$\overset{c}{\square}\overset{c}{\nabla}{}^{\gamma}\phi=\overset{c}{\nabla}{}^{\gamma}\overset{c}{\square}\phi$
independently on the weight of the scalar field $\phi$, but this
$\phi$ must be with a definite weight (and with no derivatives of
$\Omega$ parameter in the infinitesimal conformal transformation
law, so it cannot be for example $\phi'$ as used earlier). Then we
would get that
\[
\delta_{c}\left(\overset{c}{\square}{}^{3}\phi\right)=\left(w_{0}-6\right)\Omega\overset{c}{\square}{}^{3}\phi+\left(2w_{0}-10+d\right)\nabla_{\gamma}\Omega\overset{c}{\nabla}{}^{\gamma}\overset{c}{\square}{}^{2}\phi+
\]
\begin{equation}
+2\left(2w_{0}-4+d\right)\nabla_{\gamma}\Omega\overset{c}{\square}\overset{c}{\nabla}{}^{\gamma}\overset{c}{\square}\phi
\end{equation}
and even using the a priori known value of the conformal weight (which
is determined from considerations of global conformal symmetry) and
which is $w_{0}=\frac{6-d}{2}$ we remain with the formula
\begin{equation}
\delta_{c}\left(\overset{c}{\square}{}^{3}\phi\right)=\left(w_{0}-6\right)\Omega\overset{c}{\square}{}^{3}\phi-4\nabla_{\gamma}\Omega\left[\overset{c}{\nabla}{}^{\gamma},\overset{c}{\square}\right]\overset{c}{\square}\phi.
\end{equation}
Here we cannot commute derivatives and the conformal box operator
(i.e. change the order of $\overset{c}{\nabla}{}^{\gamma}$ and $\overset{c}{\square}$)
for free, because they all act on $\phi'=\overset{c}{\square}\phi$,
which does not have a transformation law without derivative of the
parameter of the conformal transformations for value of $w_{0}=\frac{6-d}{2}$,
which does not coincide with the value $\frac{2-d}{4}$ required for
the invariance of $\overset{c}{\square}\phi$ for the case of $N=1$
only. This means that the operator $\overset{c}{\square}{}^{3}$ is
not conformally covariant and we need to add modifications to it,
if we think of a six-derivative conformal differential operator. On
different vein, here we see that the space of terms that could appear
in $\left.\left(\delta_{c}\left(\overset{c}{\square}{}^{3}\phi\right)\right)\right|_{\nabla\Omega}$
is bigger, we cannot make all the independent terms there to cancel
out by solely choosing the weight $w_{0}$ and therefore the operator
$\overset{c}{\square}{}^{3}\phi$ is not conformally covariant one.

Following Wunsch, we can stick to the $\overset{c}{\square}{}^{3}$
as the leading part of the operator, containing the highest number
of conformal derivatives, here 6, and just add some subleading terms
in number of them. These other terms due to dimensional reasons must
contain powers of the conformal gravitational curvatures or of its
derivatives. For the sake of the conformal formalism here we can use
Weyl tensor (or Bach tensor -- just to be defined) and not other
GR tensors of curvatures (like Riemann tensor, Ricci tensor or Ricci
scalar) and on these we should act with conformal derivatives only.
This is a necessary condition for writing the modification in the
conformal formalism, but it is not sufficient since we must also check
that in the special modification of the operator $\overset{c}{\square}{}^{3}$,
being the combination of various terms written in the formalism, the
terms with (only first) derivatives of $\Omega$ do cancel out. The similar form for this minimal operator appeared also in works by Branson \cite{branson0,branson1,branson3,branson4,branson5,branson6,fegan}. When
the coefficient in front of $\overset{c}{\square}{}^{3}$ term in
this combination is fixed to unity (just for normalization of the
total operator), Wunsch found the following in a sense ``minimal''
 six-derivative conformal operator
\begin{equation}
D_{(6)}^{\min}\phi=\left(\overset{c}{\square}{}^{3}+\frac{16}{(d-3)(d-4)}B^{\mu\nu}\overset{c}{\nabla}_{\mu}\overset{c}{\nabla}_{\nu}\right)\phi,\label{eq: Wunsch6der}
\end{equation}
where the Bach tensor $B^{\mu\nu}$ is defined as
\begin{equation}
B_{\mu\nu}=\overset{c}{\nabla}{}^{\rho}\overset{c}{\nabla}{}^{\sigma}C_{\rho\mu\nu\sigma}.\label{eq: bachdef}
\end{equation}
The Bach tensor is symmetric in its indices. However, it is conformally
covariant tensor only in $d=4$ spacetime dimensions (with the conformal
weight $w\left(B_{\mu\nu}\right)=-2$) and there it has other interesting
properties (like covariant conservation and it is an Euler-Lagrange
tensor, so it is a variational derivative with respect to the metric
of some gravitational action functional). Here, we use it just as
a building block for construction of the operator $D_{(6)}^{\min}$
and the most important is only its infinitesimal conformal transformation
law, which reads
\begin{equation}
\delta_{c}\left(B_{\mu\nu}\right)=-2\Omega B_{\mu\nu}+2(d-4)\nabla^{\gamma}\Omega\nabla^{\rho}C_{\rho\mu\nu\gamma}.\label{eq: bachconftr}
\end{equation}

Already here we notice that in $d=3$ special odd dimension, we have
that simply $D_{(6)}^{\min}=\overset{c}{\square}{}^{3}$, because
there the Weyl and Bach tensor explicitly vanish. Hence for construction
of conformal operators we can just use conformal derivatives acting
on any object well transforming under conformal transformations. These
conformal derivatives still have to be contracted since our conclusion
relies here on complete commutativity of conformal derivatives and
the formula (\ref{eq: boxnlaw}) for the conformal variation of the
powers of the conformal box operator. These operators $\overset{c}{\square}{}^{N}$
in $d=3$ for any $N$ are still different than standard GR $\square^{N}$
differential operators, because we need to use Wunsch conformal covariant
derivatives $\overset{c}{\nabla}_{\mu}$, instead of GR-covariant
$\nabla_{\mu}$ ones, with the exploitation of the construction with
the Schouten tensor (\ref{eq: schouten}), which in $d=3$ takes some
non-zero finite values. Explicit computation in $d=3$ based on (\ref{eq: boxnlaw})
and with the weight $w_{0}=\frac{2N-3}{2}$ shows that indeed general
terms of the form $\nabla_{\gamma}\Omega\overset{c}{\nabla}{}^{\gamma}\overset{c}{\square}{}^{N-1}\phi$
in (\ref{eq: boxnlaw}) do consistently cancel out giving us the result
that $\overset{c}{\square}{}^{N}$ is the conformal operator with
the total weight $w=-\frac{2N+3}{2}$ of the expression $\overset{c}{\square}{}^{N}\phi$.

In general, the operator $D_{(6)}^{\min}$ has the leading term in
derivatives, in expansion around flat spacetime, when we can neglect
all gravitational curvatures, as $\square^{3}$, so with precisely
six derivatives. In a general framework, we have the following transformations
of this operator under infinitesimal conformal changes:
\begin{equation}
\delta_{c}D_{(6)}^{\min}\phi=\left(w_{0}-6\right)\Omega D_{(6)}^{\min}\phi=-\frac{d+6}{2}\Omega D_{(6)}^{\min}\phi
\end{equation}
with $w_{0}=\frac{6-d}{2}$ -- the conformal weight of the basic
scalar field $\phi$ here.

One notices that the expression for the minimal six-derivative conformal
operator according to Wunsch in (\ref{eq: Wunsch6der}), is singular
in the case of dimensions $d=3$ and $d=4$. We are more concerned
with the last case here. This was also confirmed by our studies in
the previous sections (\ref{s4}), where for the coefficients of expansion, in
front of some terms, we have found the singularity $\frac{1}{d-4}$
near the dimension $d=4$. Therefore, we have another confirmation
of the theorem by Graham \cite{graham1,graham2} that in $d=4$
we cannot construct a conformally covariant differential operator,
with the leading term $\square^{3}=\partial^{6}$ on flat spacetime
in Cartesian coordinates. Moreover, the same applies to the dimensions
$d=2$ where again the operator with six derivatives and the leading
term $\square^{3}$ on flat spacetime background cannot be constructed;
this time due to problems with the limit $d\to2$ in the definition
of the Schouten tensor. One can here put a conjecture (following Branson and Wunsch \cite{branson1,Wunsch1}) that the operator with $2N$ derivatives
cannot be constructed in all even spacetime dimensions smaller than
$2N$, that is for $d<2N$, while the dimension $d=2N$ is the critical.

We may comment here on the general structure of singularities in the
definition of the operators, like the generalized Hamada operator.
From the arguments of the previous section, and also from the results
derived in the tractor formalism \cite{tractor1, tractor2}, we know
that the operators are undefined in the first even subcritical dimension,
that is for $d=2N-2$. For $N\geqslant3$ we see that these dimensions
are greater than 2. In the general construction of the conformally
covariant operators we can use $\overset{c}{\square}{}^{N}$ supplemented
with some other terms built out with Weyl tensors, Bach tensors and
conformally covariant derivatives acting on them or on the scalar
field $\phi$, standing the most to the right of each expressions.
In the definition of these objects and also of Schouten tensor, we
see that there is never a place where we could generate singularities,
like $\frac{1}{d-4}$ for Hamada, or $\frac{1}{d-6}$ for the operator
with 8 derivatives, etc., since we only contract metrics which produces
positive powers of dimensions $d$ in numerators. (In the definition
of Schouten tensor we see singularities only in $d=1$ and $d=2$;
the definition of conformal covariant derivatives as in (\ref{eq: Ndersconf})
is apparently free from any singularities; the $X$ tensor does not
develop any singularity, and in the conformal weights $w_{0}$ we
see the dimensions appearing only in the numerator. The law for commutation
of conformal covariant derivatives as in (\ref{eq: commders}) contains
only $\frac{1}{d-3}$ explicit singularity. Moreover, the Weyl and
Bach tensors can be defined without any problems in any dimensions
$d\geqslant4$.) Still, in the construction of conformal operators
we must see the singularity in these first even subcritical dimensions,
because the operators are not constructible in such situations. This
singularity must show up in some rational functions of the dimension
$d$ of spacetime. Because of the nature of algebraic operations described
below, these can be only rational functions. The only reason why singularity
can appear here is due to a solution of some algebraic linear system
of equations for coefficients in front of additional terms that have
to be added to $\overset{c}{\square}{}^{N}$ to construct a minimal
conformally covariant operator with $2N\geqslant6$ derivatives in
a form $D_{(2N)}^{\min}=\overset{c}{\square}{}^{N}+\ldots$. In other
words, for operators with $2N\geqslant6$ derivatives, we need terms
besides the $\overset{c}{\square}{}^{N}$ and all these terms are
constructed in Wunsch conformal way (having transformation law only
up to the first derivatives of $\Omega$ parameter) and in their construction
we do not see any singularity for dimension $d=2N-2$. However, all
these additional terms must come with some definite coefficients in
order to construct a minimal operator $D_{(2N)}^{\min}$. These coefficients
are determined from the condition that $\left.\left(\delta_{c}\left(D_{(2N)}^{\min}\right)\right)\right|_{\nabla\Omega}=0$.
This last requirement as a tensor equation can be expanded in some
basis and therefore we are led to the linear system of equations for
apriori unknown coefficients. By solving such an algebraic system
we determine the front coefficients and in this procedure we may exploit
inverse functions, where the dimension $d$ appeared only in numerators,
hence we may end up with coefficients being rational functions of
the dimension $d$. These rational functions finally may contain the
singularities like $\frac{1}{d-4}$ or $\frac{1}{d-6}$, etc.. This
explains the reason why the singularities like $\frac{1}{d-4}$ must
appear in the subcritical dimensions and what is the algebraic origin
of them. Such singular coefficients for the definition of minimal
operators $D_{(2N)}^{\min}$ are inevitable for $N\geqslant3$. Moreover,
since the solution of some system of equations is a necessary ingredient
here, it is very probable that there does not exist a procedure or
a new definition of conformal derivative or $\overset{c}{\square'}$
that would produce $D_{(2N)}^{\min}$ directly, without algebraic
solving for the coefficients in the basis of all terms that could
be potentially added here.

Following Wunsch, we can now discuss the addition of other operators
to the minimal form of the operator $D_{(6)}^{\min}$ in arbitrary
dimension. These must be terms with subleading number of derivatives,
but still the total energy dimension of the term must agree with 6
as it is dimension of $\square^{3}$. These terms may possibly contain
still 4, 3, 2 or one conformal covariant derivative acting on the
scalar field $\phi$ and we also allow for endomorphic terms that
do not contain any derivatives at all acting on the scalar field.
Due to the fixed energy dimension of terms in such combination we
can use terms with one, two or three powers of curvatures and possibly
also with the derivatives acting on it. Actually, we see that because
of the dimensionality of curvature tensor, we cannot have 5 derivatives
on the scalar and further insights into the construction of conformally
covariant terms show that we can have only 2, 1 or no derivatives
acting on scalar in such terms. For the sake of the conformally covariant
formalism (up to first derivatives in $\Omega$) we shall use only
conformally covariant derivatives and only Weyl tensor as curvatures.
Since in the formula (\ref{eq: bachdef}) we defined Bach tensor,
we can use it although it is a fully conformal tensor only in $d=4$
due to the transformation law in (\ref{eq: bachconftr}). So we could
treat Bach tensor as another expression which transforms in general
up to first derivatives of $\Omega$. We consider the following terms:
\begin{equation}
D_{1}\phi=\overset{c}{\nabla}_{\mu}\left[\left(C^{\mu}{}_{\rho\sigma\kappa}C^{\nu\rho\sigma\kappa}-\frac{1}{4}g^{\mu\nu}C_{\rho\sigma\kappa\lambda}C^{\rho\sigma\kappa\lambda}\right)\overset{c}{\nabla}_{\nu}\phi\right]+\frac{(d-4)(d-6)}{4(d-3)^{2}}\overset{c}{\nabla}_{\mu}C^{\mu\rho\sigma\kappa}\overset{c}{\nabla}_{\nu}C^{\nu}{}_{\rho\sigma\kappa}\phi\label{eq: D1def}
\end{equation}
and
\begin{equation}
D_{2}\phi=2(d-10)\overset{c}{\nabla}_{\mu}\left[C^{\rho\sigma\kappa\lambda}C_{\rho\sigma\kappa\lambda}\overset{c}{\nabla}{}^{\mu}\phi\right]+(d-6)\overset{c}{\square}\left(C^{\rho\sigma\kappa\lambda}C_{\rho\sigma\kappa\lambda}\right)\phi.\label{eq: D2def}
\end{equation}
In the formula above, we explicitly showed the position and presence
of the scalar field $\phi$ not to avoid some ambiguities in order
and the range of derivatives, where they act only on the first tensor
they find on their right or otherwise their range of action is marked
by parentheses. We use above compact formulas with unexpanded form
of the action of differential operators, although, of course, here
we could use the Leibniz rule to write all of them in the fully expanded
form without parentheses marking the action of derivatives. Each of
these terms (namely $D_{1}$ and $D_{2}$) is separately conformally
covariant provided that $w_{0}=\frac{6-d}{2}$ and $d\geqslant4$
and moreover the total conformal weight of these two expressions is
the same and equal to $w=-\frac{6+d}{2}$. Therefore, we can consider
a following ``nonminimal'' combination as a proper more
generalized differential operator with up to six derivatives
\begin{equation}
D=D_{(6)}^{\min}+\alpha D_{1}+\beta D_{2}
\end{equation}
with arbitrary real coefficients $\alpha$ and $\beta$. Still as
a result of tedious computation we find that 
\begin{equation}
\delta_{c}\left(D\phi\right)=-\frac{6+d}{2}\Omega D\phi,\label{eq: deltacDphi}
\end{equation}
where it is important that all terms linear in the first derivative
of $\Omega$ cancel out separately in $D_{(6)}^{\min}$, $D_{1}$
and $D_{2}$. One notices that in both formulas (\ref{eq: D1def})
and (\ref{eq: D2def}) the last terms on the right hand side vanish
in dimension $d=6$. By looking at the formulas (\ref{eq: D1def})
and (\ref{eq: D2def}) one can easily understand that we have in each
combinations after Leibniz expansion terms with two, one or no conformal
derivatives on the scalar field $\phi$. It is easy to understand
why terms with four and three derivatives on the scalar field $\phi$
cannot be constructed and added here. In such a case the curvature
part of the term had to be Weyl tensor or the first derivative of
the Weyl tensor. But due to antisymmetry of the indices on the Weyl
tensor, we have that the presence of four or three derivatives acting
in a sequence on the scalar field is superfluous. This is because
the indices on these derivatives must be contracted with Weyl tensor
to form a scalar from the point of view of GR and antisymmetry (in
at least three indices found on derivatives acting on scalar we can
find a pair of them coinciding with a pair of antisymmetric indices
on Weyl) implies that we should form a commutator of these derivatives.
Finally this leads to a generation of one or more Weyl curvature tensor
and hence the number of derivatives is consequently reduced to 2,
1 or 0 in action on the scalar field. Additionally, we remind the
reader that only in the very special combinations (\ref{eq: D1def})
and (\ref{eq: D2def}) with very specific numerical frontal coefficients
(dependent on the dimension $d$ in a rational way as a result of
solving some system of algebraic equations), we find that the whole
expressions in (\ref{eq: D1def}) and (\ref{eq: D2def}) transform
covariantly under conformal transformations. Single terms there separately
do not possess such properties. 

Another observation pertaining to terms found in two combinations
above for $D_{1}$ and $D_{2}$ is that the last terms found in both,
which are as mentioned proportional to $(d-6)$, are purely endomorphic
since the derivatives there act only on Weyl tensors. These are absent
in the critical dimension $d=6$ for six-derivative operators as conjectured
earlier for any number $N$ of derivatives. We also verified in $d=4$
for the Paneitz operator that endomorphic terms appear only separately
as some non-minimal fully endomorphic additions and in this critical
dimension do not enter in the construction of the ``minimal''
conformal operator $\overset{c}{\square}{}^{2}$ with $2N=4$ derivatives
in the leading term. We see that also in the $D_{(6)}^{\min}$ operator
as chosen by Wunsch the $\overset{c}{\square}{}^{3}$ operator must
be supplemented by the non-endomorphic term proportional to $B^{\mu\nu}\overset{c}{\nabla}_{\mu}\overset{c}{\nabla}_{\nu}$
and its frontal coefficient does not vanish in $d=6$ spacetime dimensions.
This is obviously consistent with our conjecture since these terms
do contain derivatives on the scalar field $\phi$ and their addition
is crucial even in the critical dimension $d=6$ for a six-derivative
operator.

Finally, we can add here some purely endomorphism terms. They do not
contain any derivatives on the scalar field $\phi$. However, the
derivatives can be in action on Weyl tensors. We find in general dimension
$d$ three such terms, which we can write in the following combination
\begin{equation}
D_{{\rm end}}=\gamma_{1}C_{\mu\nu\rho\sigma}C^{\rho\sigma}{}_{\kappa\lambda}C^{\kappa\lambda\mu\nu}+\gamma_{2}C^{\mu\nu\rho\sigma}C_{\mu\kappa\lambda\sigma}C_{\nu}{}^{\kappa\lambda}{}_{\rho}+\gamma_{3}T.\label{eq: endomorph}
\end{equation}
They of course have very stringent relation to terms defining conformal
gravitation with six-derivatives, so in the six-dimensional spacetime.
The scalar $T$ is defined as
\begin{equation}
T=\frac{10-d}{2}\left(\overset{c}{\nabla}_{\mu}C_{\nu\rho\sigma\kappa}\overset{c}{\nabla}{}^{\mu}C^{\nu\rho\sigma\kappa}-\frac{4(d-2)}{(d-3)^{2}}\overset{c}{\nabla}_{\mu}C^{\mu\rho\sigma\kappa}\overset{c}{\nabla}_{\nu}C^{\nu}{}_{\rho\sigma\kappa}\right)-2\overset{c}{\square}\left(C_{\mu\nu\rho\sigma}C^{\mu\nu\rho\sigma}\right).\label{eq: scalarT}
\end{equation}
While the two first terms in (\ref{eq: endomorph}) have obviously
very good conformal transformation laws with the weight given by $w=-6$,
to prove that the last scalar $T$ also transforms covariantly (without
first derivatives of $\Omega)$ is a bit more involved. But at the
end of the day, one finds again that $w=-6$ and for the scalar $T$
the infinitesimal conformal transformation law is $\delta_{c}T=-6\Omega T$
with no terms with derivatives of the $\Omega$ parameter. This means
that for the expression $D_{{\rm end}}\phi$, we find the transformation
law: $\delta_{c}\left(D_{{\rm end}}\phi\right)=-\frac{6+d}{2}\Omega D_{{\rm end}}\phi$
for arbitrary real values of the parameters $\gamma_{1}$, $\gamma_{2}$
and $\gamma_{3}$. Therefore, the most general combination which will
give an expression linear in the scalar field $\phi$ is 
\begin{equation}
D_{{\rm gen}}\phi=D\phi+D_{{\rm end}}\phi
\end{equation}
and from this the most general conformal operator with leading terms
of six derivatives takes the form
\begin{equation}
D_{{\rm gen}}=D+D_{{\rm end}}=D_{(6)}^{\min}+\alpha D_{1}+\beta D_{2}+\gamma_{1}C_{\mu\nu\rho\sigma}C^{\rho\sigma}{}_{\kappa\lambda}C^{\kappa\lambda\mu\nu}+\gamma_{2}C^{\mu\nu\rho\sigma}C_{\mu\kappa\lambda\sigma}C_{\nu}{}^{\kappa\lambda}{}_{\rho}+\gamma_{3}T\label{eq: genop}
\end{equation}
with five arbitrary real constants $\alpha$, $\beta$, $\gamma_{1}$,
$\gamma_{2}$ and $\gamma_{3}$.

One could ask whether in the list of endomorphic terms in (\ref{eq: endomorph})
we could have also some with four derivatives on one Weyl tensor.
Obviously three derivatives on Weyl tensor do not match the total
energy dimensionality of the term like $\square^{3}$. The reason
for the absence of four derivatives on one Weyl tensor is the same
antisymmetry argument as put above after the formula (\ref{eq: deltacDphi}).
The term is reduced to ones having two less derivatives since all
indices on derivatives are again contracted with Weyland antisymmetry
is in action here. Eventually, here we can have terms with two Weyl
tensors and two conformal derivatives (they could act on one Weyl
tensor, then they could be contracted to form $\overset{c}{\square}$;
or they act on two different Weyl tensor), present in the scalar $T$
or terms with only three Weyl tensors with all indices properly contracted.
These last ones give rise to two possibilities in the general dimension
of construction of Weyl cube terms, which of course have obvious conformal
transformation laws. All terms present in the definition of the scalar
$T$ in (\ref{eq: scalarT}) are again written in the unexpanded form
where we could use Leibinz rule for conformal derivatives (working
the same as in the standard GR case). The terms that are present there
are the most general constructible all consistent with the symmetries
of the Weyl tensor and Bianchi identities.

When the scalar and conformally covariant term $T$ is properly densitized
and integrated over the whole spacetime volume integral, one gets
a prototypical integral for the action of conformal gravitation with
the leading term of six derivatives. Since $w(T)=-6$, then obviously
the action is conformally invariant only when $d=-w(T)=6$. Only
then we can speak of a conformal gravitation with six-derivative terms.
We have explicitly
\begin{equation}
S_{{\rm grav},6{\rm d}}=\int d^{6}x\sqrt{|g|}\alpha_{C}\left(2\left(\overset{c}{\nabla}_{\mu}C_{\nu\rho\sigma\kappa}\overset{c}{\nabla}{}^{\mu}C^{\nu\rho\sigma\kappa}-\frac{16}{9}\overset{c}{\nabla}_{\mu}C^{\mu\rho\sigma\kappa}\overset{c}{\nabla}_{\nu}C^{\nu}{}_{\rho\sigma\kappa}\right)-2\overset{c}{\square}\left(C_{\mu\nu\rho\sigma}C^{\mu\nu\rho\sigma}\right)\right),\label{eq: T2}
\end{equation}
where $\alpha_{C}$ is a positive coupling constant (for reflection-positivity
of the theory in Euclidean setup). By using the formula for \O{}rsted-Penrose
operator (\ref{eq: oersted}) and integrating by parts ordinary GR-covariant
box operator we find an equivalent but a simpler form
\begin{equation}
S_{{\rm grav},6{\rm d}}=\int d^{6}x\sqrt{|g|}\alpha_{C}\left(2\left(\overset{c}{\nabla}_{\mu}C_{\nu\rho\sigma\kappa}\overset{c}{\nabla}{}^{\mu}C^{\nu\rho\sigma\kappa}-\frac{16}{9}\overset{c}{\nabla}_{\mu}C^{\mu\rho\sigma\kappa}\overset{c}{\nabla}_{\nu}C^{\nu}{}_{\rho\sigma\kappa}\right)+\frac{4}{5}RC_{\mu\nu\rho\sigma}C^{\mu\nu\rho\sigma}\right).\label{eq: T3}
\end{equation}
We here remind that we could neglect total derivatives and total box
operators, when they are made with GR-covariant derivatives. But we
cannot proceed with the same about total conformal covariant derivatives
and total conformal box operators $\overset{c}{\square}$. However,
in the last formula not all the terms look conformally covariant (even
with the proviso that they are conformal only up to the first derivatives
of $\Omega$ parameter of conformal transformation). The presence
of the Ricci scalar is here essential, however under the action integral
it transforms up to total derivatives of the parameter $\Omega$ hence
can be used, although, of course, the expression in (\ref{eq: T2})
looks better from the point of view of conformal formalism. 

The action (\ref{eq: T2}) describes conformal gravitation in $d=6$
spacetime dimension. It can also be used in Euclidean signature where
it would define a six-dimensional conformal shape dynamics. As obvious
from the term in (\ref{eq: T2}) this is a six-derivative theory,
where the graviton kinetic term has naturally six derivatives in its
action. And hence the propagator is even more suppressed in comparison
to the four-derivative Stelle gravitation considered as a flag model
of higher-derivative gravitational theories in $d=4$ dimensions.
The last term in (\ref{eq: T3}) of the type $RCC$ obviously describes
only the non-trivial graviton interactions on the flat spacetime backgrounds.
Moreover, two terms with $\gamma_{1}$ and $\gamma_{2}$ in (\ref{eq: endomorph})
can be as well added to the 6-dimensional conformal gravitation action
with arbitrary real coefficients. They do not influence the flat spacetime
graviton propagator, however give rise to new graviton interaction
vertices. Altogether the theory defined by volume integral of (\ref{eq: endomorph})
in $d=6$ is conformally invariant pure gravitational theory that
is also fully renormalizable on the quantum level. The interested reader can peruse following references (and citations therein) to the works of Maldacena and others on 6-dimensional conformal gravitation \cite{Maldacena:2011mk,Anastasiou:2016jix,Anastasiou:2020mik}.

Here, we understand that the sense in which the operator $D_{(6)}^{\min}$
is minimal is largely arbitrary since we could add the part of operators
$D_{1}$ or $D_{2}$ or any parts from endomorphisms with whatever
real coefficients and still the resulting operator is conformally
covariant. Probably the motivation behind this choice of $D_{(6)}^{\min}$
as in (\ref{eq: Wunsch6der}) was that its formula in this shape is
more compact. As clearly visible from formulas for $D_{1}$, $D_{2}$
and $D_{{\rm end}}$ they contain terms which are quadratic in conformal
curvatures (Weyl tensors) and finally in $D_{{\rm end}}$ there are
terms cubic in curvature in terms multiplied by $\gamma_{1}$ and
$\gamma_{2}$. In the scalar $T$ we find instead only terms quadratic
in curvatures. Finally, the Bach tensor is understood as an expression
linear in Weyl curvature; actually two conformal derivatives act here
on Weyl tensor according to the definition in (\ref{eq: bachdef}).
Hence Bach tensor is also an expression linear in (conformal) curvature.
This distinguishes the term of addition to $\overset{c}{\square}{}^{3}$
in (\ref{eq: Wunsch6der}) from other terms in $D_{1}$, $D_{2}$
and $D_{{\rm end}}$ since only in the first case (with Bach tensor)
the necessary addition is linear in curvatures, while in last cases
it is quadratic or cubic in conformal Weyl tensor. Therefore the differential
operator as proposed by Wunsch $D_{(6)}^{\min}$ in (\ref{eq: Wunsch6der})
can be viewed as minimal in a sense, because it contains only terms
with no conformal curvature $\overset{c}{\square}{}^{3}$ and other
compensating terms are only linear in conformal curvature (but with
the possibility that conformal derivatives act on this curvature).
In this sense the addition of the Bach tensor in (\ref{eq: Wunsch6der})
with the prescribed coefficient, dependent on dimension $d$, is a
unique choice for the minimal operator $D_{(6)}^{\min}$ with only
linear in curvature correction terms.

Therefore, the choice of the minimal operator is never unique and
invariantly and for comparisons we can only consider the most general
operator as in (\ref{eq: genop}) and count its free parameters. We
see that in the sector of proper differential operators we have two-parametric
freedom exemplified by two constants $\alpha$ and $\beta$. Finally,
in the sector of endomorphisms we have a bigger three-parameter freedom
with three constants $\gamma_{1}$, $\gamma_{2}$ and $\gamma_{3}$.
We compare this with our findings from the previous sections and found
a complete agreement with the number of free real parameters needed
to define the most general operator $D_{{\rm gen}}$ in arbitrary
dimensions $d>4$, which contains up to 6 derivatives acting on the
scalar field $\phi$, possibly with other terms giving some curvature
corrections to this formula.

Before we emphasized the singularities in the construction for generalized conformal operators on a general gravitational backgrounds in some special even dimensions. However, here we remark that in $d=4$ the operators with six derivatives in the kinetic terms can still be well defined and considered on flat or conformally flat spacetimes, the problem is only with their generalization to properly conformally curved manifold setting, where the Weyl curvature does not vanish. In that case for the Hamada or Wunsch operators, we find the simple form, namely $\overset{c}{\square}{}^{3}$, without any conformal curvature correction as in (\ref{eq: Wunsch6der}). The expansion of this $\overset{c}{\square}{}^{3}$ differential operator in terms of standard GR-covariant derivatives, standard GR-curvatures is without any problem in dimension $d=4$ and hence it perfectly exists there. As seen in (\ref{eq: Wunsch6der}) the problem in $d=4$ was really with conformal curvature corrections needed on the general background. But when the background is conformally flat then these conformal objects do not arise and their singular frontal coefficients are here immaterial. (For example, one can take the smooth limit $d\to4$, staying all the time on conformal backgrounds $C_{\mu\nu\rho\sigma}=0$, and the expression for the operator does not develop any singularity.) Moreover, the condition of always staying  on conformally flat backgrounds is preserved by conformal transformations and hence these operators like $\overset{c}{\square}{}^{3}$ do transform well (i.e. conformally covariantly) under any conformal transformation in $d=4$ as well. The background is still conformally flat when we change the metric and transform the operator, but of course the standard curvature tensors as in GR, they are not left invariant.

\section{New construction of generalized Hamada operator}
\label{s7}

In the previous section we discussed the construction of the minimal
six-derivative conformal operator acting on the scalar field along
the lines of the developments done by Wunsch and Graham. Here in this
last main section of the paper, we plan to present a new different derivation of an
equivalent form for the Hamada operator. We accept the fact that just
the operator $\overset{c}{\square}{}^{3}$ is not conformally covariant.
In previous discussion we had decided to add correcting curvature
terms. However, there exists also another formal solution to this
problem. Since generally conformally covariant derivatives do not
commute, we may try to construct operators with six derivatives but
with a different order of them. There is a hope that in a specific
combination of such terms we will be able to cancel terms with the
first derivatives of the $\Omega$ parameter. In this way we expect
to derive all terms which are non-endomorphic in the general form
of a six-derivative operator. Of course, endomorphic terms can be
added easily just from $D_{{\rm end}}$. It is well known that curvature
terms arise as the result of doing commutation of covariant derivatives.
Here we want to somehow invert this process and see whether the six-derivative
Hamada operator can be written without curvature terms but as a combination
of terms with different orders of conformally covariant derivatives.
We also want to understand when this process is possible.

To start the explanation of this way of deriving the terms we must
remind the reader that the cube of the conformally covariant scalar
box operator $\overset{c}{\square}{}^{3}$ we view as a particular
order of six conformal covariant derivatives. Namely, we have
\begin{equation}
\overset{c}{\square}{}^{3}=g^{\mu\nu}g^{\rho\sigma}g^{\kappa\lambda}\overset{c}{\nabla}_{\mu}\overset{c}{\nabla}_{\nu}\overset{c}{\nabla}_{\rho}\overset{c}{\nabla}_{\sigma}\overset{c}{\nabla}_{\kappa}\overset{c}{\nabla}_{\lambda}=\overset{c}{\nabla}_{\mu}\overset{c}{\nabla}{}^{\mu}\overset{c}{\nabla}_{\nu}\overset{c}{\nabla}{}^{\nu}\overset{c}{\nabla}_{\rho}\overset{c}{\nabla}{}^{\rho}\label{eq: boxcubeorder}.
\end{equation}
For definiteness of our solutions, we can keep the coefficient of
this leading term $\overset{c}{\square}{}^{3}$ with the default order
of conformal derivatives $\overset{c}{\nabla}_{\mu}$ set to one in
our procedure. Due to the fact that the metric tensor is also covariantly
conserved by conformal derivatives $\overset{c}{\nabla}_{\mu}g_{\nu\rho}=0$,
the position of contracted indices and their order is not important
in (\ref{eq: boxcubeorder}).

Since we have in general six indices on six conformal derivatives,
then the number of completely arbitrary metric contractions initially
is $(6-1)!!=5!!=15$. This number of combinations is too big since
it does not take into account the symmetries that exist between various
terms, when we do not have to commute derivatives since for example
the indices on them are symmetrized by the usage of contravariant
metric tensor as in the formula above. For this one has to carefully
look at the commutation law for conformal derivatives as this was
stipulated in the formula (\ref{eq: commders}). Based on this formula
one sees that two (the most internal) conformal derivatives acting
on the scalar field $\phi$ must commute, that is we have
\begin{equation}
\overset{c}{\nabla}_{\kappa}\overset{c}{\nabla}_{\lambda}\phi=\overset{c}{\nabla}_{\lambda}\overset{c}{\nabla}_{\kappa}\phi.
\end{equation}
This is because standard GR-covariant derivative on a scalar commute
and here we cannot construct any contraction of the Weyl tensor due
to counting of the number of indices and due to energy dimensionality
reasons. However, we warn the reader that it is not true for example
that higher order conformal derivatives commute, and we have
\begin{equation}
\overset{c}{\nabla}_{\rho}\overset{c}{\nabla}_{\sigma}\overset{c}{\square}\phi\neq\overset{c}{\nabla}_{\sigma}\overset{c}{\nabla}_{\rho}\overset{c}{\square}\phi,\label{eq: notcomm}
\end{equation}
since now some contraction of the Weyl tensor can be done with the
indices on derivatives resulting from splitting of the box operator
($\overset{c}{\square}=\overset{c}{\nabla}_{\rho}\overset{c}{\nabla}{}^{\rho}$).
Moreover, since the last two conformally covariant derivatives on
the scalar do commute, then by considerations of various orders of
all 6 derivatives and by doing the commutation of them we never produce
endomorphic terms in which all derivatives would be substituted by
Weyl curvature tensors (or their derivatives). In a sense, we can
never replace the last two derivatives by Weyl tensor, since their
commutator vanishes. Therefore, the endomorphic terms, like the scalar
$T$ and two contractions of Weyl cube have to appended by hand for
the most general operator. Here we construct only the part of the
operator which is non-endomorphic, i.e. contains derivatives acting
on the scalar field $\phi$.

Another corollary from the commutation law (\ref{eq: commders}) is
that the following commutator vanishes, when acting directly on the
scalar $\phi$:
\begin{equation}
\left[\overset{c}{\square},\overset{c}{\nabla}_{\sigma}\right]=0.
\end{equation}
The first part of the commutation law does not apply here, since in
standard GR this commutator is $\left[\square,\overset{c}{\nabla}_{\mu}\right]=R_{\mu\nu}\nabla^{\nu}$
and for Weyl tensor all traces, like Ricci tensor to Riemann tensor,
vanish. The absence of the second part is again due to the reason
that we cannot contract indices in the second part of the law in (\ref{eq: commders})
when acting with derivatives on Weyl tensor, when only index $\Omega$
has to be free, but we have also constraints due to dimensionality
of possible terms. Again, in the more general situation, like for
the commutator 
\begin{equation}
\left[\overset{c}{\square},\overset{c}{\nabla}_{\sigma}\right]\overset{c}{\square}\phi
\end{equation}
we cannot simplify since the second term on the right hand side of
(\ref{eq: commders}) can be now constructed in which additional indices
are contracted with these on conformal derivatives resulting after
the splitting of the most internal $\overset{c}{\square}$ operator.

Furthermore, we have other following commutators between single conformal
derivatives that vanish here:
\begin{equation}
\left[\overset{c}{\nabla}_{\mu},\overset{c}{\nabla}_{\nu}\right]\phi=\left[\overset{c}{\nabla}_{\mu},\overset{c}{\nabla}_{\nu}\right]\overset{c}{\nabla}{}^{\mu}\phi=\left[\overset{c}{\nabla}_{\mu},\overset{c}{\nabla}_{\nu}\right]\overset{c}{\nabla}{}^{\nu}\phi=\left[\overset{c}{\nabla}_{\mu},\overset{c}{\nabla}_{\nu}\right]\overset{c}{\nabla}{}^{\mu}\overset{c}{\nabla}{}^{\nu}\phi=0
\end{equation}
and the reasons for their vanishing are all of the same type as discussed
above. Due to the commutation law and the fact expressed in (\ref{eq: notcomm})
we see that the numer of independent terms can be bigger than in the
corresponding case for standard GR-covariant derivatives, where we
have significantly more reductions. Since two conformal derivatives
do not in general commute when acting on a GR-scalar like $\overset{c}{\square}{}^{k}\phi$
for $k>0$, then such terms are independent and cannot be simplified
as it would be the case in GR. This is because in the transformation
law (\ref{eq: commders}) we can pull out some terms, derivatives
and indices from the general expression $\overset{c}{\square}{}^{k}\phi$
after splitting.

Using various symmetries and looking carefully at the commutation
law in (\ref{eq: commders}), one arrives at an irreducible basis
of terms with six derivatives, which has precisely 6 terms. Therefore
all the terms in the combinations, we are looking for, can be presented
as
\begin{equation}
{\cal O}=C^{\rho_{1}\rho_{2}\rho_{3}\rho_{4}\rho_{5}\rho_{6}}\overset{c}{\nabla}_{\rho_{1}}\overset{c}{\nabla}_{\rho_{2}}\overset{c}{\nabla}_{\rho_{3}}\overset{c}{\nabla}_{\rho_{4}}\overset{c}{\nabla}_{\rho_{5}}\overset{c}{\nabla}_{\rho_{6}},\label{eq: defoperatoro}
\end{equation}
where the $C$ tensor is
\[
C^{\rho_{1}\rho_{2}\rho_{3}\rho_{4}\rho_{5}\rho_{6}}=\alpha_{1}g^{\rho_{1}\rho_{2}}g^{\rho_{3}\rho_{4}}g^{\rho_{5}\rho_{6}}+\alpha_{2}g^{\rho_{1}\rho_{3}}g^{\rho_{2}\rho_{4}}g^{\rho_{5}\rho_{6}}+\alpha_{3}g^{\rho_{1}\rho_{4}}g^{\rho_{2}\rho_{3}}g^{\rho_{5}\rho_{6}}+\alpha_{4}g^{\rho_{1}\rho_{4}}g^{\rho_{2}\rho_{5}}g^{\rho_{3}\rho_{6}}+
\]
\begin{equation}
+\alpha_{5}g^{\rho_{1}\rho_{5}}g^{\rho_{2}\rho_{4}}g^{\rho_{3}\rho_{6}}+\alpha_{6}g^{\rho_{1}\rho_{5}}g^{\rho_{2}\rho_{6}}g^{\rho_{3}\rho_{4}}
\end{equation}
and it defines all 6 basis elements. In other words, our combination
for the conformal differential operator with six derivatives is
\[
{\cal O}=\alpha_{1}\overset{c}{\square}{}^{3}+\alpha_{2}\overset{c}{\nabla}_{\mu}\overset{c}{\nabla}_{\nu}\overset{c}{\nabla}{}^{\mu}\overset{c}{\nabla}{}^{\nu}\overset{c}{\square}+\alpha_{3}\overset{c}{\nabla}_{\mu}\overset{c}{\square}\overset{c}{\nabla}{}^{\mu}\overset{c}{\square}+\alpha_{4}\overset{c}{\nabla}_{\mu}\overset{c}{\nabla}_{\nu}\overset{c}{\nabla}_{\rho}\overset{c}{\nabla}{}^{\mu}\overset{c}{\nabla}{}^{\nu}\overset{c}{\nabla}{}^{\rho}+
\]
\begin{equation}
+\alpha_{5}\overset{c}{\nabla}_{\mu}\overset{c}{\nabla}_{\nu}\overset{c}{\nabla}_{\rho}\overset{c}{\nabla}{}^{\nu}\overset{c}{\nabla}{}^{\mu}\overset{c}{\nabla}{}^{\rho}+\alpha_{6}\overset{c}{\nabla}_{\mu}\overset{c}{\nabla}_{\nu}\overset{c}{\square}\overset{c}{\nabla}{}^{\mu}\overset{c}{\nabla}{}^{\nu}.\label{eq: explopero}
\end{equation}
As mentioned above for definiteness we can take $\alpha_{1}=1$ since
this only changes the overall normalization of the operator.

In line with what was mentioned extensively in the previous section,
in the construction of the conformally covariant operators using formalism
of Wunsch one must finally check whether the terms with precisely
the first derivative of the $\Omega$ parameter cancel out in the
combination. This will signify that the operator is conformally covariant
and transforms only linearly with $\Omega$ undifferentiated, so the
operator ${\cal O}$ receives only the conformal weight. To check
this we need to know $\left.\left(\delta_{c}{\cal O}\right)\right|_{\nabla\Omega}$.
Based on the consideration of the previous section this is simply
given by
\begin{equation}
\left.\left(\delta_{c}{\cal O}\right)\right|_{\nabla\Omega}=\nabla_{\gamma}\Omega X^{\gamma}{}_{\rho_{1}\rho_{2}\rho_{3}\rho_{4}\rho_{5}\rho_{6}}{}^{\beta_{1}\beta_{2}\beta_{3}\beta_{4}\beta_{5}}C^{\rho_{1}\rho_{2}\rho_{3}\rho_{4}\rho_{5}\rho_{6}}\overset{c}{\nabla}_{\beta_{1}}\overset{c}{\nabla}_{\beta_{2}}\overset{c}{\nabla}_{\beta_{3}}\overset{c}{\nabla}_{\beta_{4}}\overset{c}{\nabla}_{\beta_{5}},
\end{equation}
hence for cancellation we only require that the contraction of the
$C$ tensor with the $X$ tensor explicitly vanish, when this expression
with six free indices $(\gamma,\beta_{1},\beta_{2},\beta_{3},\beta_{4},\beta_{5})$
is expanded within some basis of terms. For the explicit computation
we can use the formula for the $X$ tensor (\ref{eq: Xgenexplicit})
with $N=7$ (in our notation from the previous section, where we also
have to shift indexing of indices $\rho$: from the sequence $\rho_{2}$
to $\rho_{7}$ into a different one from $\rho_{1}$ to $\rho_{6}$).
Then we are left with the expression with free indices $\gamma,\beta_{1},\beta_{2},\beta_{3},\beta_{4},\beta_{5}$.
However, since it is contracted with a sequence of $N-2=5$ conformal
derivatives on the right of it, this tensor also inherits the symmetry
of exchange of indices between $\beta$. Here again due to the properties
of conformal covariant derivatives and their commutation law (when
they act on the scalar field $\phi$ on the most right), we have some
identities between various terms, when we expand the result in the
basis of all orders of five conformal derivatives. We remark that
the tensor contraction $X$ with $C$ (apart from some numerical coefficients)
is built out only with the contravariant metric tensor of spacetime
$g^{\mu\nu}$. Hence, again the number of possible orders and the
combinations here for $6$ indices is $(6-1)!!=5!!=15$. This is before
the symmetries of five conformal derivatives are taken into account.
Similarly to the previous case one finds exactly $9$ identities and
the number of elements in the independent basis here for writing such
terms is $6$. This is not a mere coincidence since our system for
six unknown coefficients $\alpha_{i}$ for $i=1,\ldots,6$ has to
be determined by six equations to remove the possible situation of
algebraic degeneracy or inconsistency. These identities are summarized
below:
\[
1)\quad g^{\gamma\beta_{1}}g^{\beta_{2}\beta_{4}}g^{\beta_{3}\beta_{5}}\overset{c}{\nabla}_{\beta_{1}}\overset{c}{\nabla}_{\beta_{2}}\overset{c}{\nabla}_{\beta_{3}}\overset{c}{\nabla}_{\beta_{4}}\overset{c}{\nabla}_{\beta_{5}}\to g^{\gamma\beta_{1}}g^{\beta_{2}\beta_{3}}g^{\beta_{4}\beta_{5}}\overset{c}{\nabla}_{\beta_{1}}\overset{c}{\nabla}_{\beta_{2}}\overset{c}{\nabla}_{\beta_{3}}\overset{c}{\nabla}_{\beta_{4}}\overset{c}{\nabla}_{\beta_{5}}
\]
equivalent to
\[
\overset{c}{\nabla}{}^{\gamma}\overset{c}{\nabla}_{\mu}\overset{c}{\nabla}_{\nu}\overset{c}{\nabla}{}^{\mu}\overset{c}{\nabla}{}^{\nu}\to\overset{c}{\nabla}{}^{\gamma}\overset{c}{\square}{}^{2}
\]
\[
2)\quad g^{\gamma\beta_{1}}g^{\beta_{2}\beta_{5}}g^{\beta_{3}\beta_{4}}\overset{c}{\nabla}_{\beta_{1}}\overset{c}{\nabla}_{\beta_{2}}\overset{c}{\nabla}_{\beta_{3}}\overset{c}{\nabla}_{\beta_{4}}\overset{c}{\nabla}_{\beta_{5}}\to g^{\gamma\beta_{1}}g^{\beta_{2}\beta_{3}}g^{\beta_{4}\beta_{5}}\overset{c}{\nabla}_{\beta_{1}}\overset{c}{\nabla}_{\beta_{2}}\overset{c}{\nabla}_{\beta_{3}}\overset{c}{\nabla}_{\beta_{4}}\overset{c}{\nabla}_{\beta_{5}}
\]
equivalent to
\[
\overset{c}{\nabla}{}^{\gamma}\overset{c}{\nabla}_{\mu}\overset{c}{\square}\overset{c}{\nabla}{}^{\mu}=\overset{c}{\nabla}{}^{\gamma}\overset{c}{\nabla}_{\mu}\overset{c}{\nabla}_{\nu}\overset{c}{\nabla}{}^{\nu}\overset{c}{\nabla}{}^{\mu}\to\overset{c}{\nabla}{}^{\gamma}\overset{c}{\square}{}^{2}
\]

\[
3)\quad g^{\gamma\beta_{2}}g^{\beta_{1}\beta_{4}}g^{\beta_{3}\beta_{5}}\overset{c}{\nabla}_{\beta_{1}}\overset{c}{\nabla}_{\beta_{2}}\overset{c}{\nabla}_{\beta_{3}}\overset{c}{\nabla}_{\beta_{4}}\overset{c}{\nabla}_{\beta_{5}}\to g^{\gamma\beta_{2}}g^{\beta_{1}\beta_{3}}g^{\beta_{4}\beta_{5}}\overset{c}{\nabla}_{\beta_{1}}\overset{c}{\nabla}_{\beta_{2}}\overset{c}{\nabla}_{\beta_{3}}\overset{c}{\nabla}_{\beta_{4}}\overset{c}{\nabla}_{\beta_{5}}
\]
equivalent to
\[
\overset{c}{\nabla}_{\mu}\overset{c}{\nabla}{}^{\gamma}\overset{c}{\nabla}_{\nu}\overset{c}{\nabla}{}^{\mu}\overset{c}{\nabla}{}^{\nu}\to\overset{c}{\nabla}_{\mu}\overset{c}{\nabla}{}^{\gamma}\overset{c}{\nabla}{}^{\mu}\overset{c}{\square}
\]
\[
4)\quad g^{\gamma\beta_{2}}g^{\beta_{1}\beta_{5}}g^{\beta_{3}\beta_{4}}\overset{c}{\nabla}_{\beta_{1}}\overset{c}{\nabla}_{\beta_{2}}\overset{c}{\nabla}_{\beta_{3}}\overset{c}{\nabla}_{\beta_{4}}\overset{c}{\nabla}_{\beta_{5}}\to g^{\gamma\beta_{2}}g^{\beta_{1}\beta_{3}}g^{\beta_{4}\beta_{5}}\overset{c}{\nabla}_{\beta_{1}}\overset{c}{\nabla}_{\beta_{2}}\overset{c}{\nabla}_{\beta_{3}}\overset{c}{\nabla}_{\beta_{4}}\overset{c}{\nabla}_{\beta_{5}}
\]
equivalent to
\[
\overset{c}{\nabla}_{\mu}\overset{c}{\nabla}{}^{\gamma}\overset{c}{\square}\overset{c}{\nabla}{}^{\mu}=\overset{c}{\nabla}_{\mu}\overset{c}{\nabla}{}^{\gamma}\overset{c}{\nabla}_{\nu}\overset{c}{\nabla}{}^{\nu}\overset{c}{\nabla}{}^{\mu}\to\overset{c}{\nabla}_{\mu}\overset{c}{\nabla}{}^{\gamma}\overset{c}{\nabla}{}^{\mu}\overset{c}{\square}
\]
\[
5)\quad g^{\gamma\beta_{3}}g^{\beta_{1}\beta_{5}}g^{\beta_{2}\beta_{4}}\overset{c}{\nabla}_{\beta_{1}}\overset{c}{\nabla}_{\beta_{2}}\overset{c}{\nabla}_{\beta_{3}}\overset{c}{\nabla}_{\beta_{4}}\overset{c}{\nabla}_{\beta_{5}}\to g^{\gamma\beta_{3}}g^{\beta_{1}\beta_{4}}g^{\beta_{2}\beta_{5}}\overset{c}{\nabla}_{\beta_{1}}\overset{c}{\nabla}_{\beta_{2}}\overset{c}{\nabla}_{\beta_{3}}\overset{c}{\nabla}_{\beta_{4}}\overset{c}{\nabla}_{\beta_{5}}
\]
equivalent to
\[
\overset{c}{\nabla}_{\mu}\overset{c}{\nabla}_{\nu}\overset{c}{\nabla}{}^{\gamma}\overset{c}{\nabla}{}^{\nu}\overset{c}{\nabla}{}^{\mu}\to\overset{c}{\nabla}_{\mu}\overset{c}{\nabla}_{\nu}\overset{c}{\nabla}{}^{\gamma}\overset{c}{\nabla}{}^{\mu}\overset{c}{\nabla}{}^{\nu}
\]
\[
6)\quad g^{\gamma\beta_{4}}g^{\beta_{1}\beta_{2}}g^{\beta_{3}\beta_{5}}\overset{c}{\nabla}_{\beta_{1}}\overset{c}{\nabla}_{\beta_{2}}\overset{c}{\nabla}_{\beta_{3}}\overset{c}{\nabla}_{\beta_{4}}\overset{c}{\nabla}_{\beta_{5}}\to g^{\gamma\beta_{3}}g^{\beta_{1}\beta_{2}}g^{\beta_{4}\beta_{5}}\overset{c}{\nabla}_{\beta_{1}}\overset{c}{\nabla}_{\beta_{2}}\overset{c}{\nabla}_{\beta_{3}}\overset{c}{\nabla}_{\beta_{4}}\overset{c}{\nabla}_{\beta_{5}}
\]
equivalent to
\[
\overset{c}{\square}\overset{c}{\nabla}_{\nu}\overset{c}{\nabla}{}^{\gamma}\overset{c}{\nabla}{}^{\nu}=\overset{c}{\nabla}_{\mu}\overset{c}{\nabla}{}^{\mu}\overset{c}{\nabla}_{\nu}\overset{c}{\nabla}{}^{\gamma}\overset{c}{\nabla}{}^{\nu}\to\overset{c}{\square}\overset{c}{\nabla}{}^{\gamma}\overset{c}{\square}
\]
\[
7)\quad g^{\gamma\beta_{5}}g^{\beta_{1}\beta_{2}}g^{\beta_{3}\beta_{4}}\overset{c}{\nabla}_{\beta_{1}}\overset{c}{\nabla}_{\beta_{2}}\overset{c}{\nabla}_{\beta_{3}}\overset{c}{\nabla}_{\beta_{4}}\overset{c}{\nabla}_{\beta_{5}}\to g^{\gamma\beta_{3}}g^{\beta_{1}\beta_{2}}g^{\beta_{4}\beta_{5}}\overset{c}{\nabla}_{\beta_{1}}\overset{c}{\nabla}_{\beta_{2}}\overset{c}{\nabla}_{\beta_{3}}\overset{c}{\nabla}_{\beta_{4}}\overset{c}{\nabla}_{\beta_{5}}
\]
equivalent to
\[
\overset{c}{\square}{}^{2}\overset{c}{\nabla}{}^{\gamma}=\overset{c}{\nabla}_{\mu}\overset{c}{\nabla}{}^{\mu}\overset{c}{\nabla}_{\nu}\overset{c}{\nabla}{}^{\nu}\overset{c}{\nabla}{}^{\gamma}\to\overset{c}{\square}\overset{c}{\nabla}{}^{\gamma}\overset{c}{\square}
\]
\[
8)\quad g^{\gamma\beta_{5}}g^{\beta_{1}\beta_{3}}g^{\beta_{2}\beta_{4}}\overset{c}{\nabla}_{\beta_{1}}\overset{c}{\nabla}_{\beta_{2}}\overset{c}{\nabla}_{\beta_{3}}\overset{c}{\nabla}_{\beta_{4}}\overset{c}{\nabla}_{\beta_{5}}\to g^{\gamma\beta_{4}}g^{\beta_{1}\beta_{3}}g^{\beta_{2}\beta_{5}}\overset{c}{\nabla}_{\beta_{1}}\overset{c}{\nabla}_{\beta_{2}}\overset{c}{\nabla}_{\beta_{3}}\overset{c}{\nabla}_{\beta_{4}}\overset{c}{\nabla}_{\beta_{5}}
\]
equivalent to
\[
\overset{c}{\nabla}_{\mu}\overset{c}{\nabla}_{\nu}\overset{c}{\nabla}{}^{\mu}\overset{c}{\nabla}{}^{\nu}\overset{c}{\nabla}{}^{\gamma}\to\overset{c}{\nabla}_{\mu}\overset{c}{\nabla}_{\nu}\overset{c}{\nabla}{}^{\mu}\overset{c}{\nabla}{}^{\gamma}\overset{c}{\nabla}{}^{\nu}
\]
\[
9)\quad g^{\gamma\beta_{5}}g^{\beta_{1}\beta_{4}}g^{\beta_{2}\beta_{3}}\overset{c}{\nabla}_{\beta_{1}}\overset{c}{\nabla}_{\beta_{2}}\overset{c}{\nabla}_{\beta_{3}}\overset{c}{\nabla}_{\beta_{4}}\overset{c}{\nabla}_{\beta_{5}}\to g^{\gamma\beta_{4}}g^{\beta_{1}\beta_{5}}g^{\beta_{2}\beta_{3}}\overset{c}{\nabla}_{\beta_{1}}\overset{c}{\nabla}_{\beta_{2}}\overset{c}{\nabla}_{\beta_{3}}\overset{c}{\nabla}_{\beta_{4}}\overset{c}{\nabla}_{\beta_{5}}
\]
equivalent to
\begin{equation}
\overset{c}{\nabla}_{\mu}\overset{c}{\square}\overset{c}{\nabla}{}^{\mu}\overset{c}{\nabla}{}^{\gamma}=\overset{c}{\nabla}_{\mu}\overset{c}{\nabla}_{\nu}\overset{c}{\nabla}{}^{\nu}\overset{c}{\nabla}{}^{\mu}\overset{c}{\nabla}{}^{\gamma}\to\overset{c}{\nabla}_{\mu}\overset{c}{\square}\overset{c}{\nabla}{}^{\gamma}\overset{c}{\nabla}{}^{\mu}.
\end{equation}
With this system we end up with the basis of terms for contraction
of the $X$ and $C$ tensors contracted with a sequence of five conformal
derivatives
\begin{equation}
\left\{ \overset{c}{\nabla}{}^{\gamma}\overset{c}{\square}{}^{2},\,\overset{c}{\nabla}_{\mu}\overset{c}{\nabla}{}^{\gamma}\overset{c}{\nabla}{}^{\mu}\overset{c}{\square},\,\overset{c}{\nabla}_{\mu}\overset{c}{\nabla}_{\nu}\overset{c}{\nabla}{}^{\gamma}\overset{c}{\nabla}{}^{\mu}\overset{c}{\nabla}{}^{\nu},\,\overset{c}{\square}\overset{c}{\nabla}{}^{\gamma}\overset{c}{\square},\,\overset{c}{\nabla}_{\mu}\overset{c}{\nabla}_{\nu}\overset{c}{\nabla}{}^{\mu}\overset{c}{\nabla}{}^{\gamma}\overset{c}{\nabla}{}^{\nu},\,\overset{c}{\nabla}_{\mu}\overset{c}{\square}\overset{c}{\nabla}{}^{\gamma}\overset{c}{\nabla}{}^{\mu}\right\} \label{eq: basres}
\end{equation}
or equivalently for terms built with metric only (without contractions
with derivatives)
\begin{equation}
\left\{ g^{\gamma\beta_{1}}g^{\beta_{2}\beta_{3}}g^{\beta_{4}\beta_{5}},\,g^{\gamma\beta_{2}}g^{\beta_{1}\beta_{3}}g^{\beta_{4}\beta_{5}},\,g^{\gamma\beta_{3}}g^{\beta_{1}\beta_{4}}g^{\beta_{2}\beta_{5}},\,g^{\gamma\beta_{3}}g^{\beta_{1}\beta_{2}}g^{\beta_{4}\beta_{5}},\,g^{\gamma\beta_{4}}g^{\beta_{1}\beta_{3}}g^{\beta_{2}\beta_{5}},\,g^{\gamma\beta_{4}}g^{\beta_{1}\beta_{5}}g^{\beta_{2}\beta_{3}}\right\} .\label{eq: basis1}
\end{equation}
The choice of bases in (\ref{eq: explopero}) as well as in (\ref{eq: basres})
is quite arbitrary. We made one because we had to use an irreducible
set of basis elemenets to solve the system of algebraic equations
for six coefficients $\alpha_{1}$ to $\alpha_{6}$ unambiguosly.
The basis defined in (\ref{eq: explopero}) is not special, and of
course, others may be used here as well. We do not find why this basis
could be thought of as a preferred one; it is just a basis. We will
soon present some results and interesting features independent of
the choice of the basis.

Requiring that the contraction of the $X$ with $C$ tensor vanish
expanded in the basis (\ref{eq: basis1}), we get a system of 6 equations
for 6 unknown coefficients. The system of six linear equations reads
explicitly, where we also used the fact that here $w_{0}=\frac{6-d}{2}$,
\begin{eqnarray}
\frac{1}{2}\left(-8\alpha_{1}-(d+2)\left(\alpha_{2}+\alpha_{3}+\alpha_{4}+\alpha_{5}+\alpha_{6}\right)\right) & = & 0\label{eq: sysfirst}\\
\frac{1}{2}\left(-4\alpha_{2}-4\alpha_{3}-d\left(\alpha_{4}+\alpha_{5}+\alpha_{6}\right)\right) & = & 0\\
4\alpha_{1}+\frac{d-2}{2}\alpha_{2}+\frac{d-2}{2}\alpha_{3}+\alpha_{4}+\alpha_{5}+\alpha_{6} & = & 0\\
0 & = & 0\\
4\alpha_{2}+(d-1)\alpha_{4}+\frac{d-2}{2}\alpha_{5}+2\alpha_{6} & = & 0\\
4\alpha_{3}+\alpha_{4}+\frac{d+2}{2}\alpha_{5}+(d-2)\alpha_{6} & = & 0\label{eq: syslast},
\end{eqnarray}
where we also notice that in the basis (\ref{eq: basis1}) the contraction
$XC$ does not at all produce a term proportional to $\overset{c}{\square}\overset{c}{\nabla}{}^{\gamma}\overset{c}{\square}$
(the fourth equation in the system above). This is obviously not a
problem since we want a null solution to all equations here. Already
this means that our system of homogenous equation will have some freedom
and not all $\alpha_{1}$ to $\alpha_{6}$ coefficients will be determined
uniquely. This fact is consistent with the discussion from the previous
section, where we indeed found some freedom in defining the six-derivative
Hamada operator without any endomorphic parts.

System of equations (\ref{eq: sysfirst}) to (\ref{eq: syslast})
can be solved for various set of coefficients. Before this answer
we should put a question how big is the freedom in the system, which
is a statement independent of the basis and it defines the dimension
of the non-vanishing kernel of the system. We already know that the
system is not maximal, with a kernel 6-dimensional because of the
absence of terms proportional to $\overset{c}{\square}\overset{c}{\nabla}{}^{\gamma}\overset{c}{\square}$.
Actually, the matrix rank of the system (\ref{eq: sysfirst}) to (\ref{eq: syslast})
is equal to three, so we have only 3 independent equations and 3-parametric
freedom of the solutions. As we remember we can always take as for
the normalization $\alpha_{1}=1$ and then the freedom is reduced
to a 2-parametric one.

We decided to solve the system for three unknown coefficients $\alpha_{3}$,
$\alpha_{5}$ and $\alpha_{6}$ (leaving $\alpha_{1}=1$ and undetermined
$\alpha_{2}$ and $\alpha_{4}$ playing here the role of real parameters).
The explicit solutions are
\begin{equation}
\alpha_{3}=\frac{8d}{(d-4)(d+2)}+\alpha_{2},
\end{equation}
\begin{equation}
\alpha_{5}=-\frac{128}{(d-6)(d-4)(d+2)}-\frac{8}{d-6}\alpha_{2}-\frac{2(d-3)}{d-6}\alpha_{4},
\end{equation}
\begin{equation}
\alpha_{6}=\frac{32(d-2)}{(d-6)(d-4)(d+2)}+\frac{8}{d-6}\alpha_{2}+\frac{d}{d-6}\alpha_{4}.
\end{equation}
We see that these solutions are not defined in dimensions $d=4$ and
$d=6$ as well. This first singularity in all determined coefficients
$\alpha_{3}$, $\alpha_{5}$ and $\alpha_{6}$ for $d=4$ is here
obvious and consistent with the previously mentioned theorem due to
Graham that such an operator with six-derivative leading term does
not exist in dimensions $d=4$. However, we will also comment on this
in more detailed later and discuss whether such result is basis independent
and whether it depends on the choice of the set of three coefficients
for which to solve here.

Actually, the situation in dimensions $d=6$ is more interesting since
this singularity is only apparent, but still there could be some problem
or the defect of defined operator that we will comment on below. To
see that the singularity is only an artefact of the choice of coefficients
to solve for, one can choose another set of them, for example, $\alpha_{3}$,
$\alpha_{4}$ and $\alpha_{5}$. Then the system in $d=6$ of the
explicit form
\begin{eqnarray}
4\left(-\alpha_{1}-\left(\alpha_{2}+\alpha_{3}+\alpha_{4}+\alpha_{5}+\alpha_{6}\right)\right) & = & 0\nonumber \\
-2\alpha_{2}-2\alpha_{3}-3\left(\alpha_{4}+\alpha_{5}+\alpha_{6}\right) & = & 0\nonumber \\
4\alpha_{1}+2\alpha_{2}+2\alpha_{3}+\alpha_{4}+\alpha_{5}+\alpha_{6} & = & 0\nonumber \\
0 & = & 0\nonumber \\
4\alpha_{2}+5\alpha_{4}+2\alpha_{5}+2\alpha_{6} & = & 0\nonumber \\
4\alpha_{3}+\alpha_{4}+4\alpha_{5}+4\alpha_{6} & = & 0\label{eq: sysd6}
\end{eqnarray}
has solutions for $\alpha_{3}$, $\alpha_{4}$ and $\alpha_{5}$ which
are
\begin{equation}
\alpha_{3}=-3-\alpha_{2}\label{eq: solfirst},
\end{equation}
\begin{equation}
\alpha_{4}=-\frac{4}{3}-\frac{4}{3}\alpha_{2},
\end{equation}
\begin{equation}
\alpha_{5}=\frac{10}{3}+\frac{4}{3}\alpha_{2}-\alpha_{6},\label{eq: sollast}
\end{equation}
where we also used the normalization condition that $\alpha_{1}=1$.
So we see in the system of solutions (\ref{eq: solfirst}) to (\ref{eq: sollast})
that it is not singular in $d=6$, just that we have to choose a good
set to solve for. The singularity or its disappearance depends on
the set of final parameters here. However, there is another problem,
when one computes the following combination of all coefficients: $\alpha_{1}+\alpha_{2}+\alpha_{3}+\alpha_{4}+\alpha_{5}+\alpha_{6}$
for the just constructed non-singular operator in $d=6$ dimensions.
This vanishes and this implies that the operator ${\cal O}$ just
constructed in $d=6$ does not have a leading term $\square^{3}$
on flat spacetime. On flat spacetime order of conformal covariant
derivatives does not matter (actually it does not matter on any conformally
flat spacetime, where Weyl tensor explicitly vanishes $C_{\mu\nu\rho\sigma}=0$)
and then the operator reduces to GR-covariant $\square^{3}$ with
all GR-covariant derivative with the frontal coefficient which is
the sum of $\alpha_{i}$ for $i=1$ to $i=6$. On curved spacetime
(it has to be not conformally flat) the operator has other terms,
so this is not equal to zero, but in all of these terms the conformal
derivatives can be ordered and terms higher in conformal curvature
arises, and nothing out of them survives on conformally flat spacetimes.
This is a defect in $d=6$ of the proposals for the operator ${\cal O}$
that does not depend on the choice of three-element set of coefficients
for which we solve for (or in complementary way on the set of two
parameters). It also does not depend on the choice of writing our
operator as in (\ref{eq: explopero}) in particular basis of terms
with order of six covariant derivatives. This can be seen already
from the first of the equation in system (\ref{eq: sysd6}) in $d=6$
which tells us that $\sum_{i=1}^{6}\alpha_{i}=0$. This is a defect
of the operator ${\cal O}$ only in critical dimension for six derivatives.
In higher dimensions nothing like this happens since it can be proven
that independently on the choice of the three-element set of coefficients
to solve for, from the system of equations (\ref{eq: sysfirst}) to
(\ref{eq: syslast}), we derive a general expression for the sum $\sum_{i=1}^{6}\alpha_{i}$
which is independent on the choice of terms in the basis. By using
just the first equation (\ref{eq: sysfirst}) we derive
\begin{equation}
\frac{d-6}{2}\alpha_{1}=\frac{d+2}{2}\left(\alpha_{1}+\alpha_{2}+\alpha_{3}+\alpha_{4}+\alpha_{5}+\alpha_{6}\right),
\end{equation}
 so
\begin{equation}
\sum_{i=1}^{6}\alpha_{i}=\frac{d-6}{d+2}\alpha_{1}\label{eq: sumall}
\end{equation}
and for normalization $\alpha_{1}\neq0$ we have that the sum is non-zero
in every dimension $d\neq6$. This confirms that the problem only
persists in critical dimension $d=6$ and that it is independent on
the basis of how we write the operator in (\ref{eq: explopero}).

One can also analyze the system of equations (\ref{eq: sysfirst})
to (\ref{eq: syslast}) in dimensions $d=4$ in a basis-independent
way. The system is explicitly in the form 
\begin{eqnarray}
-4\alpha_{1}-3\left(\alpha_{2}+\alpha_{3}+\alpha_{4}+\alpha_{5}+\alpha_{6}\right) & = & 0\nonumber \\
-2\left(\alpha_{2}+\alpha_{3}+\alpha_{4}+\alpha_{5}+\alpha_{6}\right) & = & 0\nonumber \\
4\alpha_{1}+\alpha_{2}+\alpha_{3}+\alpha_{4}+\alpha_{5}+\alpha_{6} & = & 0\nonumber \\
0 & = & 0\nonumber \\
4\alpha_{2}+3\alpha_{4}+\alpha_{5}+2\alpha_{6} & = & 0\nonumber \\
4\alpha_{3}+\alpha_{4}+3\alpha_{5}+2\alpha_{6} & = & 0.
\end{eqnarray}
One can clearly see, for example, by analyzing the first and second
equation in the above system that the inevitable conclusion here is
that $\alpha_{1}=0$, which means that we cannot normalize our operator
${\cal O}$ by setting the coefficient in front of $\overset{c}{\square}{}^{3}$
to one. It must vanish. Since this happens, again we stay with the
same problem that $\sum_{i=1}^{6}\alpha_{i}=0$, because in the formula
(\ref{eq: sumall}) we must use unfortunately that $\alpha_{1}=0$.
When one uses a different set of coefficients to solve for (different
than $\alpha_{3}$, $\alpha_{5}$ and $\alpha_{6}$), a singularity
in $d=4$ is not seen any more. Instead we have as solutions
\begin{equation}
\alpha_{1}=0,\quad\alpha_{2}=\frac{1}{4}\left(-3\alpha_{4}-\alpha_{5}-2\alpha_{6}\right),\quad\alpha_{3}=\frac{1}{4}\left(-\alpha_{4}-3\alpha_{5}-2\alpha_{6}\right)
\end{equation}
with the values of $\alpha_{4}$, $\alpha_{5}$ and $\alpha_{6}$
real and completely arbitrary. We see that in this case we still have
a three-parameter freedom. Following \cite{Wunsch1}
we can expand comments on this case. It was noticed that in $d=4$
six-derivative operator with leading term $\square^{3}$ on flat background
is not constructible; however there exists an operator which is a
part of the Wunsch construction for $D_{(6)}^{\min}$ which is conformally
covariant in this dimension and whose expansion starts with terms
linear in curvature (and with conformally covariant derivatives acting
upon it). This operator is unique and it equals to $D_{{\rm Bach}}\phi=B^{\mu\nu}\overset{c}{\nabla}_{\mu}\overset{c}{\nabla}_{\nu}\phi$
when in action on the scalar field. Here we can forget about the general
coefficient dimension-dependent which was present in (\ref{eq: Wunsch6der}).
In special dimension $d=4$ the Bach tensor is itself conformally
covariant with the weight $w\left(B_{\mu\nu}\right)=-2$ according
to (\ref{eq: bachconftr}). However, the operatorial action of $B^{\mu\nu}\overset{c}{\nabla}_{\mu}\overset{c}{\nabla}_{\nu}$
on a scalar field with the weight $w_{0}=1$, brings another
conformally covariant expression $D_{{\rm Bach}}\phi$, when the coefficient
in front can be here completely arbitrary. The resulting conformal
weight of this expression is $w\left(B^{\mu\nu}\overset{c}{\nabla}_{\mu}\overset{c}{\nabla}_{\nu}\phi\right)=-5$.
In general, in this dimension we can construct a general expression
with the same resulting weight $w=-5$, which is an arbitrary
linear combination of $D_{{\rm Bach}}$, $D_{1}$ and $D_{2}$ operators
acting on a scalar field $\phi$. According to the definitions in
(\ref{eq: D1def}) and (\ref{eq: D2def}), in $d=4$ the form of these
operators simplify to
\begin{equation}
D_{1}=\overset{c}{\nabla}_{\mu}\left[\left(C^{\mu}{}_{\rho\sigma\kappa}C^{\nu\rho\sigma\kappa}-\frac{1}{4}g^{\mu\nu}C_{\rho\sigma\kappa\lambda}C^{\rho\sigma\kappa\lambda}\right)\overset{c}{\nabla}_{\nu}\right]
\end{equation}
and
\begin{equation}
D_{2}=-12\overset{c}{\nabla}_{\mu}\left[C^{\rho\sigma\kappa\lambda}C_{\rho\sigma\kappa\lambda}\overset{c}{\nabla}{}^{\mu}\right]-2\overset{c}{\square}\left(C^{\rho\sigma\kappa\lambda}C_{\rho\sigma\kappa\lambda}\right).
\end{equation}
This is exactly the same three-parameter freedom that we find with
arbitrary real values of $\alpha_{4}$, $\alpha_{5}$ and $\alpha_{6}$
for the combination of different orders of derivatives in the operator
${\cal O}$ in (\ref{eq: explopero}). The explicit form of this operator
${\cal O}$ (with the three real parameters $\alpha_{4}$, $\alpha_{5}$
and $\alpha_{6}$) can be presented as 
\[
{\cal O}=\frac{1}{4}\left(-3\alpha_{4}-\alpha_{5}-2\alpha_{6}\right)\overset{c}{\nabla}_{\mu}\overset{c}{\nabla}_{\nu}\overset{c}{\nabla}{}^{\mu}\overset{c}{\nabla}{}^{\nu}\overset{c}{\square}+\frac{1}{4}\left(-\alpha_{4}-3\alpha_{5}-2\alpha_{6}\right)\overset{c}{\nabla}_{\mu}\overset{c}{\square}\overset{c}{\nabla}{}^{\mu}\overset{c}{\square}+\alpha_{4}\overset{c}{\nabla}_{\mu}\overset{c}{\nabla}_{\nu}\overset{c}{\nabla}_{\rho}\overset{c}{\nabla}{}^{\mu}\overset{c}{\nabla}{}^{\nu}\overset{c}{\nabla}{}^{\rho}+
\]
\begin{equation}
+\alpha_{5}\overset{c}{\nabla}_{\mu}\overset{c}{\nabla}_{\nu}\overset{c}{\nabla}_{\rho}\overset{c}{\nabla}{}^{\nu}\overset{c}{\nabla}{}^{\mu}\overset{c}{\nabla}{}^{\rho}+\alpha_{6}\overset{c}{\nabla}_{\mu}\overset{c}{\nabla}_{\nu}\overset{c}{\square}\overset{c}{\nabla}{}^{\mu}\overset{c}{\nabla}{}^{\nu}.
\end{equation}
Here we just can express it or as different order of conformal derivatives
or by an expression explicitly with conformal curvatures (Weyl tensors
or Bach tensors) and conformal derivatives on them. We just have to
remember that in $d=4$ it is impossible to have the operator with
the leading term $\overset{c}{\square}{}^{3}$ and all our operators
in whatever form of writing them start with terms linear in conformal
curvatures.

Finally, we comment that in dimension $d=2$ the operator ${\cal O}$
cannot be defined as the results of the known and previously discussed
problems with the definition of Schouten tensor in this dimension.
So in all even dimensions $d=2$ and $d=4$ smaller than the critical
we have the impossibility to construct such operators with six derivatives
according to a conjecture that such constructions are impossible in
all even dimension. In the critical dimension in turn we see the problem
with the conformally flat spacetime limit of the operator constructed
only as a combination of different orders of conformal derivatives.
For all dimensions -- all odd and even ones above the critical one,
we do not see any problem with such a construction of the generalization
of the Hamada operator.

We come back now to the discussion of the situation within the critical
dimension $d=6$ for a generalized Hamada operator. From our previous
discussion we understand now that the minimal operator $D_{(6)}^{\min}$
as proposed by Wunsch cannot be reproduced as results of commutation
of different orders of conformal derivatives, from a general form
of the operator ${\cal O}$ in (\ref{eq: explopero}). Its explicit
form in $d=6$ reads
\begin{equation}
D_{(6)}^{\min}=\overset{c}{\square}{}^{3}+\frac{8}{3}B^{\mu\nu}\overset{c}{\nabla}_{\mu}\overset{c}{\nabla}_{\nu}.\label{eq: Dmind6}
\end{equation}
We see that also in this case we can only reproduce a linear combination
of $D_{1}$ and $D_{2}$ as acting on an conformally inert scalar
field $\phi$ (with the vanishing weight since this is a situation
in critical dimension). Therefore the operator ${\cal O}$ given explicitly
by (where $\alpha_{2},\alpha_{6}\in\mathbb{R}$)
\[
{\cal O}=\overset{c}{\square}{}^{3}+\alpha_{2}\overset{c}{\nabla}_{\mu}\overset{c}{\nabla}_{\nu}\overset{c}{\nabla}{}^{\mu}\overset{c}{\nabla}{}^{\nu}\overset{c}{\square}+\left(-3-\alpha_{2}\right)\overset{c}{\nabla}_{\mu}\overset{c}{\square}\overset{c}{\nabla}{}^{\mu}\overset{c}{\square}+\left(-\frac{4}{3}-\frac{4}{3}\alpha_{2}\right)\overset{c}{\nabla}_{\mu}\overset{c}{\nabla}_{\nu}\overset{c}{\nabla}_{\rho}\overset{c}{\nabla}{}^{\mu}\overset{c}{\nabla}{}^{\nu}\overset{c}{\nabla}{}^{\rho}+
\]
\begin{equation}
+\left(\frac{10}{3}+\frac{4}{3}\alpha_{2}-\alpha_{6}\right)\overset{c}{\nabla}_{\mu}\overset{c}{\nabla}_{\nu}\overset{c}{\nabla}_{\rho}\overset{c}{\nabla}{}^{\nu}\overset{c}{\nabla}{}^{\mu}\overset{c}{\nabla}{}^{\rho}+\alpha_{6}\overset{c}{\nabla}_{\mu}\overset{c}{\nabla}_{\nu}\overset{c}{\square}\overset{c}{\nabla}{}^{\mu}\overset{c}{\nabla}{}^{\nu}
\end{equation}
corresponds to a two-parameter freedom of defining the operator as
a linear combination of $D_{1}$ and $D_{2}$. Again we find for consistency
the same degree of freedom. In $d=6,$ we have the following simplification
for the explicit form of the operators $D_{1}$ and $D_{2}$, according
to (\ref{eq: D1def}) and (\ref{eq: D2def}):
\begin{equation}
D_{1}=\overset{c}{\nabla}_{\mu}\left[\left(C^{\mu}{}_{\rho\sigma\kappa}C^{\nu\rho\sigma\kappa}-\frac{1}{4}g^{\mu\nu}C_{\rho\sigma\kappa\lambda}C^{\rho\sigma\kappa\lambda}\right)\overset{c}{\nabla}_{\nu}\right]
\end{equation}
and
\begin{equation}
D_{2}=-8\overset{c}{\nabla}_{\mu}\left[C^{\rho\sigma\kappa\lambda}C_{\rho\sigma\kappa\lambda}\overset{c}{\nabla}{}^{\mu}\right].
\end{equation}

One can understand that in a special dimension $d=6$ we cannot produce
the operator (\ref{eq: Dmind6}), specifically its part with Bach
tensor, as the result of commutation of conformal derivatives. This
is a particular realization of the fact that not all expressions with
curvatures can be reproduced as commutators of covariant derivatives.
Another example is the square of Weyl tensor $C^{2}=C^{\rho\sigma\kappa\lambda}C_{\rho\sigma\kappa\lambda}$
acting by scalar multiplication on some sequence of conformal derivatives,
like for example in the expression $C^{2}\overset{c}{\square}\phi$.
Such expressions we indeed find in the definitions of the operators
$D_{1}$ and $D_{2}$ in (\ref{eq: D1def}) and (\ref{eq: D2def}).
However, there we see that covariant derivatives in some terms after
Leibniz expansion act on these invariants like $C^{2}$ and hence
lead to a splitting of the contraction of all indices and some curvature
tensors there can be really ``uncommmuted''. In the
sole expression like $C^{2}\overset{c}{\square}\phi$ again, curvatures
cannot be ever written as a square of the commutator since such operation
of commutation of conformal derivatives never leads to a full contraction
of indices between two commutators. Some indices have to be internally
contracted between the commutator and the tensor on which it acts.
(For example in GR we write $\left[\nabla_{\mu},\nabla_{\nu}\right]v_{\rho}=R_{\mu\nu\rho}{}^{\sigma}v_{\sigma}$
and here $\sigma$ is this internally contracted index and $v_\rho$ is arbitrary covariant vector field.) Therefore,
it is not a surprise that some terms in the expansion of the Hamada
operator cannot be expressed as the commutator of conformal derivatives,
so in this case the ``uncommutation'' procedure is unsuccessful,
like for example for the full $D_{(6)}^{\min}$ operator. One must
remember that in general curvature tensors are objects in differential
geometry that live their own life, independently how they were defined.
For example, they can be introduced as the result of some variational
derivative, when we effectively differentiate with respect to a tensor
on which they act upon. Then the curvature is properly defined and
we can abstract and forget on which particular tensor in differential
geometry this commutator was evaluated. The idea of one expression
with a different order of derivatives is that in such expression the
operator acts on precisely on one tensor on the right. However, when
we take a square of the curvature, for example like $C^{2}$ term,
then if we would like to write them back all these curvatures as resulting
from commutations of derivatives, then we would have to take variational
derivatives two times and with respect to different arguments. But
in the original expression there was precisely one tensor as the argument.

In general dimension, it is however important that we find the same
degree of freedom in number of free parameters needed to define the
non-endomorphic parts of the six-derivative operator. No matter, whether
we use explicitly Wunsch construction with $D_{(6)}^{\min}$, $D_{1}$
and $D_{2}$ operators or the operator ${\cal O}$ with arbitrary
real parameters ($\alpha_{2}$ and $\alpha_{4}$ for example). We
have all the time two-parameter freedoms in construction of the derivative
part of the conformal operators. By analysis of various solutions
of the system (\ref{eq: sysfirst}) to (\ref{eq: syslast}) in arbitrary
dimension $d$ for different three dependent coefficients, we find
that even if we completely simplify the choice of the independent
parameters such that $\alpha_{1}=1$ and two free real parameters
are set to zero, therefore in this sense looking for some minimal
operator, the combination in the operator ${\cal O}$ always contains
four terms. And this number of different orders for conformal derivatives
cannot be reduced further. In general, the non-endomorphic parts of
the generalized Hamada operator, can have $4,$ $5$ or $6$ terms
with different orders of conformal derivatives. Therefore the result
of ``uncommutation'' technique produces generally more
terms and does not lead to a direct one simple linear combination
of orders of six derivatives that would give a tentative answer for
``minimal'' conformal operator. It is generally believed
that commuting the derivatives reduces the number of terms and this
is true, for example, for the construction of $D_{(6)}^{\min}$ operator
in (\ref{eq: Wunsch6der}), which is built only with two terms, with
only term of curvature corrections. With different orders of all six
conformal derivatives we in principle, need more terms to write and
construct the conformal operator even in its shortest form.

With the procedure of studying of various order of conformal derivatives
outlined above, the search for such combinations also for higher number
of derivatives, higher than $2N=6$, is possible, since this is now
revealed as only an algebraic task of solving a big system of linear
equations. However, we envisage that the number of terms and of possible
orders of conformal derivatives grows quite fast with the number $2N$
of derivatives (rough estimate is that it grows as $(2N-1)!!$, when
we do not take into account any type of special identity discussed
at the beginning of this section). Hence, the size and the complexity
of the matrix of linear equations that has to be diagonalized here
is also big for even relatively small $N$ and this is left as the
task only for symbolic and numerical algorithms on computers. We believe
that even ``minimal'' form of conformal operators for
higher $N$ is not so short since they may contain many necessary
curvature correction terms, or many necessary different orders of
a sequence of $2N$ conformal derivatives.

\section{Discussion}
\label{s8}

One could also perform a comparison of three methods for finding the
explicit expression for the generalized Hamada operator and their
algebraic complicacy. In all these methods we need to solve some big
system of algebraic linear equations, hence the important signal of
the effectiveness of the methods (and new methods that we have proposed
in this paper) would be the size of the algebraic system involved.
As we remarked earlier there does not exist any signigicant hope that
the expression for the generalized Hamada operator can be found without
solving such system of equations for linear coefficients of terms
and as a result of using of some new powerful conformal covariant
derivative or the conformally covariant box operator.

Here we add another comment that we find useful regarding the construction of conformal operators and why they have to be built out with a linear combination of some terms and not just one single term of a very special structure (with conformal tensors and with conformal derivatives $\overset{c}{\nabla}$). Actually, it is difficult to imagine a construction of a single term that for each consequent dimension the operator is singular, i.e. that for 6-derivatives it is singular in $d=4$, so it contains the overall factor $\frac{1}{d-4}$, or for 8-derivative one the factor $\frac{1}{d-6}$, or for 10-derivative the factor $\frac{1}{d-8}$, etc. These singularities are in accord with our findings in section \ref{s5}. Therefore, for example, the initial construction of \cite{tractor1,tractor2} that is $\overset{c}{\nabla}_{i_{1}}\overset{c}{\nabla}_{i_{2}}\cdots\overset{c}{\nabla}_{i_{n}}\square\overset{c}{\nabla}^{i_{n}}\cdots\overset{c}{\nabla}^{i_{2}}\overset{c}{\nabla}^{i_{1}}$ does not seem to work as for the candidate of a good conformal operator at the end. Since terms here in commutation can have only $\frac{1}{d-3}$ denominators, other denominators cannot be created just by commutation of conformal derivatives. The operator should be somehow power or a sequence (of other more simpler operators), but it is difficult to understand how these singularities could arise in such powers, if they were not present for smaller number of derivatives, and how to build up $\frac{1}{d-8}$ denominator, for example. Only algebraic combination with other terms with conformal curvatures and other combinations with conformal derivatives $\overset{c}{\nabla}$ could solve this issue via linear system of equations where the denominator of the type like $\frac{1}{d-8}$ could be generated as numerical solutions to it. This explains that we still need to use linear combinations of perfectly conformally-looking terms for construction of conformal operators that transform completely well and we cannot hope to just use some single operator written in a smart way as just one single term.

In the first method, when we use the basis of all terms that can be
used in a construction of generally covariant scalar invariants for
the expression $D_{(6)}\phi$, we use all three types of possible
curvature tensors ($R$, $R_{\mu\nu}$ and $R_{\mu\nu\rho\sigma}$)
and normal GR-covariant derivatives acting on them (so we use operators
$\nabla_{\mu}$ or $\square$). This basis of writing all possible
terms in $D_{(6)}\phi$ has many elements as we have seen in previous
chapters. The reason is that all these terms are not designed for
having good conformal transformation properties. They are just all
possible scalar GR-invariants. Hence they are not specially adjusted
for the sake of finding some conformal invariant operator like $D_{(6)}$.
The Riemann tensor, Ricci tensor and Ricci scalar do not have simple
conformal transformation laws (even in the infinitesimal form it contains
up to second derivatives of the $\Omega$ parameter). Had we used
here in the basis Weyl tensor, this task of curvature choice would
be simplified, since the last is conformal curvature tensor and it
is covariantly transforming under conformal transformations. Moreover,
the covariant derivatives $\nabla_{\mu}$ and $\square$ are not selected
in any way to facilitate the conformal transformation laws of the
object on which they act. Again, these differential operators do not
transform in any way that would be in line with conformal transformations. 

In conclusion, this basis of writing terms is not well prepared from
the point of view of looking for conformally invariant expressions
and conformally covariant operators like $D_{(6)}$. This is also
why it has many terms. This implies that we have to solve the system
of algebraic equations for many coefficients of terms in a general
combination giving at the end conformally invariant expression. Moreover,
since we use the basis of writing terms which is only covariant from
the point of view of GR, the requirement for good conformal transformation
law is equivalent here to checking that the terms with first and all
higher level of derivatives on the $\Omega$ parameter must cancel
out in the combination for generalized conformal operator $D_{(2N)}\phi$.
In the case of Hamada operator we need to check up to the sixth derivative
of the $\Omega$ parameter and this is still on the infinitesimal
level, so all terms are linear in $\Omega$ or its derivative. Luckily
this also guarantees that the finite transformations will not generate
problem and the combination for the operator $D_{(6)}\phi$ will be
conformally covariant, if it is shown that on the infinitesimal level
the expression $D_{(6)}\phi$ transforms only up to the terms linear
in $\Omega$ and completely without derivatives of it. This again
implies that we have many equations: cancellation of all terms at
the order $\nabla\Omega$, cancellation of all terms at the order
$\nabla^{2}\Omega$, etc., so we have to have here many linear equations
with many unknowns and the system can be quite complicated.

When we used the second method with basis constructed only with conformal
curvature tensors (Weyl and Bach tensors and conformal derivatives
acting on them) and we exploited the conformal derivatives there $\overset{c}{\nabla}_{\mu}$
and $\overset{c}{\square}$, we generically have a much smaller basis
of all possible terms that we have to use to construct the conformal
operator $D_{(6)}$ when acting on the scalar field $\phi$. This
is because we use the adjusted conformal curvature tensors specially
built to facilitate conformal transformation laws and they are also
conformal tensors, meaning that they transform covariantly -- without
derivatives of the $\Omega$ parameter. The same arguments also applies
to usage of conformally covariant derivatives like $\overset{c}{\nabla}_{\mu}$
and $\overset{c}{\square}$ since with them the law of transformation
of any object is only up to the first derivatives of the $\Omega$
parameter. Therefore, here we expect to have less number of unknown
coefficients in the combinations giving conformal operators. (For
example, in the case of the minimal operator in (\ref{eq: Wunsch6der})
we needed to find only one relative coefficient in front of the Bach
tensor.) Finally, since the usage of conformal curvature tensors and
other basic conformal objects (like the scalar field with the assigned
conformal weight $w_{0}$) and of conformally covariant Wunsch derivatives
creates other tensorial expressions that under conformal transformations
change only up to the first derivatives of the $\Omega$ parameter,
we at the end of the computation need to check only if terms linear
in $\nabla\Omega$ do cancel out in our proposed combination for conformal
operators. Other higher derivatives like $\nabla^{2}\Omega$ never
arise in this formalism of conformal derivatives and other conformal
tensors. Hence we conclude that here generically we expect to have
less number of linear equations and also smaller number of unknown
coefficients, so algebraically speaking this system is more effective.
Of course, this is like that here since our building blocks are already
conformally prepared. This reduction of algebraic complication is
clearly visible when one compares the computations of sections \ref{s4} and
\ref{s6} and \ref{s7} of our paper.

With the advantages of the second method (with conformal curvatures
and conformal derivatives), there come also a few restrictions or
shortcomings. They are the most visible when instead of creating operators
like $D_{(6)}$ one considers conformally covariant Lagrangians of
the form $\phi D_{(2N)}\phi$, which after the proper densitization
give rise to conformally invariant Lagrangian densities $\sqrt{|g|}\phi D_{(2N)}\phi$,
which are at the basis for conformally invariant scalar actions. In
the case of the basis of general GR-covariant curvatures and GR-covariant
derivatives, on the level of action one could without problems use
the property of integration by parts. Moreover, for the standard Riemann
tensor with covariant derivatives we could use also the differential
Bianchi identities (possibly also in the contracted forms) to simplify
and reduce some scalar invariant terms in the action. These were the
advantages for GR-adapted basis. We remark that these nice properties
are not anymore present in the formalism with all conformal objects.
We cannot so simply integrate by parts conformal covariant derivatives
under the spacetime volume integrals, there are correction terms with
the Schouten tensor which on this side take us out of the conformal
formalism. Similarly, we cannot use any form of the conformal Bianchi
identities (with conformal derivatives on the conformal Weyl tensor)
to reduce terms further. Still we can use first Bianchi identities
(cyclicity) because they work equally well for the Weyl tensors and
also originally for Riemann tensors as in standard differential geometry. 

Therefore, roughly understanding the construction of invariant scalar
Lagrangian densities with conformal objects is hampered and one does
not have many options for reduction of terms, and one would conclude
that the number of possible different scalar invariants is bigger
than in the corresponding GR case. However, this is only a naive thinking
since all the examples show that the usage of conformal objects instead
of just GR-adapted ones is beneficial for construction of conformal
operators, tensors, expressions and conformally invariant actions
as well. Of course, the last sentence is a very natural conclusion:
for general conformal objects use basic conformal objects as their
building elements. One could compare the lost freedom of integration
by parts and differential Bianchi identity in the conformal case with
the additional complications that one inevitably meets in the GR basis.
In the last case, one also has to consider terms with the contractions
of Riemann tensor (Ricci tensor and Ricci scalar) since the trace
is not removed there from the basic curvature Riemann tensor. The
advantage here of the Weyl tensor in the conformal case is that it
is completely trace-free so we do not have to worry about these additional
contractions. Barring these differences the construction of actions
in conformal or GR case are exactly the same and they lead to roughly
the same structures (when in the conformal case we use conformal derivatives
instead of GR ones and conformal Weyl curvatures instead of the Riemann
ones). The estimate is that however, in the conformal case we can
create less number of terms and then this conformal basis for the
invariant actions is smaller and hence more effective. An example
here is the basis of all scalar invariant terms with six derivatives
and constructed exactly with three powers of curvatures. In general
dimension, we have $8$ terms in the GR basis, while in the conformal
one we have only $2$ terms. The effectiveness of using only the trace-free
objects is clearly seen here.

As a kind of point in the middle we can understand our new method
when we construct all possible terms in $D_{(2N)}$ as different orders
in a sequence of precisely $2N$ conformal covariant derivatives.
In this way we do not explicitly use conformal curvature tensors and
we only consider combinations of various orders of derivatives. As
explained above this method has some advantages and disadvantages.
From the point of view of counting number of unknown coefficients
and of linear algebraic equations to solve, again this method stays
in the middle between the case of very efficient conformal objects
(derivatives and curvatures) and the GR case. As seen in the explicit
example for the generalization of Hamada operator here we had $5$
(after normalization) unknown coefficients and effectively only two
equations for them giving us a remaining three-parameter freedom which
was totally consistent with the previous findings. We compare it with
the case of Wunsch minimal operator $D_{(6)}^{\min}$, where only
one relative coefficient needed to be determined from effectively
one linear equation. This minimal form of the operator could be seen
as a result of commutation of conformal derivatives and of various
ordering of terms there; this is possible in every dimension different
than the critical one for a given $2N$ number of derivatives. However,
by the analysis of solutions to the system in (\ref{eq: sysfirst}),
we see that it is impossible to write such a minimal term as a combination
with less than three different orders of conformal derivatives. Again,
the expression provided by Wunsch as minimal in (\ref{eq: Wunsch6der})
is the shortest in the number of conformal terms and then in this
case the algebraic system to solve is the smallest one with smallest
number of unknown coefficients to determine. That is why in that case
the method is the most effective.

Here we also discuss other generalizations of the conformally invariant actions with scalar fields.
When the scalar theory with the action functional 
\begin{equation}
S_{(2N)}=\!\int\! d^{d}x\sqrt{|g|}\phi D_{(2N)}\phi\label{eq: actiongend}
\end{equation}
is considered on flat spacetime, then the theory is free (but certainly
with higher derivatives -- $2N$ derivatives in the scalar propagator
from the leading term when all curvatures vanish) and hence there
are no interactions and no divergences generated at the quantum level.
This is because in this case we have that $D_{(2N)}=\square^{N}=(\partial^{2}){}^{N}$
in Cartesian coordinates. One could here analyze a slightly more general
case, when we add conformal interaction terms to the action with a
form
\begin{equation}
S_{{\rm int}}=\!\int\! d^{d}x\sqrt{|g|}\lambda\phi^{\alpha},\label{eq: Sintgend}
\end{equation}
where the exponent $\alpha=\frac{2d}{d-2N}=2+\frac{4N}{d-2N}$ can
be defined in any dimension different than a critical one $d_{{\rm crit}}=2N$.
In (\ref{eq: Sintgend}) $\lambda$ is a dimensionless parameter of
self-interaction of the scalar field. The action (\ref{eq: Sintgend})
or the total action (\ref{eq: actiongend})+(\ref{eq: Sintgend})
are both classically conformally invariant. This way one adds non-linearities
and classical interactions to the conformal action of the theory.
Above we have added non-linearities without derivatives, where only
generalized powers of the scalar field are present, however the more
complicated other ones are also still possible here and they could
potentially include derivatives and higher than quadratic monomials.
For example in $d=4$ and when $N=1$ this is a standard two-derivative
theory giving rise to classically scale-invariant Klein-Gordon model
on flat spacetime with $\phi^{4}$-type of scalar quartic self-interactions
-- therefore it is a conformal Higgs model. Of course, this has vast
applications to the conformal Standard Model of particle physics,
where one knows that only the classical phenomenon of spontaneous
symmetry breaking destroys the classical conformal symmetry due to
the mass of the Higgs particle. But as the zero-order approximation
(or for very high temperatures, or very early Universe) one can take
a conformal model as a very good and symmetric classical starting
point of further considerations.

\section{Conclusions}
\label{s9}

The six-derivative conformal scalar operator was originally found by Hamada in its critical dimension of spacetime, $d=6$.
We showed that it is possible to generalize this construction to arbitrary even dimensions of spacetime, provided that this is bigger than the critical dimensions $d_{\rm crit}=6$. We gave the explicit form of its coefficients of expansion in the basis of GR scalars (invariant terms with respect to diffeomorphisms). The critical case is the situation for the Hamada operator with six derivatives in six spacetime dimensions.
 As the second main part of this work, we also proved a general theorem that a scalar conformal operator with $n$ derivatives in $d=n-2$ dimensions is impossible to construct. One notices that the critical dimension of the operator equals to the number of derivatives present in the leading term $\square^n$ of the expansion in curvature terms. We can conjecture extending our theorem that for any even dimension $d<n=d_{\rm crit}$ such construction is impossible since the coefficients blow up. The operators in question can be constructed only in dimensions $d\geqslant d_{\rm crit}$, when also the conformal weight of the basic scalar field $w$ must be non-positive. This is an interesting pattern for the construction of various conformal operators that perhaps we can explore also for other vectors and tensors (higher rank representations of the Lorentz group) \cite{Erdmenger:1997wy}.

In this article, we showed different ways of how to generalize the
original construction of Hamada conformal operator when acting on
scalar fields. We started in the situation with the critical dimension
for this operator and since it must contain six derivatives in the
leading part, then this situation was in $d=6$ spacetime dimensions.
We extended the construction to arbitrary dimension $d$ but also
found that in dimensions $d=2$ and $d=4$ such construction is impossible
due to various singularities. Our form of the generalized Hamada operator
works naturally in spacetimes with Minkowskian signature as well as
for the completely Euclidean spaces useful for purely geometrical
applications of conformal symmetry. In the course of our article,
we first discussed the explicit and straightforward method of finding
coefficients in front of all terms containing up to six derivatives
and built with standard gravitational curvatures and standard GR-covariant
derivatives acting on them. This required solving some big system
of linear algebraic equations. Finally there we encountered the 3-parameter
freedom since the generalized form of Hamada operator is not unique
and there are various parameters or terms that could be added to its
``minimal'' form for free.

In the last sections of this article, we embarked on explaining and
applying the formalism of conformal covariant derivatives as constructed
by Wunsch. With this theoretical tool, we showed that the construction
can be a bit simplified for the case of six-derivative operator, while
it was completely clear for the cases of studies of \O{}rsted-Penrose
and Paneitz operator. In the last two simplifying cases, the usage
of conformal covariant derivatives $\overset{c}{\nabla}_{\mu}$ and
the conformally covariant box operator $\overset{c}{\square}$ give
directly answers for the conformal operators containing explicitly
2 or 4 derivatives in their leading term respectively. However, for
the case of six-derivative operator (and all higher) this construction
is not fully and completely successful since even the minimal operators
must also contain correction terms which are higher in conformal curvatures
(Weyl or Bach tensors). We also discussed the possible ambiguities
and freedom that we found for construction using these lines for generalization
of Hamada operator. By comparison with the results from the previous
sections, we found a full agreement numerically for the form of constructed
conformally covariant operators as well as for a qualitative description
of the level of freedom that we have in such constructions. Of course,
using both methods we found that the freedom for the generalized Hamada
conformal operator is contained in three arbitrary real parameters. 

Finally, we showed that using a combination of different orders of
six conformal covariant derivatives, one reproduces the same results
as before for the generalization of the conformal Hamada operator.
However, this last construction is not completely successful in the
critical dimension ($d_{{\rm crit}}=6$ for Hamada operator), but
works in any dimensions $d\neq2,4,6$. Therefore we provide some \emph{new}
results written in a \emph{new} and distinct way. We also showed
the roadmap how one can think about looking for generalization for
operators containing more than six derivatives in their leading terms.

We also explained using algebraic methods and arguments why the conformal
operator with six derivatives in its action must necessarily contain
curvature correction terms (except the case of special dimension $d=3$)
and why the coefficient of these additional correcting terms arise
only as a solution of system of algebraic equations and why never
they can be obtained from some other constructions using different
definition of conformal covariant derivative or conformal box operator.
For this last assertion, we used and studied the fact that the generalization
of the conformal Hamada operator cannot be defined in $d=4$ and traced
it back to the singularity of the explicit coefficients of curvature
correction terms which must contain in the denominators factors $(d-4)$.
This was of course, a special case of a general theorem, that we have
also proved here using algebraic and infinitesimal conformal transformation
methods, stating that the conformal operator with the leading term
containing precisely $2N$ derivatives cannot exist in spacetime dimension
$d=2N-2$.

Finally, we comment that the operators considered on the classical
level that we have studied in this paper, can bring a lot of applications.
First with them we can construct conformally invariant actions for
scalar field theories containing precisely $2N$ derivatives and in
any dimensions \cite{Osborn:2015rna}. These actions are necessarily quadratic in scalar
fields (so they give contributions only to the propagator on the quantum
level of quantum scalars), but we can easily also add here some interaction
terms (without derivatives, which examples of were given above, or
containing also derivatives). Such conformal actions then define good
scalar models, where the conformal symmetry is preserved and realized
explicitly on the classical level. (As it is expected and proved by
various quantum computations these pure scalars models (but coupled
to external gravitation) unfortunately cease to be conformal on the
quantum level of considerations.) Obviously, we can use them for the
definition of new conformal field theories, which are put on curved
spacetime backgrounds, where the conformal curvature does not vanish.
These in turn, may have new applications for the famous AdS/CFT correspondence
in the new framework where the boundary theory is placed, not on usual
flat background, but when it is characterized by some non-zero conformal
curvature tensors. 
From the mathematical point of view these new conformal operators are relevant for the problems of coupling of scalar fields to external geometry (in particular to conformal geometry). They are also essential for the study of 6-dimensional Yamabe problem, analysis of Huyghens principle and conformal wave equations as well as to the problems  of the analytic properties of conformal Laplacians operators (their generalizations and their powers), their determinants, logarithmic traces, spectra and positive eigenvalues. Some of the mentioned topics are discussed in \cite{branson0,branson1,branson3,branson4,branson5,branson6,fegan} and in the further literature cited there.

On the other hand, we can also mention applications
to recent studies of conformal anomaly induced action, where these
examples of generalized Paneitz and Hamada operators play significant
roles \cite{Ferreira:2015lna,Ferreira:2017wqz,Ferreira:2018utt}. These actions succeed in trying to capture all important non-local
quantum effects directly related to the conformal anomaly (and therefore
neglecting other less important effects in the realm of all full quantum
corrections \cite{Barvinsky:2023aae,Camargo:2022gcw,Asorey:2022ebz,Gusynin:1986kz}). In recent studies they were for example applied to study quantum
gravitational perturbative corrections in cosmology \cite{CesareSilva:2020ihf}. As different
from standard physical applications, the methods of conformal symmetry
in Euclidean space are also quite powerful and they are already exploited
in theoretical considerations in computer graphics (CG). 

Eventually, we can discuss the possible future applications of the found form of the Hamada operators in any dimensions $d$ of space(spacetime). Generally such scalar conformal operators with six derivatives constitute a basis for CFT algebras of primary scalar operators coupled to external geometry (so on non-trivial background spacetimes) hence they will give rise to new CFT's coupled to gravity. Such operators may have long ranging applications from studies of conformal anomalies (i.e. integrating various nonlocal terms like in \cite{Ferreira:2015lna,Ferreira:2017wqz,Ferreira:2018utt}), through applications of scalar fields in cosmology. There could be applications also for CFT's: like the description near fixed points (FP) of RG flows of theories with six-derivative scalars coupled to background geometry, so related to asymptotic safety program in gravity (where there is a conformal symmetry in the vicinity of FP). One could also think more phenomenologically of extended Higgs models, where the scalar fields come with higher derivatives and are conformally coupled to the background curved spacetime geometries. Therefore, these scalar operators are very important in theoretical high energy physics. Some obvious applications can be probably found in condensed matter physics, in the situations, where there are some fundamental scalars around like for example for an excitation of graphene curved sheets varying in time ($d=3$ case) \cite{Iorio:2010pv,Iorio:2012xg,Iorio:2013ifa}.

All these
present (and also hopefully expanded in future) motivations ensure
us that the studies that we have undertaken in this paper about the
generalization of the Hamada conformal operator are important both
from the mathematical as well as physical points of view.

\section*{Acknowledgements}

L.R. wishes to thank the Department of Physics, Federal University of Juiz de Fora for kind hospitality. P.R.B.R.V. is grateful to CAPES for supporting his Ph.D. projects.

\end{document}